\documentclass[smallextended,twoside]{article}   
\usepackage[margin=2cm]{geometry}
\pdfoutput=1 
\usepackage[hidelinks]{hyperref}
\usepackage{graphicx, amsmath, amssymb, amsfonts}
\usepackage[round]{natbib}
\usepackage{authblk}
\usepackage{fancyhdr}
\newcommand{\degree}{\ensuremath{^\circ}}
\providecommand{\keywords}[1]{\textbf{Keywords} #1}

\begin{document}

	\pagestyle{fancy}
	\fancyhead{}
	\fancyhf[cf]{}
	\fancyhead[LE,RO]{\small{\thepage}}
	\fancyhead[RE]{\small{Creech et al}}
	\fancyhead[LO]{\small{Simulations of an offshore windfarm (postprint)}}
	\fancyfoot{}

	\title{\textbf{Simulations of an offshore wind farm using \\ large eddy simulation and a torque-controlled \\ actuator disc model (postprint)}}

	\author[1]{Angus Creech%
		\thanks{Email: a.creech@ed.ac.uk (corresponding author)}}
	\author[2]{Wolf-Gerrit Fr\"uh}
	\author[3]{A. Eoghan Maguire}
	
	\affil[1]{Institute of Energy Systems, University of Edinburgh, Scotland}
	\affil[2]{School of Engineering and Physical Sciences, Heriot-Watt University, Scotland}
	\affil[3]{Vattenfall Wind Power, Edinburgh, Scotland}
	
	\date{12\textsuperscript{th} December 2014}

	\maketitle
	\thispagestyle{empty}

\begin{abstract}
We present here a computational fluid dynamics (CFD) simulation of Lillgrund offshore wind farm, which is located in the {\O}resund Strait between Sweden and Denmark. The simulation combines a dynamic representation of wind turbines embedded within a Large-Eddy Simulation CFD solver, and uses hr-adaptive meshing to increase or decrease mesh resolution where required. This allows the resolution of both large scale flow structures around the wind farm, and the local flow conditions at individual turbines; consequently, the response of each turbine to local conditions can be modelled, as well as the resulting evolution of the turbine wakes.  This paper provides a detailed description of the turbine model which simulates the interaction between the wind, the turbine rotors, and the turbine generators by calculating the forces on the rotor, the body forces on the air, and instantaneous power output. This model was used to investigate a selection of key wind speeds and directions, investigating cases where a row of turbines would be fully aligned with the wind or at specific angles to the wind.  Results shown here include presentations of the spin-up of turbines, the observation of eddies moving through the turbine array, meandering turbine wakes, and an extensive wind farm wake several kilometres in length. The key measurement available for cross-validation with operational wind farm data is the power output from the individual turbines,  where the effect of unsteady turbine wakes on the performance of downstream turbines was a main point of interest. The results from the simulations were compared to performance measurements from the real wind farm to provide a firm quantitative validation of this methodology. Having achieved good agreement between the model results and actual wind farm measurements, the potential of the methodology to provide a tool for further investigations of engineering and atmospheric science problems is outlined.


\end{abstract}

\section{Introduction}\label{S:Intro}

\subsection{Background}
\label{s:introduction-background}

Substantial offshore wind farms with many tens of turbines over 100~m tall are being built at an increasing pace, which leads to a number of challenging and interesting problems for engineering and atmospheric sciences, as much as for the electricity industry.   In this article, we will investigate some of these by comparing operational data from Lillgrund offshore wind farm with a computational model of that wind farm.   Lillgrund wind farm consists of 48 turbines, each with a rated power output of 2.3~MW, in a compact array in the waters between Denmark and Sweden just south of the {\"O}resund bridge.  

Modern offshore wind turbines often have a rotor diameter in excess of 100~m, sampling the wind from typically 50~m to 150~m above the sea surface.  They are therefore sampling a dynamically active part of the turbulent planetary boundary layer with a typical wind shear profile of the mean wind increasing with height, as well as turbulent eddies of length scales comparable with the turbine rotor blades, and time scales including that of the typical inertial time scale of the rotor of a few seconds.  For these reasons, considerable research is being carried out to characterise and understand the turbulence structures, the transport phenomena in the boundary layer, and their interactions with the turbines~\citep{Abkar:2013qa,Banta:2013mi,Kalvig:2014xq,Rajewski:2013ee}. These are of great importance to the design and performance of wind turbines.

Conversely, while a single wind turbine would only affect the atmosphere locally in the form of a wake decaying over the length scale of around ten rotor diameters, the cumulative effect of a whole wind farm on the atmosphere is much greater.  For example, the effect on vertical mixing through the turbulence generation by the rotor blades can lead to warming near the surface in stable atmospheric conditions, and cooling in unstable conditions~\citep{Roy:2004dz,FitchLund2013}.  Satellite and airborne observations of winds in the lee of wind farms suggest that wind farm wakes modify the atmospheric flow for many tens of kilometres downstream of the turbine array~\citep{Christiansen2005,Christiansen2006,Hasager:2008dk}.   The effect of wind farms is not only noticeable behind the turbine array but also above, as the wind farm induces its own developing boundary layer~\citep{Wu:2013gb} with significant upwelling observed at heights well above the turbines. Even flow in the upper layer of the oceans is reported to be affected by large offshore wind farms~\citep{Brostrom2008585}.

Both the horizontal and vertical scales of large wind farms have increased to the point that their presence can be expected to affect weather and climate~\citep{Keith:2004la,Wang:2011rw}, and should therefore be included in climate models through a suitable parameterisation.  While early parameterisation approaches were based on modifying the surface roughness \citep{Barrie:2010kl,Ivanova:1998fy,WangPrinn2010,Wang:2011rw},  \citet{fitch2013} demonstrated that those approaches lead to a very different result when compared to a parameterisation which models the wind farm as a momentum sink not at the surface, but at the rotor height.  Momentum and heat fluxes were significantly affected throughout the depth of the planetary boundary layer and at length scales of 100~km.  This demonstrates that the momentum exchange and turbulent energy production within the wind farm must be well understood, to develop wind farm parametrisation schemes of wind farms for NWP and climate prediction.

With wind farms easily reaching installed capacities of hundreds of megawatts, the reliable estimations of their electricity production is becoming increasingly important for the electricity industry. A key factor affecting the performance is that turbines in the array may lie in path of the wakes generated by others, whereby they experience substantially lower wind speeds than their upwind neighbours~\citep{BPF10}. The result of this is that the farm as a whole produces less electricity than the same turbines would in isolation.  

\begin{figure}
\begin{centering}
\includegraphics[width = 0.55\columnwidth]{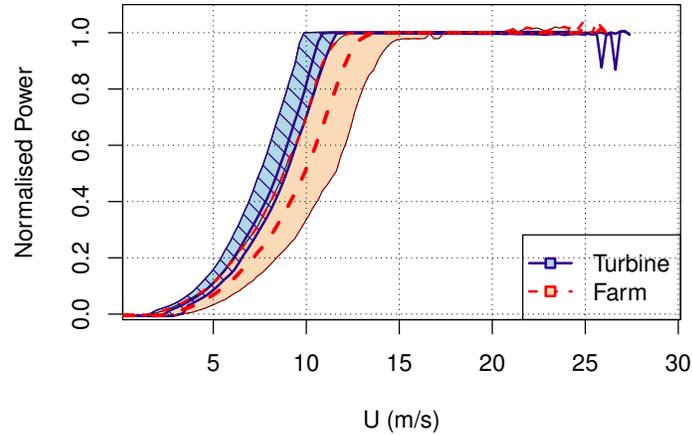} 
	\caption{Power coefficients for a wind turbine (blue line and shaded area with diagonal hatching) and for the entire wind farm (red line and plain shading). The lines denote the median of the observed output within a $0.25$ m/s wind speed window and the extent of the shaded regions denotes the 90\% of the observations (for details see section \ref{S:Lillgrund:FarmPerf}). }
\label{fig:PerformanceCurve}
\end{centering}
\end{figure}

The wind farm effect is easily illustrated by comparing the power output from the entire wind farm investigated here with that from a single turbine in the front row.  The blue shaded area in Figure~\ref{fig:PerformanceCurve} shows the power coefficient, that is the power output divided by the rated power, from the front turbine against the wind speed measured from the anemometer on that turbine's anemometer.  This shows the typical features of power generation starting at a `cut-in' wind speed of around 3~m/s, increasing with approximately $\propto U$ until the rated power is reached at the rated wind speed of around 11~m/s, above which the power output remains constant until the `cut-out' wind speed of around 25~m/s, at which point the turbine is switched off for safety reasons.  Compared with that is the total power output from all normally operating turbines in that wind farm, at the same reference wind speed measured at the front turbine.
The important point here is that the farm's power coefficient is significantly suppressed when compared to that of the front turbines, where the 90\%-  ranges do not overlap over the entire range below the rated wind speed. Only when hub height wind speeds exceed 15 m/s does the wind farm reach its full potential.  Whilst Lillgrund is an extreme case due to its turbines being closely spaced, it nevertheless highlights the issue, and the resulting need for being able to predict the wakes and wind farm performance in the planning of offshore wind farms. 

A great deal of research has also focussed upon modelling and parameterising wind turbines. Common approaches to modelling wind turbines use linear wake theory, such as Jensen's Park model \citep{BPF10,Ainslie88,Jensen83}, and it is recognised that the simple wake models lose accuracy when applied to multiple wakes interacting. Recent research has combined simple turbine models with computational fluid dynamics, with turbines often represented as simple porous discs~\citep{Espana2011}, actuator discs \citep{Creech2009}, actuator lines~\citep{Churchfield2012, Machefaux2012} or actuator surfaces~\citep{shen2007}. These can be embedded in RANS fluids solvers~\citep{cabezon2011}, pseudo-spectral solvers~\citep{calaf2010, wu2011}, fixed-mesh LES finite difference~\citep{jimenez2008} and finite volume codes~\citep{Churchfield2012}, or an LES finite element solver with an unstructured, hr-adaptive mesh~\citep{CFC11}. 

It should be mentioned that RANS and LES represent alternate approaches to the problem of modelling turbulence, and that each has its own benefits and shortcomings. In RANS, any temporal fluctuations in the fluid velocity are represented by an additional viscous term called the `eddy viscosity'. In LES the turbulence is treated explicitly, except for turbulent eddies smaller than the grid size of the CFD simulation, which are modelled as `sub-grid eddy viscosity'. The main advantage of RANS is that it is computationally inexpensive and capable of being run on desktop computers; however, details of temporal fluctuations in the flow are lost, since they are treated implicitly. On the other hand, LES provides a greater level of fidelity by preserving both temporal and spatial fluctuations on the flow, to grid resolution level; it is also much more computationally intensive, and can require supercomputer-scale resources. One option here is the use of  hr-adaptivity to reduce these demands. This meshing strategy can both move the computational meshes (r-adaptivity) and/or change the local mesh resolution (h-adaptivity) to minimise error in the solution, but also allows the mesh to track unsteady flow features \citep{piggott2004}. For a more detailed overview on RANS, LES, and their use within wind turbine modelling, see \citet{creechfruh2014}. 

Presently, detailed wind turbine and wind farm models are limited to a restricted domain around the turbines while the interaction between wind farms and the environment require much larger domains.  Turbine scales are on the order of hundreds of metres in the horizontal and 100 to 200 m in the vertical, which extends to a few kilometres in the horizontal for wind farms.  Yet atmospheric models need to resolve the planetary boundary layer of depth up to a kilometre, and tens to hundreds of kilometres in the horizontal. While one approach would be to link the two scales through nested models, computational resources are beginning to allow domain sizes in a single model which are substantially larger than the wind farm alone. This moves towards a situation where a full wind farm could be modelled in a domain, which eventually will be able to include the planetary boundary layer and a horizontal extent to investigate the wind farm wake.  This study presents the methodology aimed at this.  Given the computing resources available at the time, this study demonstrates the approach in a model which will lead to the full vertical and horizontal extent needed for the full planetary boundary layer and full wake farm.

\subsection{Aims and outline}
With the aim of demonstrating and validating time-dependent wind farm modelling, this study provides a detailed analysis of the observed wind farm performance, together with a high-resolution computational model of the wind farm for a selection of key wind conditions. This begins with section~\ref{S:Lillgrund}, which introduces Lillgrund wind farm.  Sections~\ref{S:Method} to \ref{s:full-farm} introduce the modelling approach and implementation, starting with the overall modelling methodology in section \ref{S:Method}, which describes in detail how the turbines and their response to the wind are represented. Section~\ref{S:turbine-parameterisations} describes how the model was configured for the Lillgrund turbines.  Section~\ref{S:Empty} details the modelling of the domain without turbines, used to produce a realistic background flow structure and then, in section \ref{s:full-farm}, to the full domain with the wind turbines positioned for different wind direction to simulate key wind conditions as identified from the results in section \ref{S:Lillgrund}.  The results from the CFD model and the corresponding performance data from the SCADA record are described separately in section~\ref{S:Results}, which is then followed by a comparison and validation in section~\ref{S:Validation}. To conclude, some of the findings and issues are discussed and summarised in sections~\ref{S:Discussion} and~\ref{S:Conclusion}, respectively.

\section{Lillgrund Wind Farm} \label{S:Lillgrund}
\subsection{Description of wind farm}

\begin{figure}
\centering
  \includegraphics[width=0.75\linewidth]{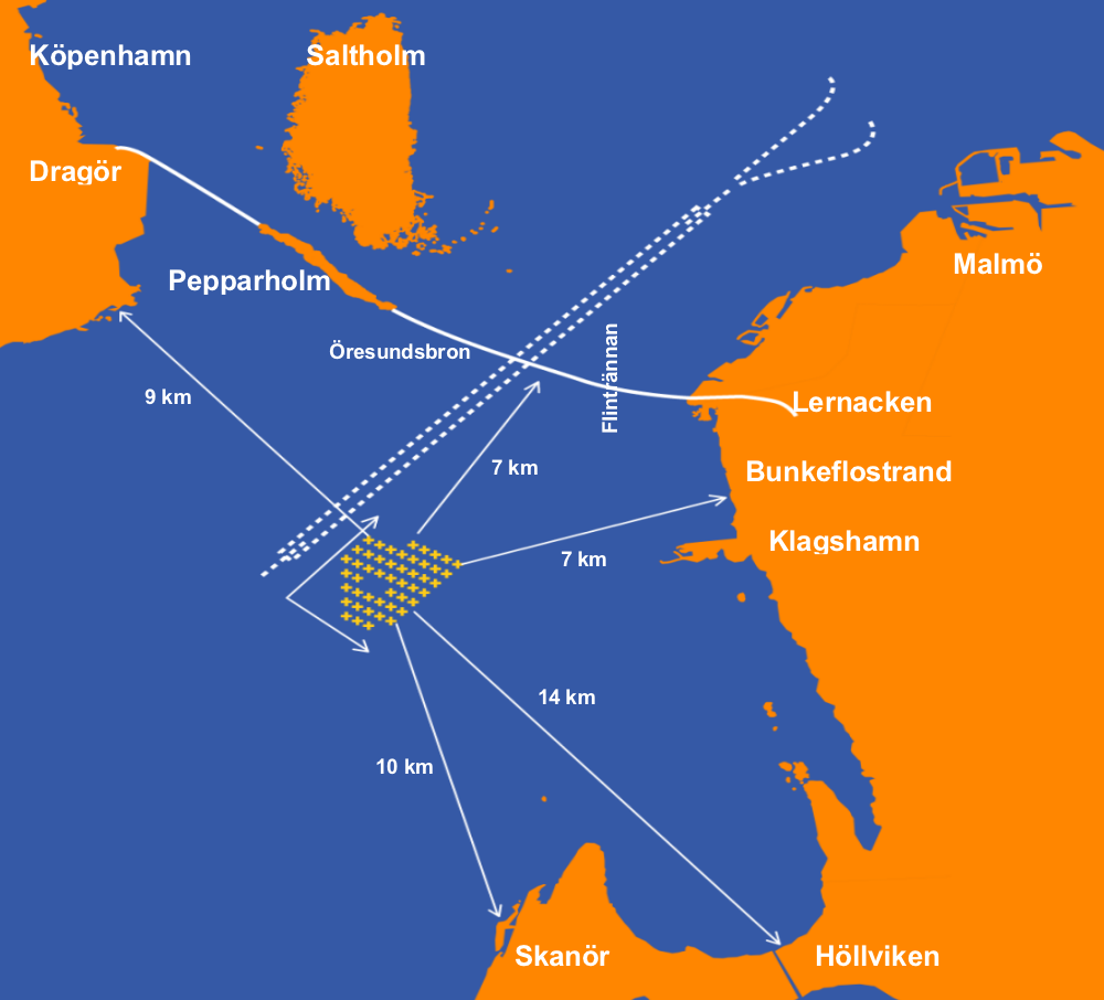} 
	\caption{Location of Lillgrund wind farm. Sweden is to the right, and Denmark to the upper left. Courtesy of \citet{Lillgrund2}.}
	\label{fig:lillgrund-location}
\end{figure}

Lillgrund offshore wind farm is located 7~km south of the {\"O}resund bridge between Copenhagen in Denmark and Malm\"o in Sweden, as shown in Figure~\ref{fig:lillgrund-location} (55\degree 31' N, 12\degree 47' E).  While it sits in a region fairly well enclosed by land, the prevailing south-westerly wind coincides  with the longest wind fetch of between 25~km and 50~km, and the effects of land topography on air flow can be ignored.  It has been operated by Vattenfall Vindkraft AB since December 2007 \citep{Lillgrund2}. 

\begin{figure}
  \centering
  \includegraphics[width=.90\linewidth]{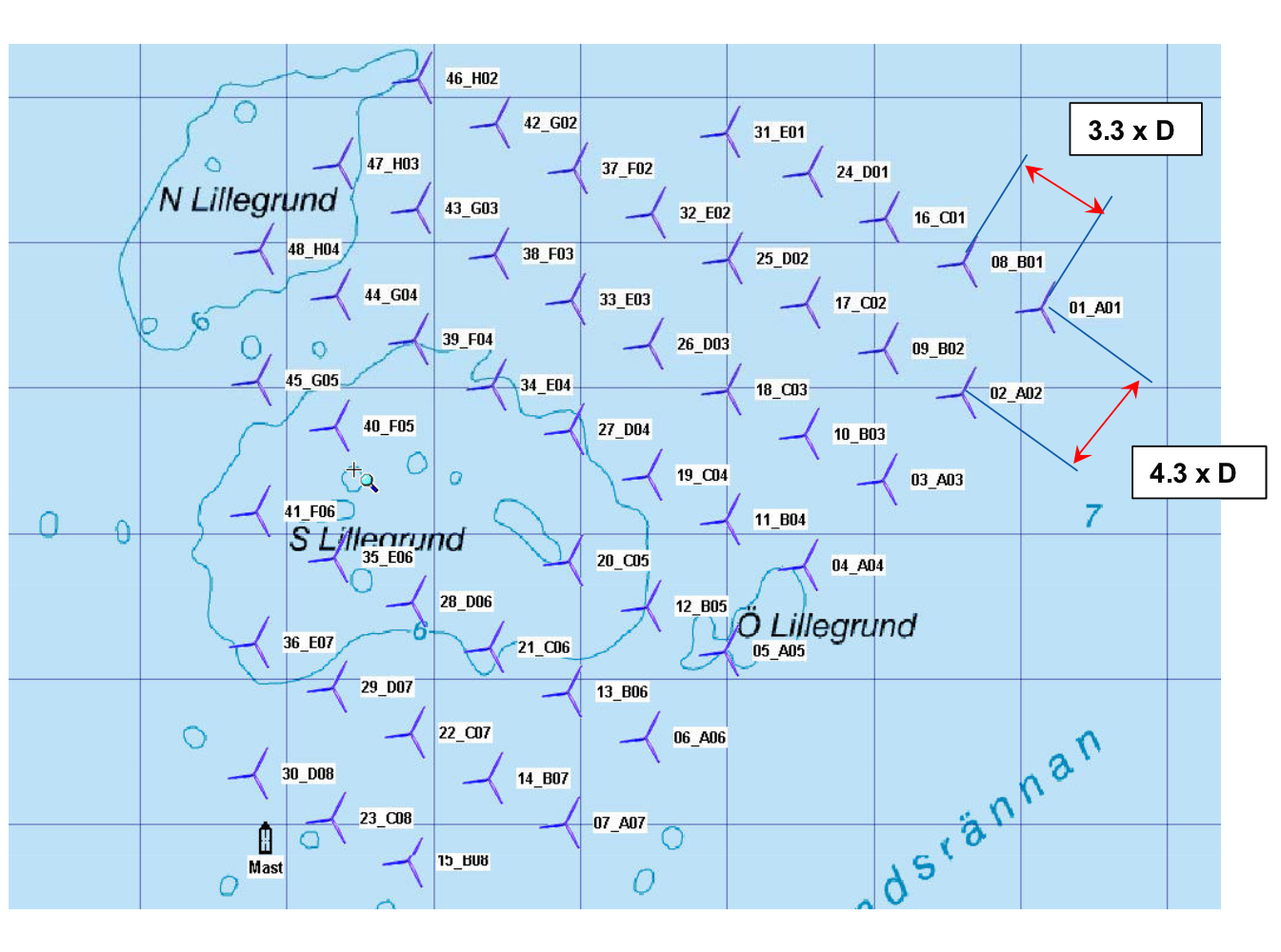}
	\caption{Detailed plan view of Lillgrund. The turbines are labelled from 01\_A01 through to 48\_H04. The grid lines have a spacing of 500~m. Courtesy of \citet{Lillgrund2}.}
	\label{fig:Map2}
\end{figure}

The array consists of 48 Siemens 2.3 MW Mk II wind turbines, each with a rotor diameter of $D=93\rm~m$ and a hub height of 65~m, in a regular lattice-type array as shown in Figure~\ref{fig:Map2} where each turbine is given a number as well as a grid-name using column letters A to H and row numbers 1 to 8\@.  There is a gap within the array where turbines D05 and E05 would have been, but the water there is too shallow for installation vessels to operate. The turbines are close to each other, with a spacing of $4.3D = 400\rm~m$ in the prevailing wind direction, SW -- NE direction (43$^\circ$ / 223$^\circ$), and $3.3D= 307\rm~m$ in the NW -- SE direction (120$^\circ$ / 300$^\circ$). Originally smaller turbines had been planned for, but by the time the turbines were being installed these larger turbines were available, and it was decided to opt for the larger turbines without changing the layout.  Overall, the extent of the wind farm is up to 2.9~km in the prevailing wind direction and 2.25~km across, covering a total area of around $6\rm~km^2$.

\subsection{Meteorological conditions}\label{S:Meteo}

\begin{figure}
a) \hspace*{0.48\textwidth} b) \\
\includegraphics[width = 0.48\textwidth]{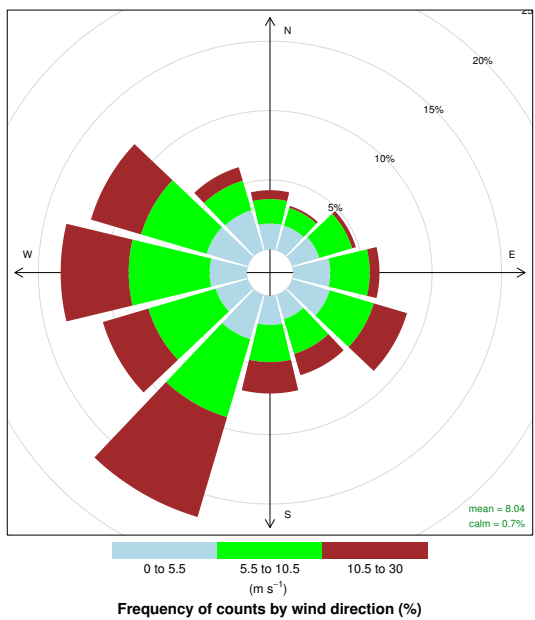} 
\includegraphics[width = 0.48\textwidth]{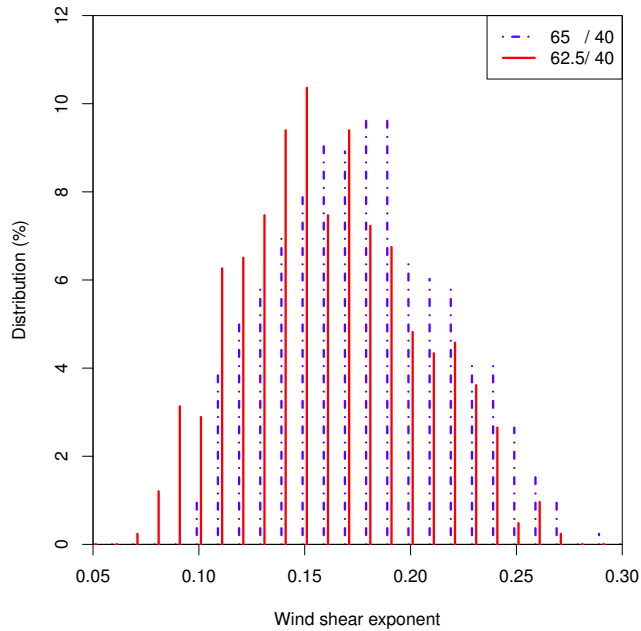} 
\caption{Wind conditions: a) wind rose of the wind speed at hub height for the site of the wind farm covering the analysis period, using \cite{Carslaw2012}; b) Histogram of the distribution of wind shear exponent within the wind bin investigated here.  The legend indicates the heights of the pair of anemometers used.}
\label{fig:Rose}
\end{figure}

The meteorological conditions at Lillgrund were monitored during the planning and construction with a meteorological mast south-west of the turbine array and are reported in \citet{Lillgrund16}.  This analysis was repeated from available data covering part of the operational phase.  The wind rose in Figure~\ref{fig:Rose}~(a), using the later operational data, shows the typical pattern of prevailing winds from the south-westerly direction.   The met mast was equipped with anemometers at three heights, 25, 40, 62.5 and 65~m, wind vanes for wind direction at 23 and 61 m, and temperature at 8 and 61 m height.

\citet{Lillgrund16} reported a correlation of the wind shear exponent, using a power-law profile, of 
$$
\frac{U_{65}}{U_{25}} = \left(\frac{65\rm m}{25\rm m}\right)^{\alpha}
$$
with an exponent of $\alpha=0.108$ for the entire period of their analysis covering the entire range of wind conditions experienced between September 2003 and February 2006\@.  The analysis was repeated with the later operational data, using all possible pairs of anemometers on the met mast. When calculated from the the ratio of the 10-minute wind speed averages and the ratio of the height of pairs of anemometers, this showed a large range in the wind shear exponents, with a slight preference for either an exponent significantly less than the mean or an exponent closer to neutral conditions ($\alpha\approx 0.14$).  Considering the focus of our study, we only present the results for those data where the wind direction was between 180\degree and 260\degree and the wind speed at 65~m between 5~m/s and 12~m/s.  The correlation in the results between those involving the upper level was very good (correlation coefficient $0.76 < r < 0.994$) but the correlation between the results of the pair 40~m and 25~m and all other pairs was poor ($r\approx 0.46$).  For that reason, Figure~\ref{fig:Rose}~(b) shows how frequent a particular instantaneous wind shear exponent occurred, using the two upper anemometers against that at 40~m.  Both show a distribution with a clear maximum though with a bias among the two pairs despite the close spacing of the upper two anemometers, one suggesting a most common wind shear exponent of $0.14 \lesssim \alpha \lesssim 0.16$ and the other of $0.15 \lesssim  \alpha \lesssim  0.2$.  

These results highlight two challenges, namely the difficulty of obtaining reliable measurements from routinely deployed instruments and of adequately describing wind conditions by common, fixed wind shear profiles, whether they have a power-law or logarithmic form.    Nevertheless, for modelling wind farms through CFD, it is necessary to represent 'the wind conditions' by typical and well-defined approximations. The results in Figure~\ref{fig:Rose}~(b) indicate that common wind shear profiles are satisfactory approximations at least at heights occupied by the turbine rotors and, in particular that wind shear profiles associated with neutral conditions of the atmosphere are sufficiently common to be a valid scenario to demonstrate the capabilities of the modelling approach and to validate its results against observations before embarking on the next step of including convection or stratification effects.

\subsection{Lillgrund diagnostics}
The analysis data set was derived from the output of turbine diagnostics from the SCADA (supervisory control and data acquisition) system at an interval of 1 minute covering a period of 480 days, starting in December 2007; however, this analysis only uses data from January 2008 when all turbines were finally connected to the system. Furthermore, the analysis only included instances when at least 40 turbines were operating normally, to ensure that the data reflected the farm as a whole while allowing for scheduled or unscheduled downtime of some turbines. Turbines with a curtailed output were also filtered out, to exclude those not operating according to their normal performance characteristics. The resulting set of valid data covered 323 days. The available data from the met mast overlapped with that period, but did not cover the full range of valid operational data.  This necessitated the use of nacelle data to infer wind speed and direction and a further validation stage to test the correspondence between nacelle data and met mast data.  The first stage in this is to identify the `\emph{front}' turbine to use as the provider for the proxy wind speed and direction measures.

\subsubsection{Front turbine selection}
To construct the turbine's performance curve, first the wind direction and representative `front' turbine had to be determined. This was achieved by selecting three turbines from each edge of the wind farm associated with a wind direction sector spanning 45$^\circ$.  At each time step, the appropriate sector was identified by finding instances where the three front turbines for that sector had a yaw direction consistent with that wind direction.   From those instances, the representative front turbine was chosen as that having the median of the nacelle wind speed, yaw direction, and active power output. The final selection was then inspected for consistent behaviour across sectors.
Having thus identified the turbine to represent the free-stream conditions, the actual consistency between the nacelle-based measures and the met mast could be carried out.

\subsubsection{Wind speed and nacelle anemometer}
\begin{figure}
a) \hspace*{0.48\textwidth} b) \\
  \includegraphics[width=.48\linewidth]{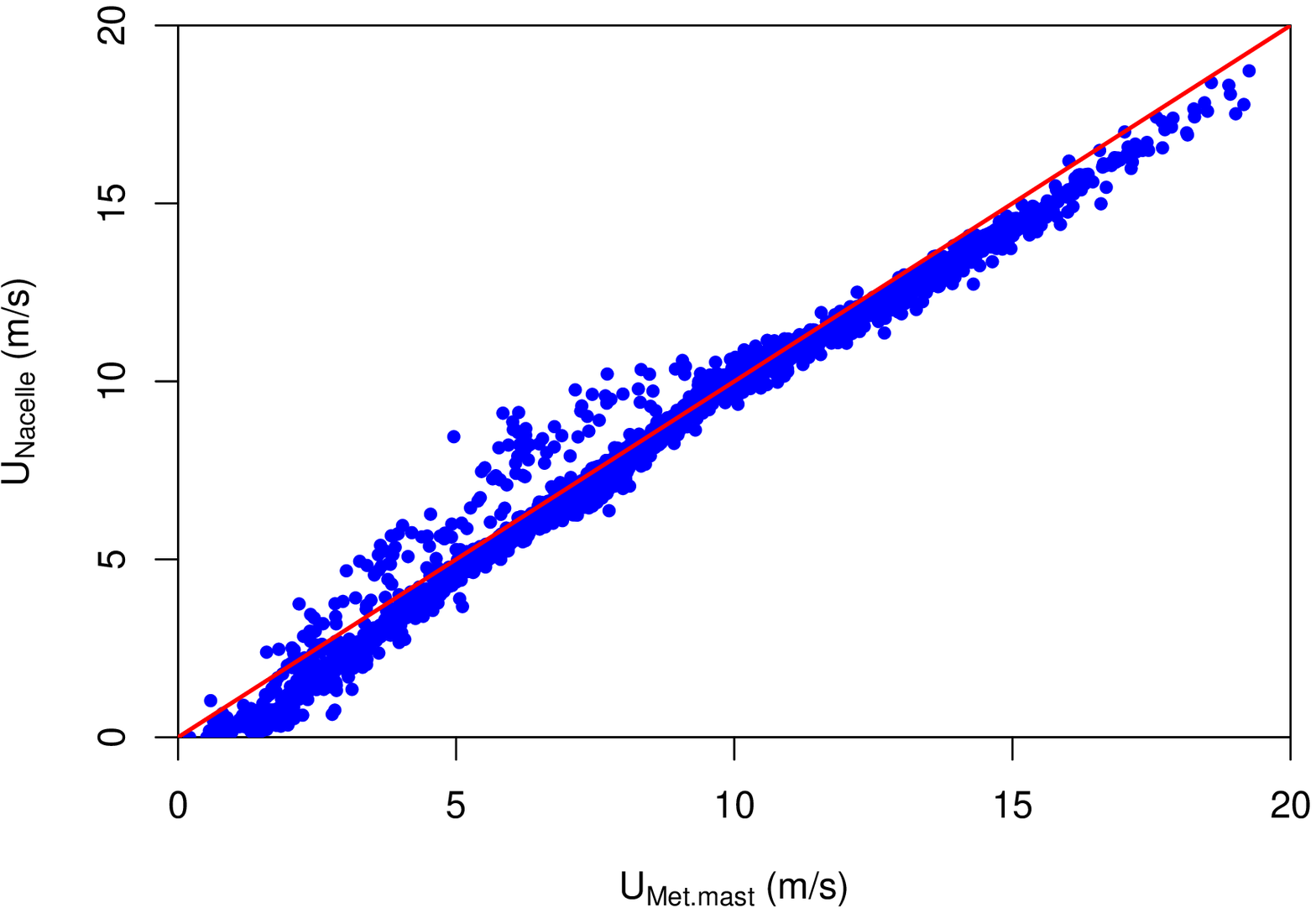}
  \includegraphics[width=.48\linewidth]{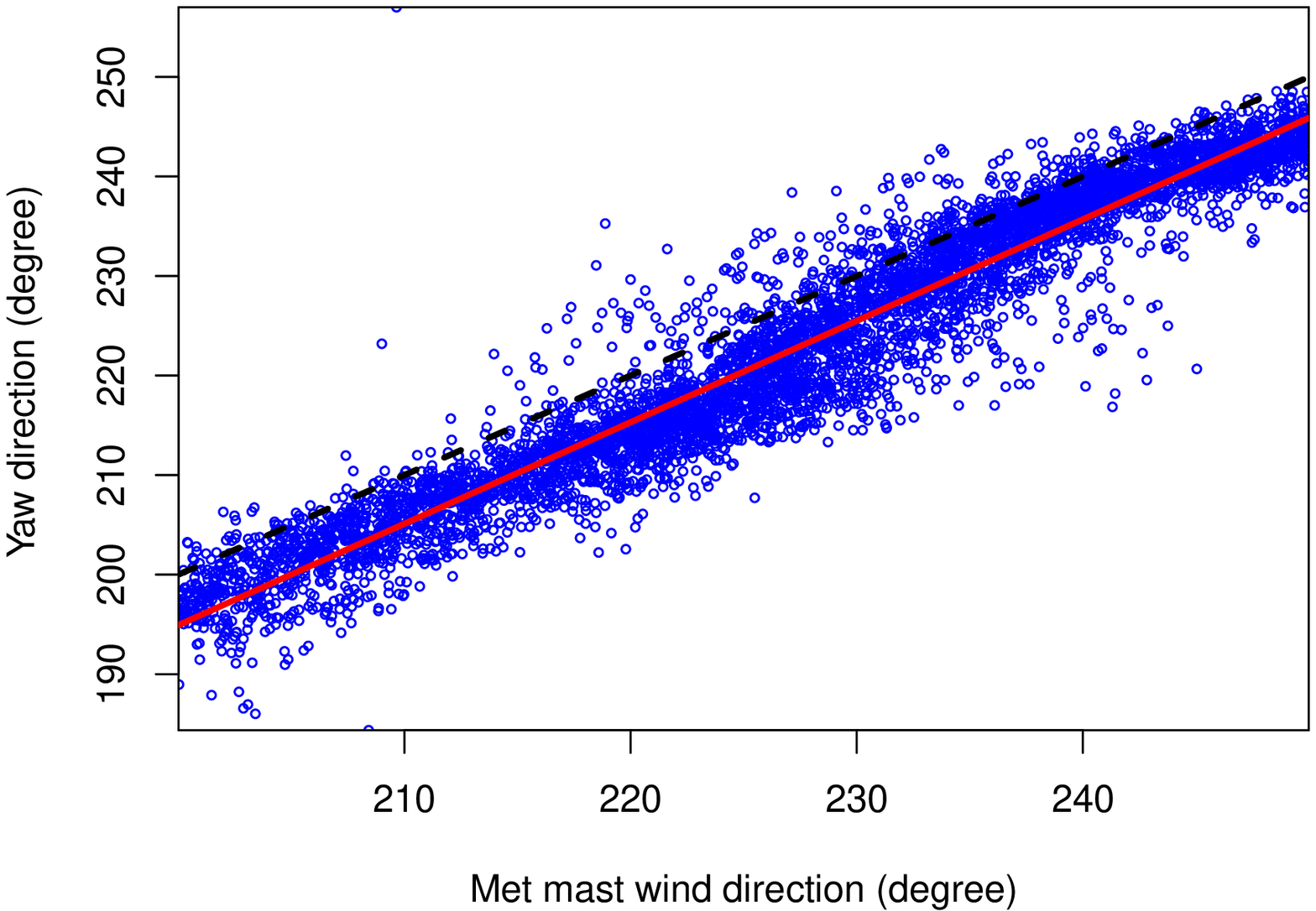}
	\caption{Comparison of direction information from the met mast and the nacelle yaw direction over the directional range investigated in detail.}
	\label{fig:Yaw}
\end{figure}

While a met mast anemometer is designed to measure the true local wind speed, in contrast a nacelle anemometer sits behind the rotor but is calibrated to estimate the free-stream velocity, and that calibration has to vary with the turbine's action.    A further complication is that Lillgrund has only a single met mast south-west of the farm but very close to the turbines.  For most of the wind directions other than south-westerlies, the anemometer is affected by turbine wakes and certainly within the wind farm wake for wind directions from the northerly sectors, so that the met mast instruments no longer measure the free-stream conditions.  For that reason, it is deemed that the most reliable measure of the free wind speed is the calibrated output from the anemometer at the top of that turbine which is most exposed to the wind.  \protect Figure~\ref{fig:Yaw}(a) shows the wind speed readings from nacelle anemometer against that from the anemometer at 63~m above the sea on the met mast for wind directions between 180\degree and 270\degree.  While there is some variation, both random and systematic, the agreement between the two measures is good enough to be able to use the nacelle wind speed as an indicator of the free-stream wind speed, especially in the range between the cut-in wind speed of the turbines and the rated wind speed.

\subsubsection{Wind direction and nacelle yaw}
As with the wind speed, a measure of the wind direction based on available turbine data had to be determined.  In ideal conditions, the nacelle yaw should follow the wind direction, but this only happens with a delay given by the yaw control mechanism of the turbine.  Furthermore, identifying the current wind direction and actuating the rotor and nacelle yaw appropriately are not trivial.  \protect\citet{Lillgrund15} presented some evidence that the nacelle yaw of the front turbine did follow the wind direction from the met mast, albeit with a slight delay, filtering out the faster fluctuations, and a with small but persistent bias.  A more complete re-analysis of the relationship between the two measures across the entire range showed both a random variation and a systematic variation over the range investigated.  This suggests that the yaw mechanism is effected by the flow induced by the other turbines in the front line affecting the selected turbine.  However over the more restricted range to be investigated in this study, that systematic variation is very small, leaving only the random variation and an offset of around 9\degree~between the met mast wind direction and the nacelle yaw, as shown in Figure~\protect\ref{fig:Yaw}. The nacelle yaw is on average $9\degree\pm 7\degree$ less than the met mast.

\subsection{Wind farm performance}\label{S:Lillgrund:FarmPerf}

The two performance curves shown in Figure~\ref{fig:PerformanceCurve} compare that of a turbine exposed to the wind (in the blue shading with the cross-hatching) with that of the entire wind farm (the red shaded area) against the`free wind speed' at hub height. In both cases, the shading captures 90\% of all valid data.  The wind turbine curve in Figure~\ref{fig:PerformanceCurve} aggregates the data from only those turbines which are on the edge of the farm facing the wind at any time.  For the wind farm in Figure~\ref{fig:PerformanceCurve}, the sum of total power output from the normally operating turbines was divided by that number of turbines and their rated power to calculate the normalised power output, normalised against the active installed capacity of the wind farm.  For both curves, the power coefficient is plotted against the wind speed recorded at the front turbines.

\subsubsection{Relative wind farm performance}
\begin{figure}
  \centering
	\includegraphics[width =0.7\linewidth]{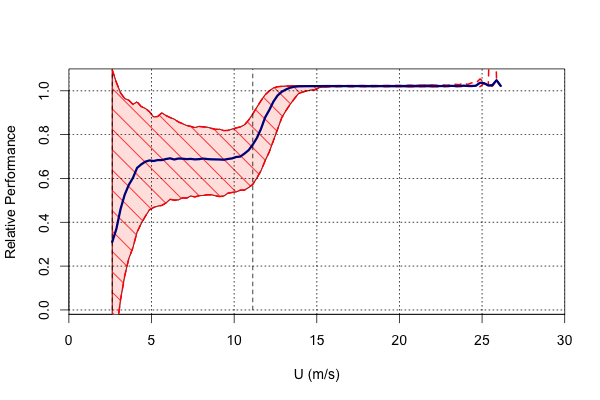} 
	\caption{Relative performance of wind farm, $\Pi$, against nominal wind speed averaged over all wind directions.  The solid line is the median, and the shaded area indicates the range from the $5^{th}$ to the $95^{th}$ percentile.  The dashed line indicates the extent over which the median relative performance is relatively constant.}
	\label{fig:RelPerformance_U}
\end{figure}

\begin{figure}
  \centering
\includegraphics[width = 0.7\linewidth]{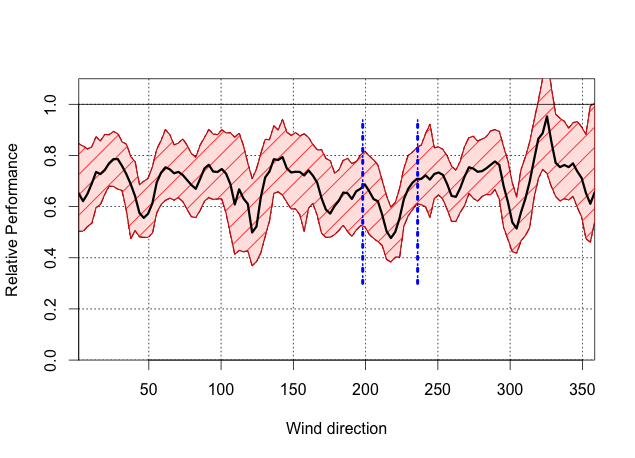} 
\caption{Relative performance of wind farm, $\Pi$, against the wind direction, for all data within the wind speed band, $5.5{\rm m/s} < U < 10.5\rm~m/s$. The solid line is the median, and the shaded area indicates the range from the $5^{th}$ to the $95^{th}$ percentile. The two vertical dot-dashed lines indicate the wind speed range focussed on for the detailed analysis.}
\label{fig:RelPerformance_Dir}
\end{figure}

A previous assessment of the wind farm performance \citep{Lillgrund15} subdivided the variable range into three zones.  To refine their analysis, a power deficit or relative wind farm performance can be defined as
\begin{equation}
\Pi \equiv P_{Farm}/(N_T P_{Front})
\end{equation}
where $P_{Farm}$ is the sum of the power output from all normally operating turbines and $P_{Front}$ is the power output from a 'front' turbine identified as being on the windward edge of the wind farm. $N_T$ is the number of normally operating turbines which excludes turbines operating at a curtailed level or turbines which have been turned off.  Figure~\ref{fig:RelPerformance_U}, which shows the median of that ratio together with the range covering 90\% of the data, demonstrates that the relative farm performance is constant over an extended wind speed range from around 5~m/s to 11~m/s (indicated by the dashed line).  As these results include all wind directions, the range is substantial within that wind speed band. 

\subsection{Identification of cases to be simulated}
When combining all relative power coefficients within that wind speed band but resolving the wind direction over small wind directional bins, $3^\circ$ in the case shown in Figure~\ref{fig:RelPerformance_Dir}, pronounced peaks and troughs can be seen as a result of the lattice structure of the wind farm layout, as turbines in the second and third row move in and out of the wake from the upwind turbines. This clear sensitivity of the power deficit to the wind direction in the wind speed band $U_{\rm cut-in} < u < U_{\rm rated}$ motivated this investigation, in which specific wind directions are analysed in more detail.  In particular, the focus is on the narrow wind direction sector indicated by the dot-dashed lines in Figure~\ref{fig:RelPerformance_Dir}, which covers the two extreme cases of the turbines in second row fully shaded and fully exposed to the free stream, and some intermediate scenarios.

\begin{table}
\caption{Cases of wind directions investigated and schematics how the wind direction and turbine layout relate to each other.}
\label{tab:Cases}
\centering
\begin{tabular}{ l l l l l }
\hline\noalign{\smallskip}
Direction & Characteristics & Direction & Characteristics  \\ 
\hline\noalign{\smallskip}
198\degree & Maximum exposure of second row & 202\degree & Second row exposed, third row in first row wake\\
& \includegraphics[width = 0.3\textwidth]{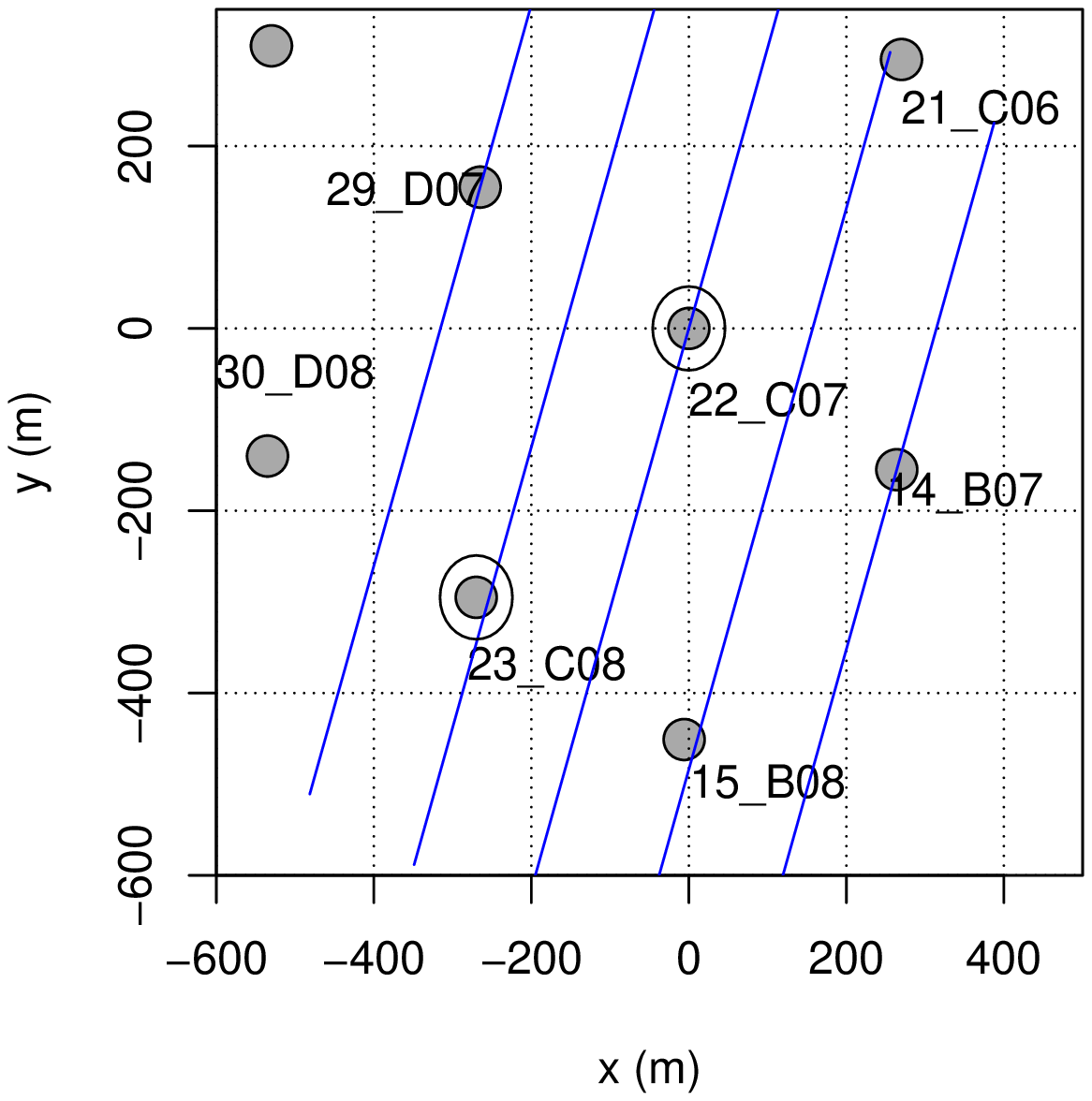} 
&& \includegraphics[width = 0.3\textwidth]{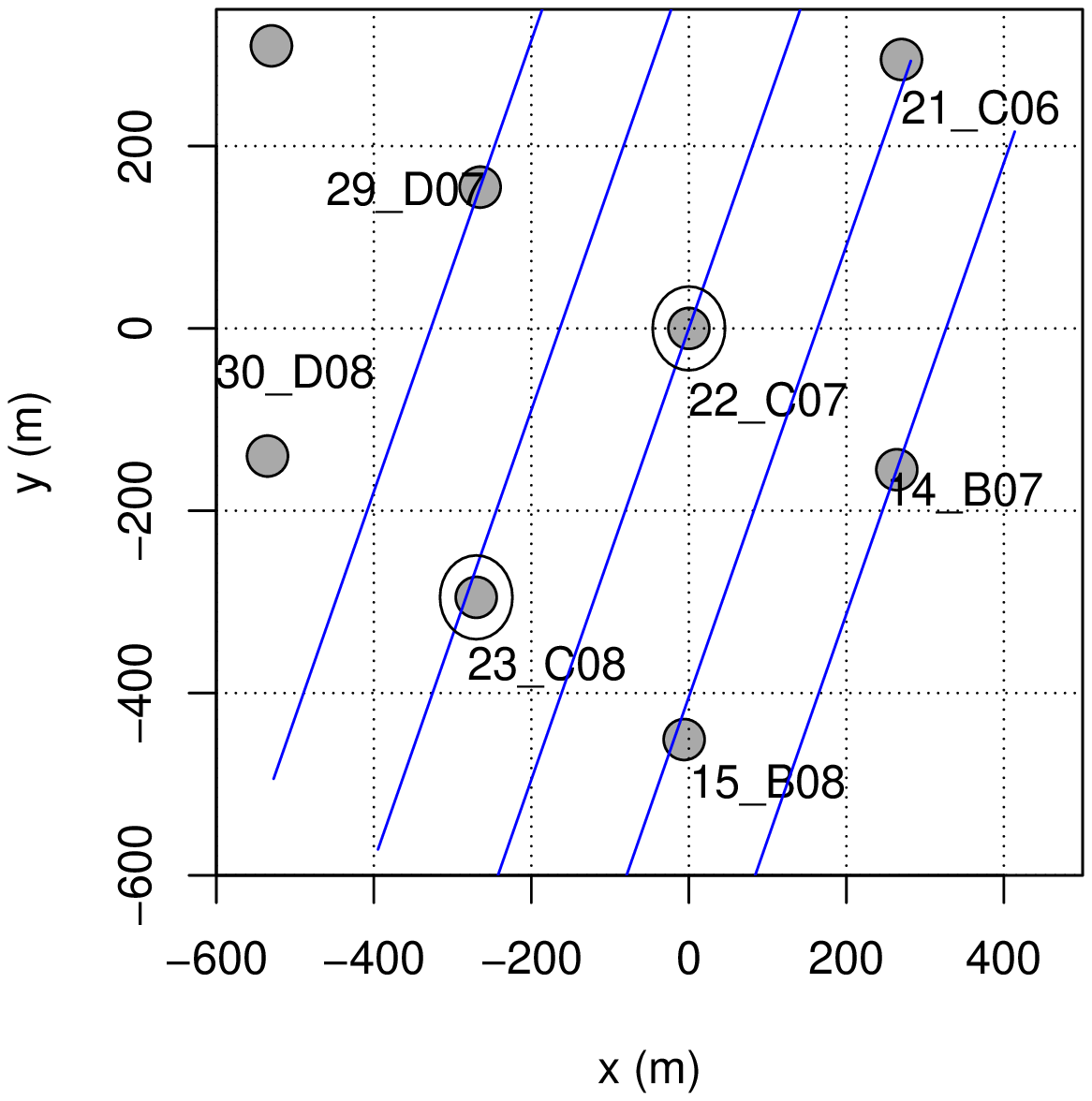}  \\
207\degree & Second and third rows not shielded & 212\degree & Second and third rows not shielded \\
& \includegraphics[width = 0.3\textwidth]{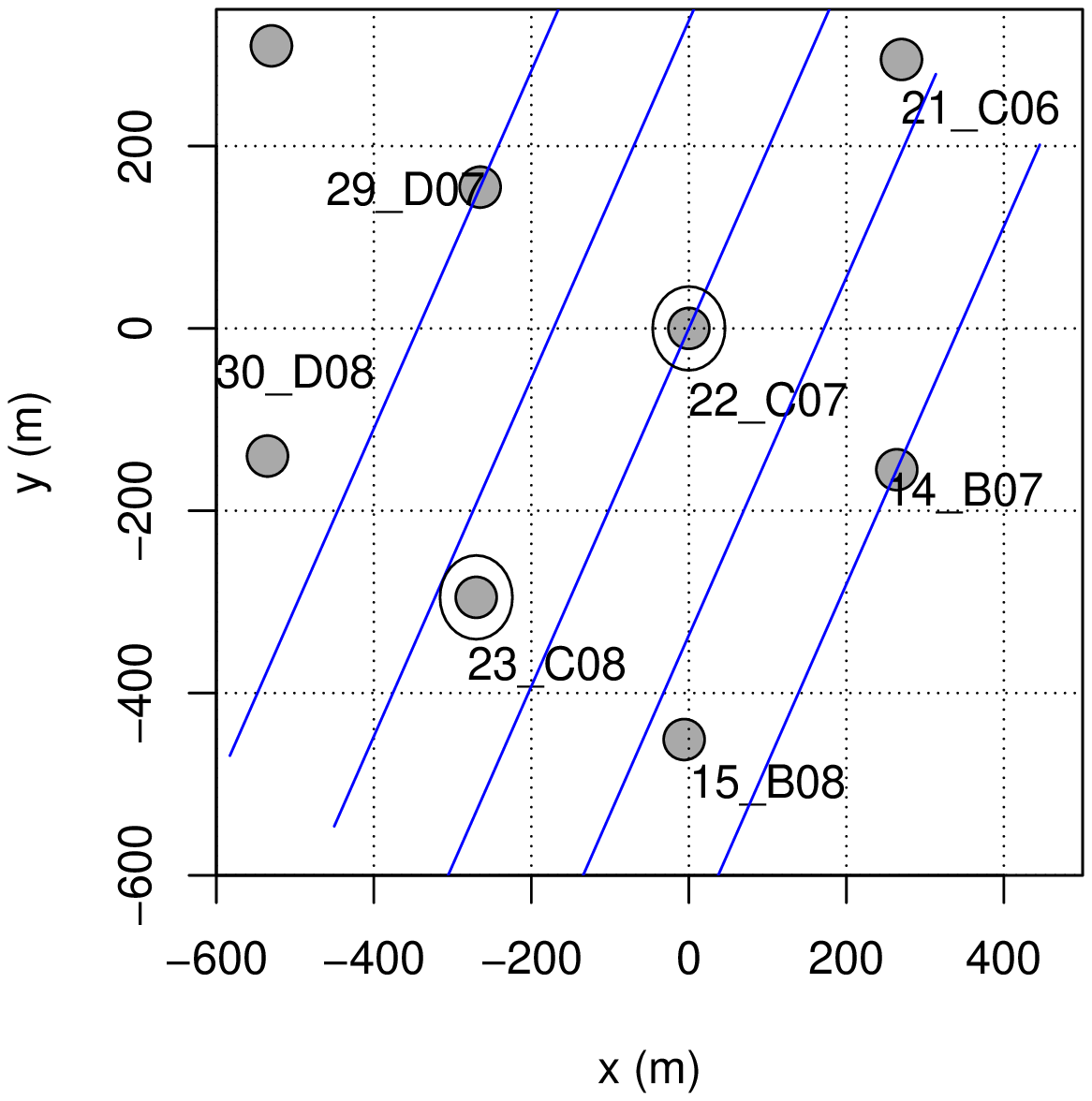} 
&& \includegraphics[width = 0.3\textwidth]{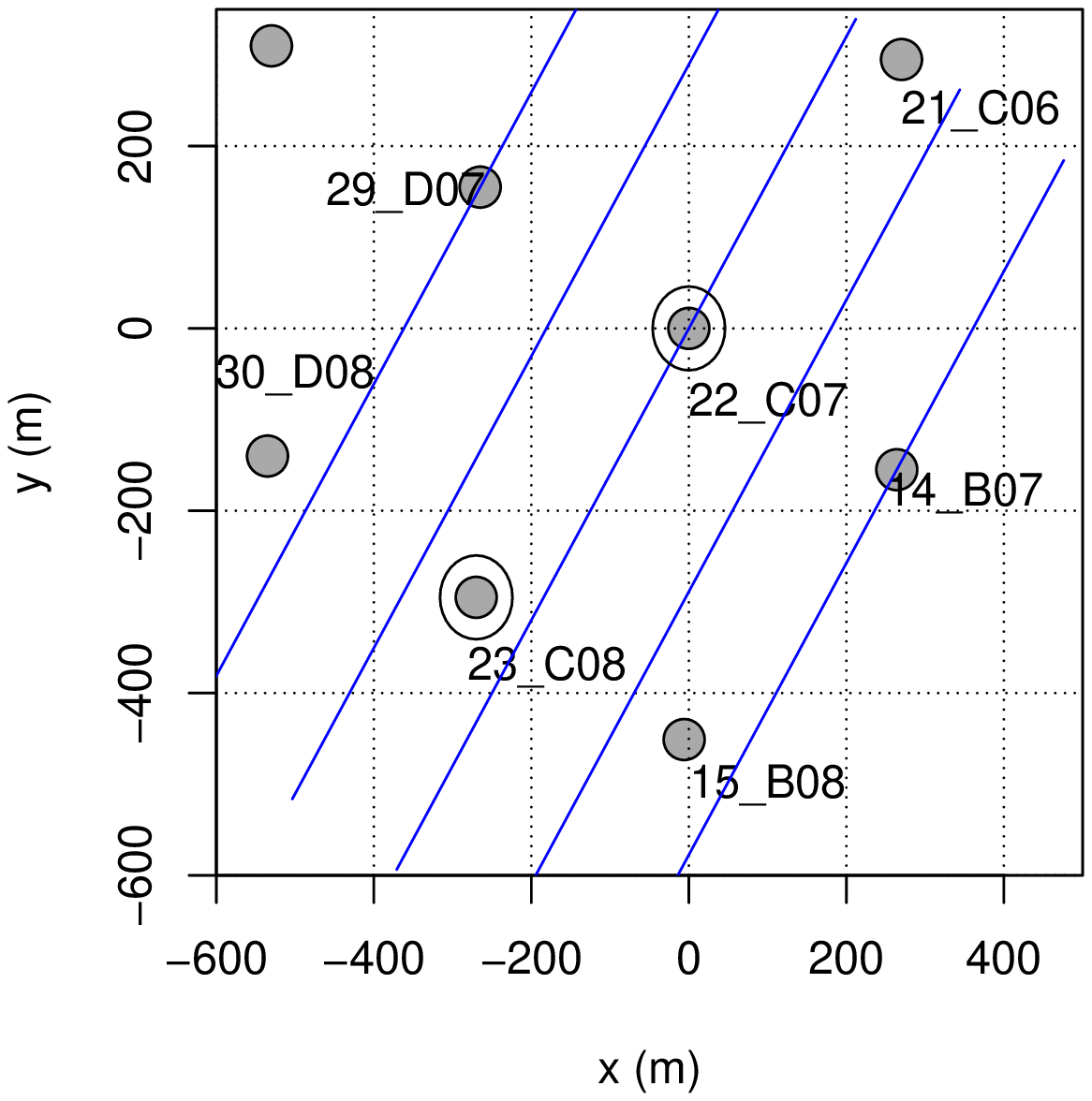}  \\ 
217\degree & Second row partially shielded & 223\degree & Turbines fully aligned with wind \\
& \includegraphics[width = 0.3\textwidth]{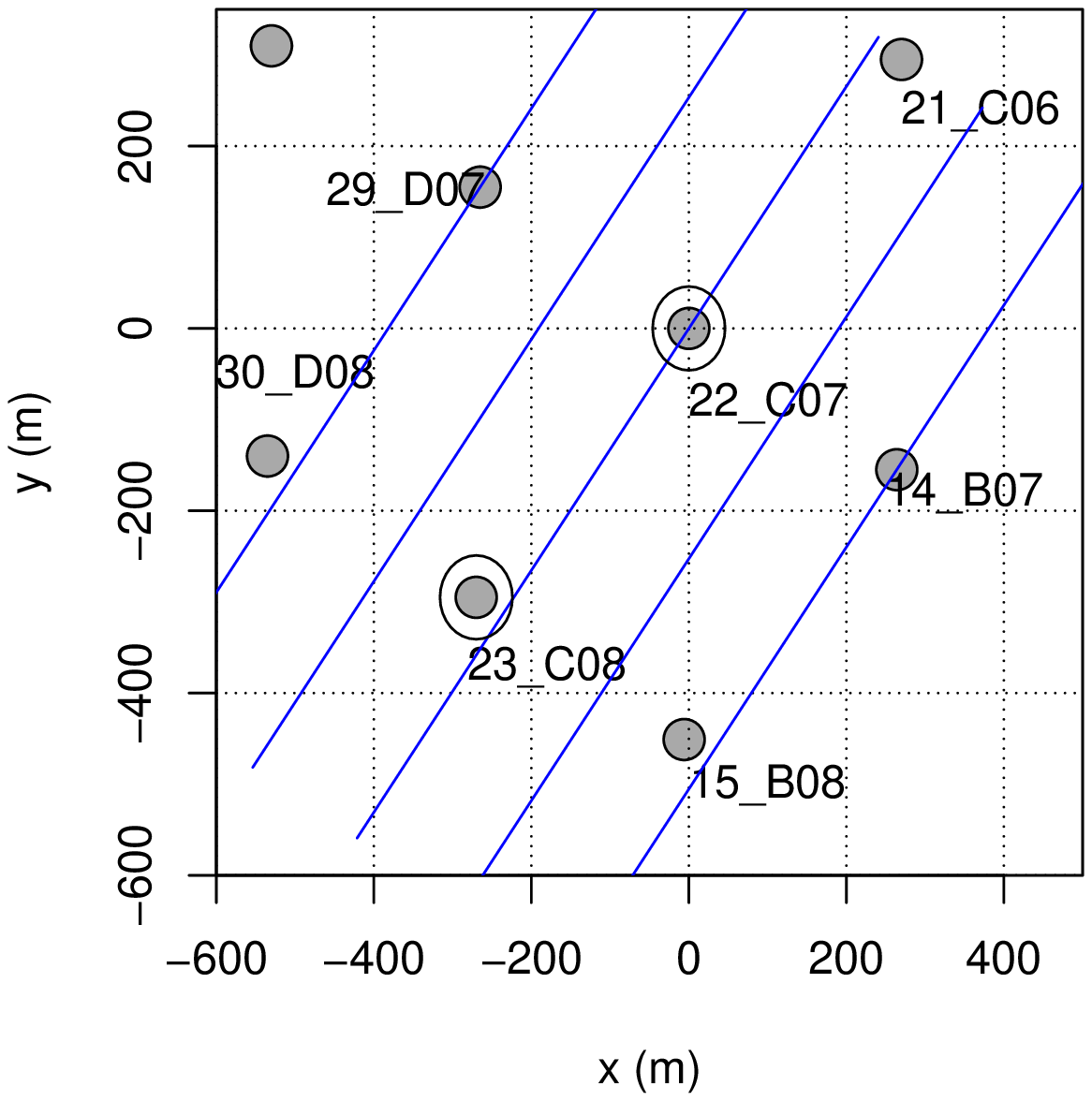} 
&& \includegraphics[width = 0.3\textwidth]{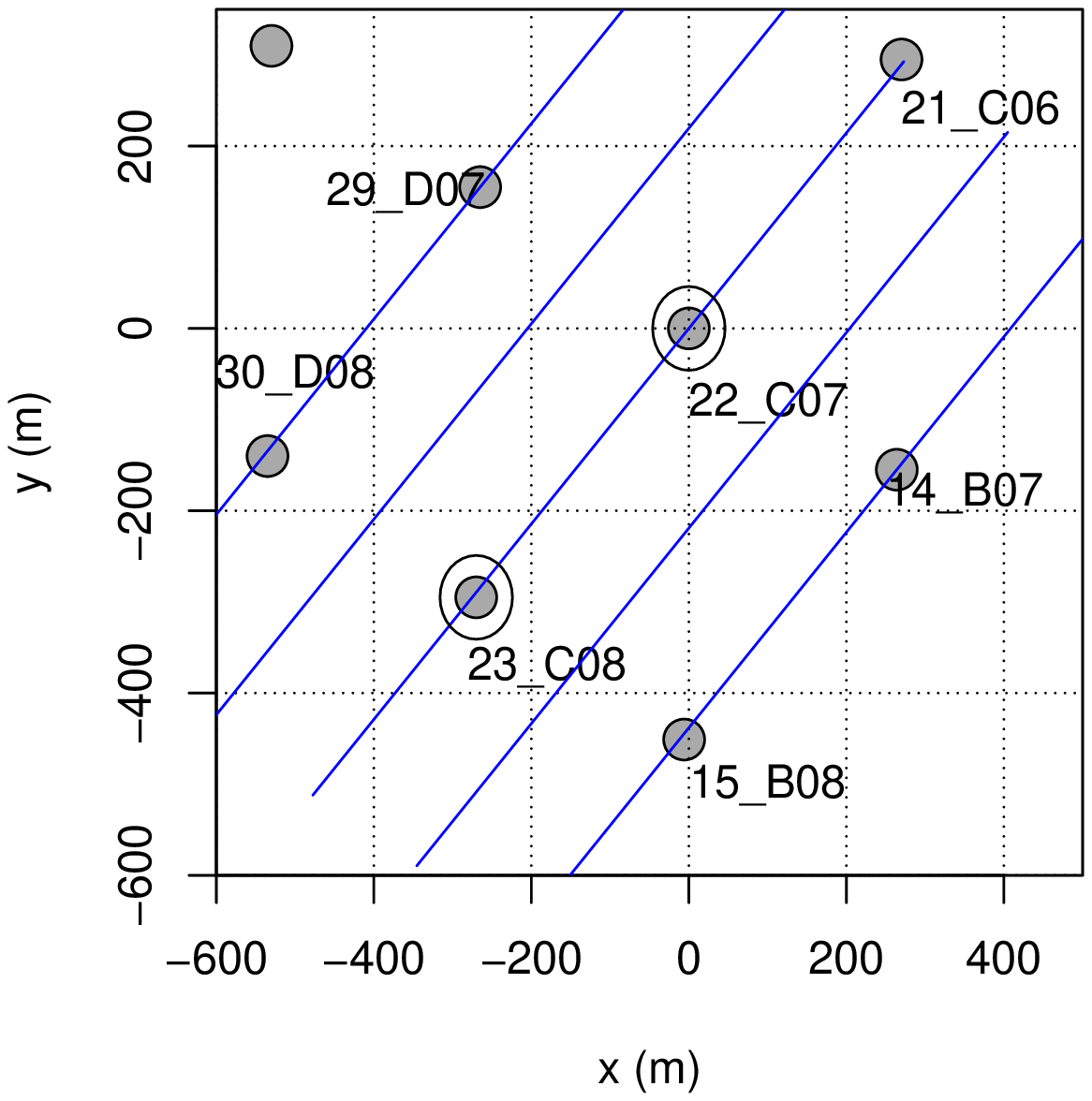}  \\ 
229\degree & Second row partially shielded & 236\degree & Second row between wakes, oblique opening \\
& \includegraphics[width = 0.3\textwidth]{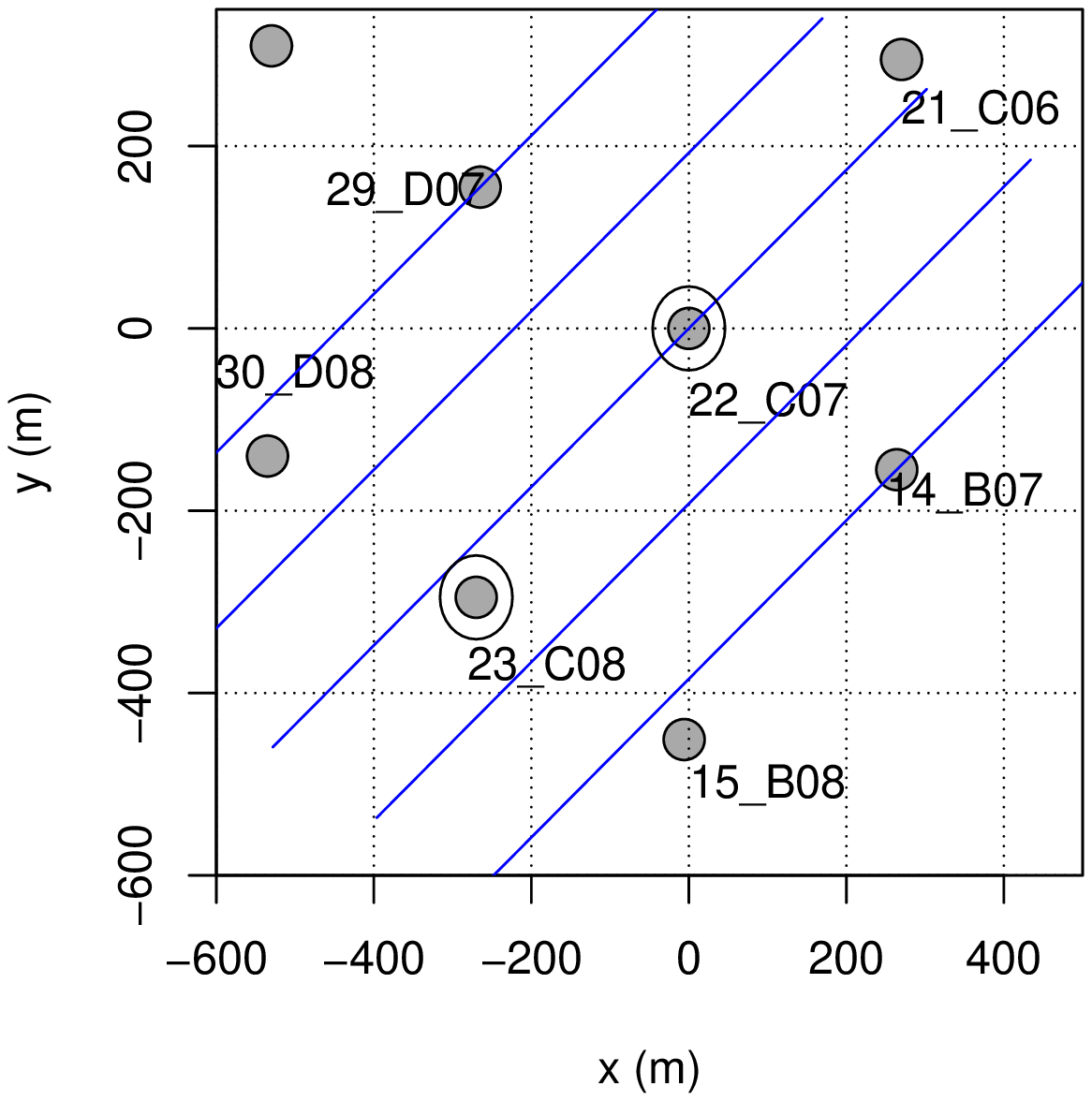} 
&& \includegraphics[width = 0.3\textwidth]{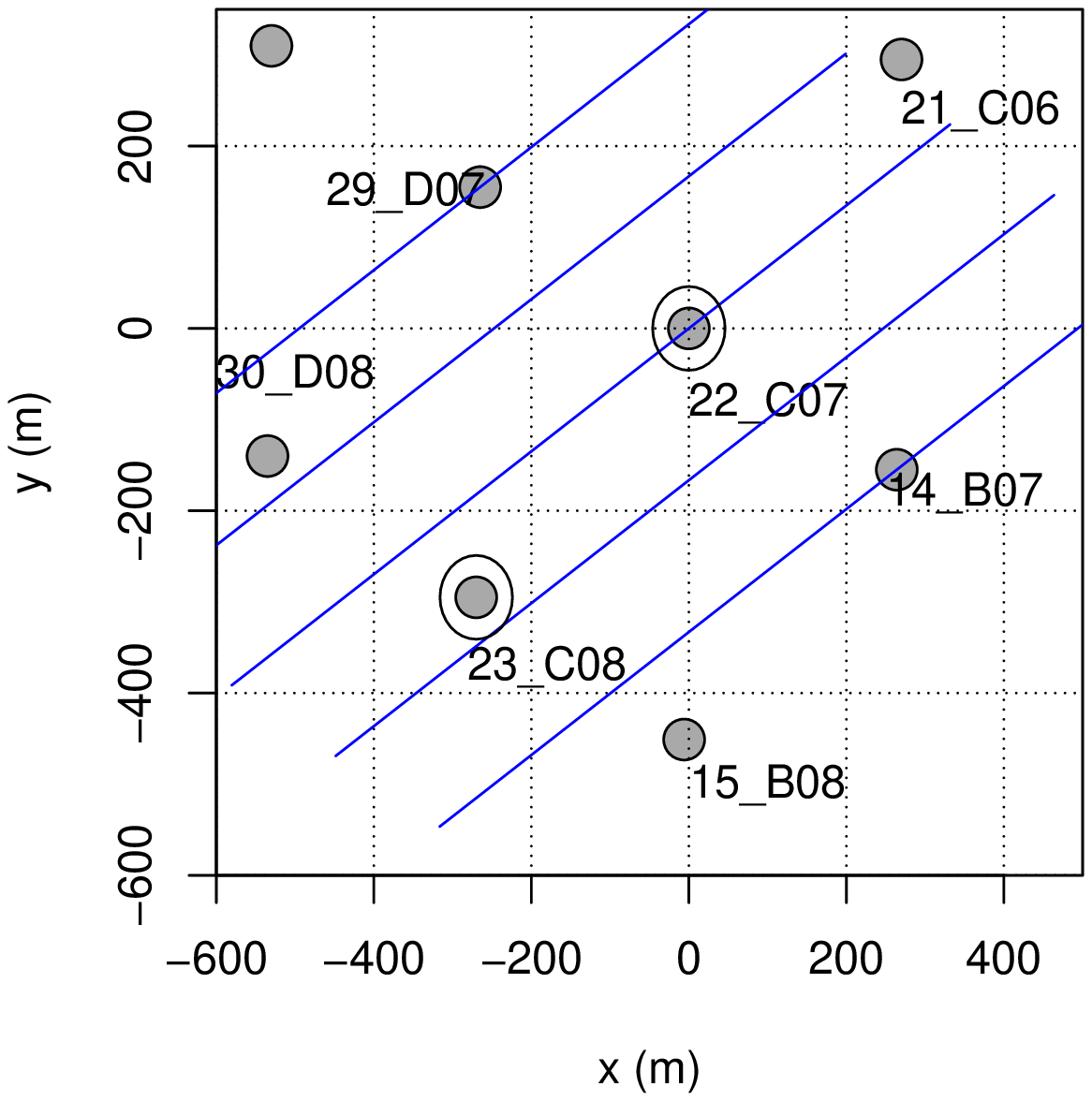} \\
\hline\noalign{\smallskip}
\end{tabular}
\end{table}

The relative power deficit, $\Pi$, of the wind farm is clearly function of the wind direction but, on average, it is constant within the reference wind speed range between 5.5~m/s and 10.5~m/s. Because of this, we chose to investigate the wind farm effect for typical wind conditions with a free stream velocity at hub height of 10~m/s for a set of south-westerly directions centred around that where turbines are fully aligned with the wind.  Based on the turbine coordinates provided by Vattenfall in local Euclidean North-East coordinates, this occurs at a wind direction of 223$^\circ$, and the cases analysed here centre around this wind direction and extend either side to that case.  The wind directions chosen and how they relate to the turbine positions are listed in Table~\ref{tab:Cases}.  For Lillgrund wind farm, the chosen wind directions are key cases, as they are within the sector of the prevailing winds as shown by the wind rose in Figure~\ref{fig:Rose}. As neutral stability conditions are sufficiently frequent and, in the absence of sufficient atmospheric stability information from the SCADA data, this set of simulations was restricted to neutral conditions.

\section{Computational methodology}
\label{S:Method}

As with previous work \citep{CFC11}, the turbine model described below in section \ref{s:turbine-formulation} is broadly derived from blade-element momentum theory. Rather than use axial induction factors however, lift and drag are calculated from tabular aerofoil data, and applied to the incompressible Navier-Stokes momentum equation as body forces, with CFD used to resolve the flow. This is a common approach utilised in wind turbine modelling \citep{jimenez2008, meyers2010, lu2011, Churchfield2012}; for a summary of techniques, see \citet{creechfruh2014}.

Fluidity, an open-source, finite-element hr-adaptive CFD solver from Imperial College \citep{piggott2004},
was used to solve the Navier-Stokes equations with LES turbulence modelling. This solver has a long history in coastal and oceanographic modelling \citep{ford2004, pain2005, piggott2008, funke2011, kimura2013, hill2014},
but has also been used to study atmospheric boundary layer turbulence \citep{aristodemou2009, pavlidis-springer-2010, pavlidis2010}.

Following on from \citet{CFC11}, the mesh for velocity and pressure was adaptive and unstructured; resolution was concentrated near the cylindrical volumes representing the turbines, to ensure that there were sufficient mesh nodes within the turbine. Furthermore, as the meshes were adaptive, the mesh nodes within these volumes had to be gathered at each timestep, since there was no guarantee that the mesh would be identical between timesteps.

Section \ref{s:fluid-equations} will briefly detail the main fluid dynamics equations, while section \ref{s:turbine-formulation} deals with the turbine model itself.

\subsection{Fluid equations}
\label{s:fluid-equations}
\begin{figure}
\begin{centering}
\includegraphics[width = 0.5\columnwidth]{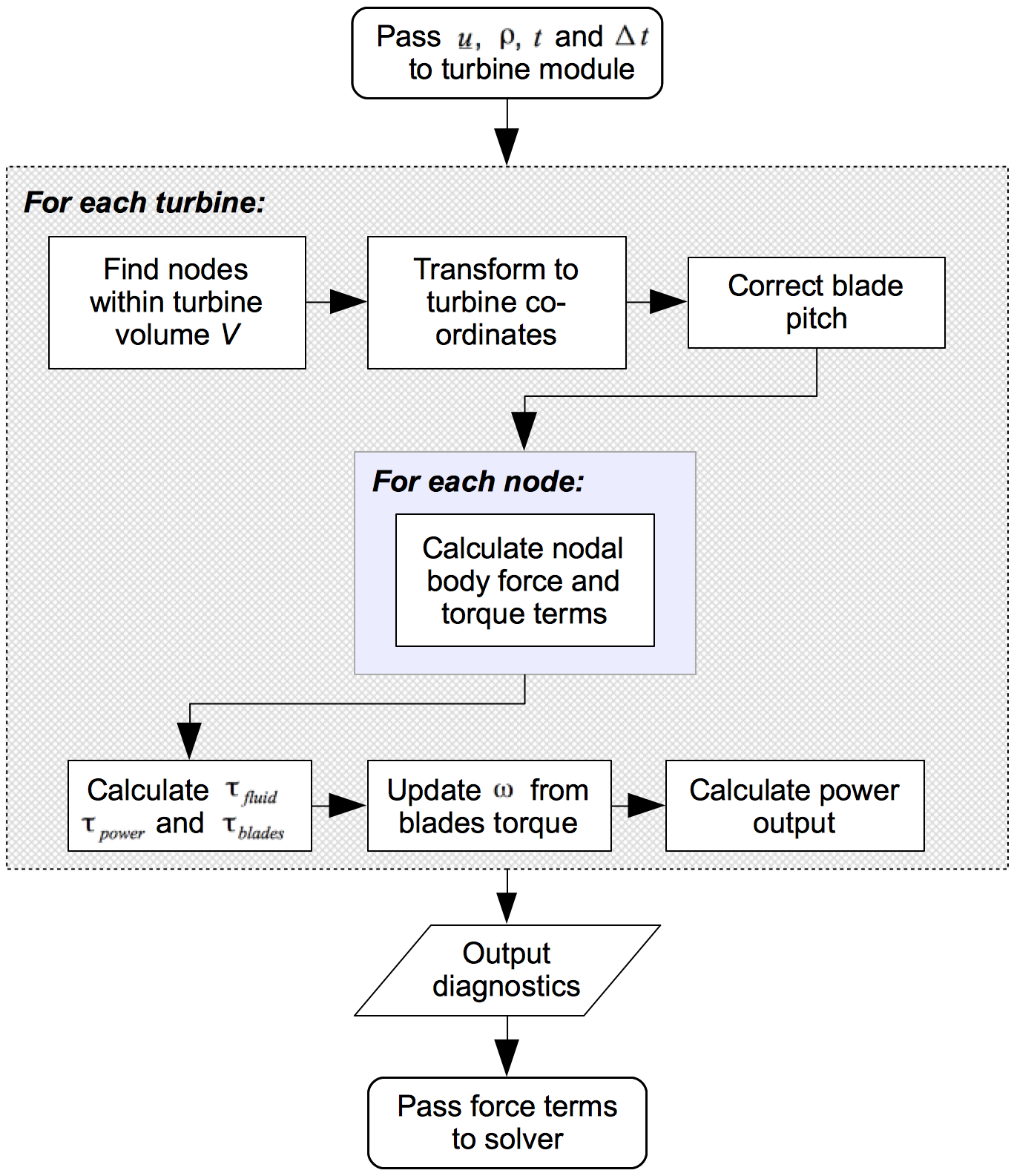} 
\caption{Overview of the calculation procedure at each time-step.}
\label{fig:main-routine-flowchart}
\end{centering}
\end{figure}

Our starting point is the Navier-Stokes momentum equation for an incompressible Newtonian fluid, i.e.

\begin{equation}
\label{eqn:basic-ns-mom}
\frac{D \mathbf{u}}{Dt} = - \frac{1}{\rho} \nabla p + \nu \nabla^2 \mathbf{u} + \frac{1}{\rho} \mathbf{F}
\end{equation}

where $\mathbf{u}$ is the velocity field, $\rho$ is the density of air, $p$ is pressure, $\nu$ is the kinematic viscosity of air, and $\mathbf{F}$ is a vector representing the body forces exerted on the air by the wind turbines. The body forces are calculated by the turbine model, and only exist within the cylindrical volumes each turbine occupies, described in more detail in section \ref{s:turbine-formulation}. At each timestep, the CFD solver passes velocity, time, and time-step size to a seperate turbine module, which then calculates the turbine performance and passes back the body force terms to the solver, which then solves the equations. This process is represented as a flowchart in Figure \ref{fig:main-routine-flowchart}; further details can be found in Section \ref{s:turbine-formulation}.

Within the Fluidity solver, equation (\ref{eqn:basic-ns-mom}) was discretised into a finite element P1-P1 element pair, with a wall-adapting local eddy-viscosity (WALE) variant of the LES turbulence model \citep{ducros1998, nicoud1999} for subgrid-scale turbulence.  In tensor notation this becomes

\begin{equation}
\frac{D \overline{u}_i}{D t} = - \frac{1}{\rho} \frac{\partial p}{\partial x_i} 
    + \frac{\partial}{\partial x_j} \left[ \left( \nu + \nu_t \right) \left( \frac{\partial \overline{u}_i}{\partial x_j} + \frac{\partial \overline{u}_j}{\partial x_i} \right) \right] + F_i
\end{equation}

The overbar denotes the velocity field filtered above the filter lengthscale $\Delta$, and $\nu_t$ represents the additional viscosity due to subgrid-scale turbulence, i.e. at lengthscales below $\Delta$. Standard Smagorinsky models define this as $\nu_t = C^2_S \Delta^2 \left| \overline{S} \right|$, where $C_S$ is the Smagorinsky coefficient, and $\left| \overline{S} \right|$ the strain-rate tensor. However this performs poorly near wall boundaries, since the eddy viscosity increases as soon as there is a velocity gradient, whereas the turbulence should drop away rapidly near the wall. With WALE LES, a new formulation of $\nu_t$ was developed to account for this phenomenon. The Smagorinsky coefficient is still required, and was set to $C_S=0.18$ for the simulations.

\subsection{Turbine formulation}
\label{s:turbine-formulation}

In a development comparable to the recent studies by \protect\citet{Archer2013} or \protect\citet{Nilsson2014}, our goal was to incorporate the dynamic response of the turbine to the local flow conditions. This builds upon \citet{CFC11}, which describes a torque-controlled actuator disc model of a fixed-pitch turbine, adding active blade pitch control and rotor yaw. Other torque-controlled models such as \citet{breton2014, wu2015}, appear to parameterise the turbine behaviour as a relaxed iteration of the rotor angular velocity, using tabulated steady-state torque data based upon manufacturers' turbine specifications. Whilst this is certainly a practical solution, our approach is aimed at modelling the physical processes and control actions to achieve 
the desired power output. Ultimately this will also encompass the mechanical inertia of the drive train, the full electro-mechanical response of the generator, and the associated frictional losses. 

\subsubsection{Frame of reference}
\begin{figure}[t]
\begin{centering}
\includegraphics[width = 0.5\columnwidth]{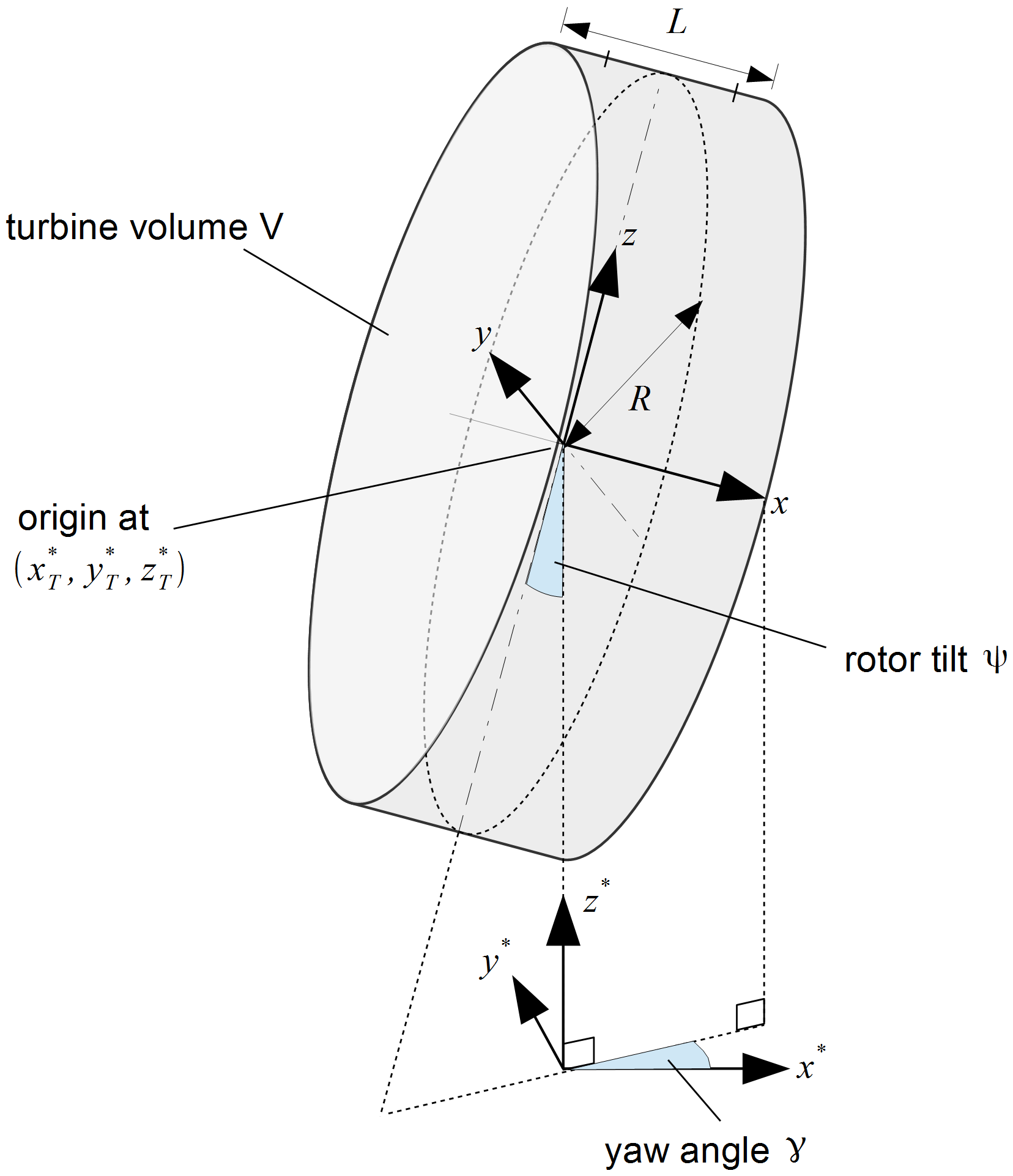} 
\caption{The cylindrical turbine volume V, with radius $R$ and length $L$. The transformation between the two coordinate systems are shown, with the axes $x^{*}$, $y^{*}$, $z^{*}$ representing the simulation coordinate axes, and $x$, $y$, $z$ representing those of the turbine reference frame. The yaw angle $\gamma$ is an anti-clockwise rotation about the $z^{*}$ axis, and the rotor tilt $\psi$ is a clockwise rotation about the $y$ axis.}
\label{fig:3d-volume-skew}
\end{centering}
\end{figure}

In order to calculate body forces due to lift and drag, the coordinates and velocity of nodes 
on the mesh must be translated to the frame of reference of each turbine rotor, i.e.~a
coordinate system local to that turbine, which must take into account the position, 
yaw and tilt of the turbine rotor. Here, we use a common technique in computer graphics used to transform between reference frames \protect\citep{foley1997}. If we indicate coordinates within the simulation reference frame with $(^{*})$, then for a wind turbine hub at position $\mathbf{x}^{*}_T=(x^{*}_T, y^{*}_T, z^{*}_T)$,
a yaw angle of $\psi$ and an upward rotor tilt of $\gamma$, 
then the coordinates of a mesh node $\mathbf{x}=(x,y,z)$ in the turbine reference frame will be

\begin{equation}\label{eqn:rotation}
\mathbf{x} =  \mathbf{M}_{-\gamma} \mathbf{M}_{-\psi} 
\begin{bmatrix}x^{*}-x^{*}_T \\ y^{*} - y^{*}_T \\ z^{*} - z^{*}_T\end{bmatrix}
\end{equation}
where 
$$
\mathbf{M}_{-\gamma} = \begin{bmatrix}
	\cos \gamma & 0 & -\sin \gamma \\
	0 & 1 & 0 \\
	\sin \gamma & 0 & \cos \gamma 
\end{bmatrix}
 \mbox{\;and\;} 
\mathbf{M}_{-\psi}= \begin{bmatrix}
	\cos \psi & -\sin \psi & 0 \\
	\sin \psi & \cos \psi & 0 \\
	0 & 0 & 1
\end{bmatrix}
$$
are the rotation matrices for rotor tilt and yaw, respectively. Figure {\ref{fig:3d-volume-skew}} shows the transformation between coordinate systems. Similarly, the velocity at a mesh node in the simulation frame of reference $\mathbf{u}^{*}=(u^{*}, v^{*}, w^{*})$ can be transformed to $\mathbf{u}=(u, v, w)$ in the turbine reference frame by

\begin{equation}
\mathbf{u} = \mathbf{M}_{-\gamma} \textbf{M}_{-\psi} \: \mathbf{u}^{*}
\label{eqn:velocity-rotation}
\end{equation}

Only nodes within a cylindrical turbine volume $V$ generate body forces. Nodes are considered to be within $V$ where $-L/2 < x < L/2$ and $r (=\sqrt{y^2+z^2}) < R$, with $L$ being the length of the cylinder, and $R$ the radius of the turbine rotor.

\subsubsection{Calculating lift and drag}
\label{sub: liftdrag}

Now that the coordinates and flow field have been transformed to the turbine rotor's reference frame, blade element momentum (BEM) theory can be applied to calculate the lift and drag forces acting on the blades. Fundamental to this are the calculated lift and drag coefficients, $C_L$ and $C_D$, which are dependent upon angle of attack $\alpha$ and the Reynolds number $Re$ of the flow over the blade. The tabulated data for $C_L$ and $C_D$ are specific to each aerofoil, and are discussed in section~\ref{S:turbine-parameterisations}.

Following the approach in \citet{CFC11}, the lift and drag forces on the blades per span unit length are

\begin{equation}
\label{fL_def}
f_L = C_L(\alpha, Re) \, \frac{1}{2}\rho u^2_{rel} c(r)
\end{equation}

\begin{equation}
\label{fD_def}
f_D = C_D(\alpha, Re) \, \frac{1}{2}\rho u^2_{rel} c(r)
\end{equation}

where $\rho$ is the density of air, $u_{rel}$ is the speed of the air relative to the blades, and $c(r)$ is the chord length of the blade at radial distance $r$ from the rotor centre.  This approach assumes a steady state response of the aerofoil to flow conditions, ignoring transient effects such as dynamic stall~\citep{CFC11} or tower shadow \citep{fruh2008}. Furthermore, rotational augmentation \citep{SchSoRo07,FruhCreech2015ICREPQ} is omitted at this stage as it is expected to be a minor correction at the operational conditions used here.

The relative speed $u_{rel}$ is calculated at each mesh node in $V$, and takes into account both rotation of the blades and of the incoming flow. For a node at a radial distance of $r$ from the rotor centre, this is written as

\begin{equation}
\label{eqn:urel}
	u_{rel} = \sqrt{(r\omega_{rel})^2 + u^2}
\end{equation}

The angular velocity component $\omega_{rel}$ is the angular velocity of the blade relative to the local angular velocity of the air, i.e. 

\begin{equation}
	\omega_{rel} = \omega - \omega_{air} 
\end{equation}

where $\omega$ is the angular velocity of the blades, and $\omega_{air}$ is the angular velocity of the air within the turbine volume $V$:
\begin{equation}
	\omega_{air} = \frac{1}{r^2} \left( yw - zv \right)
\end{equation}

\begin{figure}
\begin{centering}
\includegraphics[width = 0.45\columnwidth]{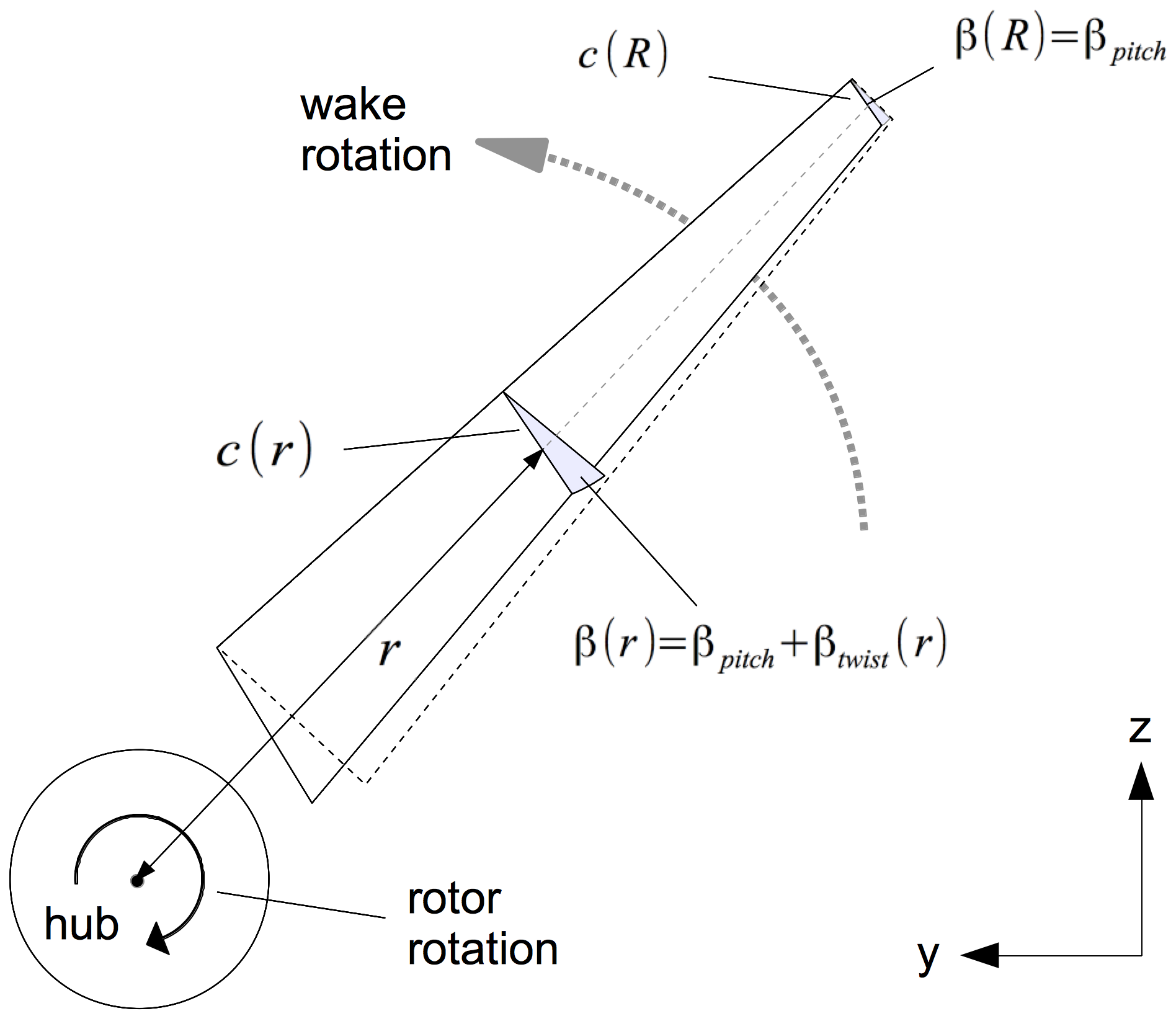} 
\caption{Diagram of a turbine blade showing chord, pitch and paths of motion. The dashed line represents a turbine blade with no twist parallel to the rotor plane, and the solid line the actual blade incorporating both pitch and twist. $\beta_{pitch}$ is taken clockwise from the rotor plane at the blade tip; $\beta$ incorporating blade twist is shown at distance $r$ from the hub centre. The rotor rotates in the opposite direction to the wake. The directions of the $y$ and $z$ axes are shown on the bottom right.}
\label{fig:blade-pitch-chord}
\end{centering}
\end{figure}

The inclusion of $\omega_{air}$ is due to Newton's third law. As lift and drag forces act to turn both the blades and the generator, so must an equal and opposite reaction force act on the flow, causing the air to rotate in the opposite direction of the blades, as can be seen in figure \ref{fig:blade-pitch-chord}. This, in turn, increases the magnitude of $u_{rel}$ quadratically, and so generating larger lift and drag forces, shown by equations (\ref{fL_def}) and (\ref{fD_def}). 

Whilst it has been demonstrated {\citep{sorensen2002}} that the azimuthal induction factor is small (5\%) for the most part of the blade under normal operating conditions and can be generally ignored, equation~(\ref{eqn:urel}) also remains valid near the blade root, and during start-up conditions where $\omega$ is small and the condition $u >> r\omega_{rel}$ cannot be guaranteed.

\begin{figure}
\begin{centering}
\includegraphics[width = 0.50\columnwidth]{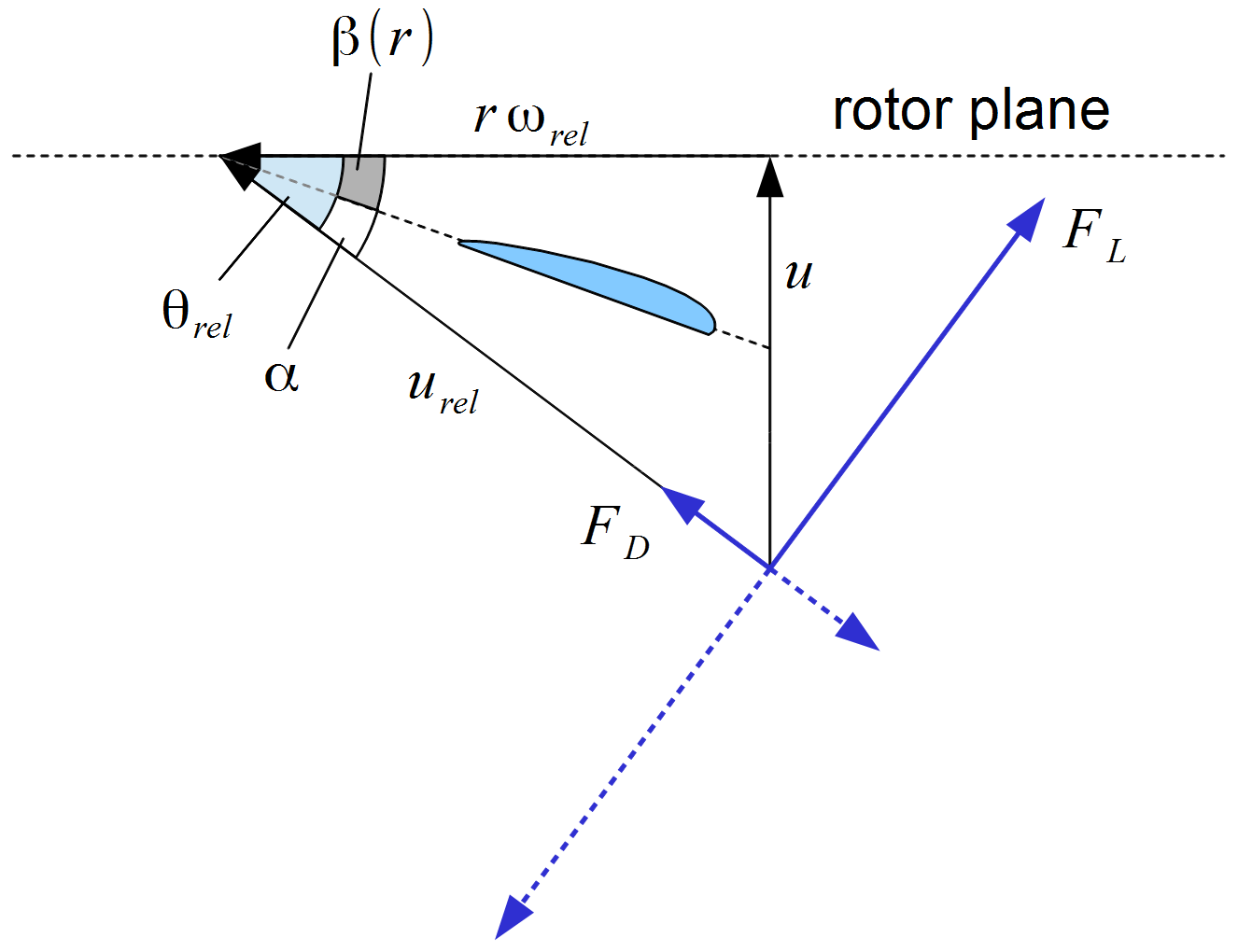} 
\caption{The relationship between the axial velocity component of the incoming air, $u$, the relative speed of the air, $u_{rel}$, and the relative azimuthal velocity $r \omega_{rel}$. The relative angle of the air flow to the rotor plane, $\theta_{rel}$, is the sum of the angle of attack $\alpha$ and local blade twist $\beta(r)$. The forces on the blades, $F_L$ and $F_D$, are indicated by solid blue lines; the dotted blue lines are the reaction forces acting on the air, which are opposite in direction but equal in magnitude. Note that $\beta$ can become negative when $r \omega_{rel}$ is large, so that an optimum angle of attack is maintained across the blade.}
\label{fig:alpha-beta-theta}
\end{centering}
\end{figure}

The relative flow angle of the air to the rotor plane, shown in Figure~\ref{fig:alpha-beta-theta},  is given as
\begin{equation}
	\theta_{rel} = \tan^{-1} \left(\frac{u}{r\omega_{rel}}\right)
\end{equation}

This allows us to compute the angle of attack as
\begin{equation}
	\alpha = \theta_{rel} - \beta
\label{eqn:alphabeta}
\end{equation}

where the local blade angle $\beta = \beta_{pitch} + \beta_{twist}$. The local blade twist angle $\beta_{twist}$ is a function of $r$, and calculated from the known turbine geometry; the methodology for determining this will be discussed in section~\ref{S:turbine-parameterisations}. The blade pitch angle $\beta_{pitch}$ is specified from the tip as shown in figure \ref{fig:blade-pitch-chord}, and is a dynamic variable altered through a blade pitch control mechanism -- this will be discussed in section~\ref{S:pitch}.

Returning to the lift and drag forces, we transform the lift and drag per unit length into body forces, i.e.~force per unit volume, so that they can be applied as force terms in the Navier-Stokes momentum equation. This gives

\begin{equation}
	F_{L} = \eta(x) \left( \frac{N_{blades}}{2\pi r}\right) f_L
\end{equation}

\begin{equation}
	F_{D} = \eta(x) \left( \frac{N_{blades}}{2\pi r}\right) f_D
\end{equation}

where $N_{blades}$ is the number of blades, and $\eta(x)$ is a Gaussian regularisation function similar to \citet{sorensen1998} and \citet{sorensen2002}. This only operates in the axial direction, as we are dealing with actuator discs and the influence of the blades is already spread azimuthally in a series of infinitely thin rings.
We define the regularisation function as
\begin{equation}
	\eta(x) = \frac{1}{\sqrt{2 \pi  \sigma^2}} e^{-\frac{1}{2} \left(\frac{x}{\sigma}\right)^2}
\end{equation}

where the standard deviation $\sigma$ controls the width of the filter. Smaller values of $\sigma$ gave greater accuracy in the axial force distribution, but as velocity and force interpolation was linear, this also required a prohibitively large increase in mesh resolution near the disc. During the turbine wind tunnel simulations detailed in section~\protect\ref{S:turbine-parameterisations}, iterative testing demonstrated that $\sigma=\frac{1}{2}L$, where $L$ is the length of the cylindrical volume, gave realistic turbine performance, whilst also allowing mesh resolutions that would permit the large domains necessary for wind farm simulation. Using explicit tip-loss correction is not necessary here; the use of CFD means the flow field upstream of the turbine is changed by the presence of the actuator forces, and so changes to the induction happen automatically \citep{sanderse2011}.

Writing down the azimuthal and axial components of the body force acting on the fluid,
which are in the opposite direction to the forces acting on the blade, we have 

\begin{equation}
	F_{azim} = - \left( F_{L} \sin \theta_{rel} - F_{D} \cos \theta_{rel} \right)
\end{equation}

\begin{equation}
	F_x = - \left( F_{L} \cos \theta_{rel} + F_{D} \sin \theta_{rel} \right)
\end{equation}

From $F_{azim}$ we can write the $y$ and $z$ components of these force terms as
\begin{equation}
	F_y = \frac{z}{r} \, F_{azim}
\end{equation}

\begin{equation}
	F_z = - \frac{y}{r} \, F_{azim}
\end{equation}

The force terms are then transformed back from the turbine reference frame to the simulation
reference frame, in an inverse operation of (\ref{eqn:velocity-rotation}) via 

\begin{equation}
\mathbf{F}^{*} =  \textbf{M}_{\psi} \textbf{M}_{\gamma} \: \mathbf{F}
\end{equation}

The body forces can now be applied to the momentum equation.

\subsubsection{Power, torque and thrust}\label{S:torque}

As the lift and drag exert forces on the blade, Newton's third law of motion dictates that there must be an equal and opposite reaction on the air; this reaction force is present at each point within the rotor volume $V$. This can be used to calculate the instantaneous power output of the turbine at time $t$, as shown in this section.  We start with the total torque acting on the fluid, i.e.

\begin{equation}
	\tau_{fluid} = \int^V \mathbf{r} \times \mathbf{F} \, dV = \int^V r F_{azim} \,  dV
\end{equation}

This torque must be balanced by $\tau_{power}$, the torque that turns the generator to create power, and $\tau_{blades}$, the torque due to the momentum of inertia of the blades.  These are resistive, i.e.~they are in the opposite direction of $\tau_{fluid}$, therefore

\begin{equation}
	\tau_{fluid} = - \left( \tau_{power} + \tau_{blades} \right)
\end{equation}

From \citet{CFC11} we use a simple model for the generator torque based on dimensional analysis:

\begin{equation}\label{eqn:torque-omega}
	\tau_{power} = k \omega^2
\end{equation}
where $k=\frac{P_{max}}{\omega_{max}^3}$ is a constant, $P_{max}$ is the maximum power output (eg. the rated power), and $\omega_{max}$ is the maximum angular velocity of the blades.  This gives us an expression for the instantaneous power output of the turbine

\begin{equation}
	P = \tau_{power} | \omega |
\end{equation}

Note that this formulation does not include any efficiency losses or
active generator control mechanisms, and assumes a direct relationship between blade angular velocity and power output. \citet{Hansen2012} show that for a Vestas V80, the maximum blade RPM is reached at $10\,\mathrm{ms}^{-1}$, whereas rated power is reached between $12.5-15\,\mathrm{ms}^{-1}$. For this paper our interest is in hub-height freestream wind speeds of $10\,\mathrm{ms}^{-1}$ and below, and in that regime the simple generator model is acceptable. Clearly a more realistic and manufacturer-specific formulation is required for higher wind speeds. This should be relatively straightforward once the generator behaviour is defined, requiring the replacement of the RHS of equation (\ref{eqn:torque-omega}) with a new model.

With the generator torque defined, we can return to the torque that accelerates the blades. Firstly, we define the moment of inertia of the blades
\begin{equation}
	I = N_{blades} \int^R r^2 m(r) \, dr
\end{equation}
where $m(r)$ is the mass-per-unit-span of the turbine blade. This is expressed as
\begin{equation}
\label{eqn:mass-per-unit-span-def}
	m(r) = \rho_{m} A(r)
\end{equation}
Where $A(r)$ is the cross-sectional area of the aerofoil, and $\rho_{m}$ is the mean blade material density. As both $c(r)$ and the aerofoil profile will be already known, we can numerically integrate to find $A(r)$, eg. by the trapezoidal rule.

The moment of inertia can now be determined, so we calculate the angular acceleration of the blades

\begin{equation}
	\dot{\omega} = \frac{\tau_{blades}}{I}
\end{equation}

With $\dot{\omega}$ we can then update $\omega$ at each time-step. In this paper, the simulations used an explicit two-step Adams-Bashforth integration method to calculate $\omega$ for the next time-step. The order of calculation from time-step $n$ to time-step $(n+1)$ can be described as

\begin{equation*}
\overbrace{\omega^{(n)} \longrightarrow \left( \tau^{(n)}_{fluid}, \, \tau^{(n)}_{power} \right) \longrightarrow \tau^{(n)}_{blades} \longrightarrow {\dot{\omega}}^{(n)}}^{\mathrm{time-step} \, n} \longrightarrow \overbrace{\omega^{(n+1)} \longrightarrow ...}^{\mathrm{time-step} \, (n+1)}
\end{equation*}

Lastly, as graphs of wind speed versus blade thrust are readily available for a number of wind turbines, they give us a useful measure of the model's correctness.  The thrust is obtained by integrating the axial body forces across the turbine volume, i.e.

\begin{equation}
	T = \int^V F_x(\mathbf x) \, dV
\end{equation}

This will be used in comparison with figures from an actual wind turbine in section~\ref{S:Turbine-validation}.

\subsubsection{Active pitch control}\label{S:pitch}

Like most modern utility-scale wind turbines, those at Lillgrund feature active pitch control, and so blade pitching was incorporated  into the turbine model to mimic this behaviour. Our wind farm simulations would only consider wind speeds below the power knee, i.e.~below speeds at which blade feathering occurs, so the active pitch algorithm would only need to optimise the blade pitch (abbreviated in this section only from $\beta_{pitch}$ to just $\beta$) for maximum lift. It is a complex optimisation problem, as the only \textit{a priori} variable is $\beta$.  The angle of attack $\alpha$ is \textit{a posteriori}, as it is a function of the time-dependent blade pitch, turbine performance and local flow conditions. This means that the optimal blade pitch $\beta_{opt}$ cannot be known beforehand without prior empirical measurements or calculation, neither of which are assumed to be available. The methodology below adapts the core arguments from \citet[Ch.4]{Creech2009} insofar as treating the blade pitching as damped harmonic oscillation, giving the solution not only of $\beta$ for maximum performance, but also that the rate of $\beta$ at $\beta_{opt}$ to be zero. The second condition ensures stability, by avoiding negative feedback between changes in $\beta$ and $\alpha$.

The first step is to define $\alpha_{opt}$, the optimum angle of attack at which the maximum lift occurs for minimum drag. This is straightforward to calculate from graphs of $C_L$ and $C_D$ for a particular aerofoil as

\begin{equation}
	\alpha_{opt} = \max\left[ C_L(\alpha, Re) - C_D(\alpha, Re)\right]
\end{equation}

This is related to the more traditional definition of optimal attack, $\alpha_{trad} = \max\left[ C_L / C_D\right]$, conventionally used for the design of the blade twist, but it is not equal, as the target is here used to determine best blade pitching in a situation where the actual angle of attack varies across the rotor area.

For this reason, the next step is to calculate the weighted average of the angle of attack across the blades, $\alpha_{wt}$.
The weighting is necessary as the aim is to maximise lift rather than simply ensuring that the mean angle of attack 
$\alpha$ across the blades is as close to $\alpha_{opt}$ as possible, which could plausibly result in sub-optimal 
performance. The weighting must consider the factors that increase lift, such as chord length and relative air speed, 
so at each mesh node $i$ within $V$ it is defined as

\begin{equation}
	w_i = c(r_i) u_{rel, i}^2
\end{equation}

Using the sum of weights, $W$, 
\begin{equation}
	W = \sum_i w_i
\end{equation}

gives the weighted average as

\begin{equation}
	\alpha_{wt} = \frac{1}{W} \sum_i w_i \alpha_i
\end{equation}

This emphasises the values that currently give greatest lift. Now we define the desired angle of attack $\alpha_d$, i.e. the angle of attack that the algorithm will aim for.  As we are not considering blade feathering in these simulations, where the lift is reduced by lowering the angle of attack below the optimum for lift, we set this to~$\alpha_d = \alpha_{opt}$. 

We define the maximum pitching rate of the blade, $|\dot{\beta}|_{max}$ below, by setting the shortest time a blade can pitch through one full rotation, $t_{pitch}$:

\begin{equation}
	|\dot{\beta}|_{max} = \frac{2 \pi}{t_{pitch}}
\end{equation}

The value of $t_{pitch}$ had to be chosen with care, as too small a value would cause unstable oscillations in blade pitch. For all simulations in this paper, $t_{pitch}=10 \, \mathrm{s}$. 
If we assume that as the timestep $\Delta t \to 0$, so $|\Delta \beta| \to |\Delta \alpha|$,
i.e.~over small periods of time, changes in the blade pitch $\beta$ lead to a change of equal magnitude in
the angle of attack $\alpha$. From eq. (\ref{eqn:alphabeta}) these changes are opposite in sign, so in the 
limit, we also state generally that rate of change of pitch  $\dot{\beta}$ is equivalent to the 
negative of the rate of change of angle attack $\dot{\alpha}$, i.e.

\begin{equation}
	\dot{\beta} \approx -\dot{\alpha}
\label{e:dotbetaalpha}
\end{equation}

The desired rate of change of attack  $\dot{\alpha}_d$ is stated as

\begin{equation}
	\dot{\alpha}_d = \frac{\alpha_d - \alpha_{wt}} {t_{pitch}}
\end{equation}

This ensures that smaller differences between $\alpha_d$ and $\alpha_{wt}$ result in smaller changes in the angle of attack, i.e. aiming for no change in angle of attack at $\alpha=\alpha_d$. If we write the desired change in blade pitch as an equal weighting of the current, known rate of change of pitch $\dot{\beta_{k}}$, and the desired rate $\dot{\beta}_d$, we can write

\begin{equation}
	\Delta \beta_d = \Delta t \left( \frac{\dot{\beta_k} + \dot{\beta_d}}{2} \right)
\end{equation}

Through our equivalence relation in (\ref{e:dotbetaalpha}), we define $\dot{\beta}_d \approx -\dot{\alpha}_d$.

As a final precaution, the rate of change in the pitch is limited by $|\dot{\beta}|_{max}$, 
so defining the maximum change in pitch as 
$\Delta \beta_{max} =  \mbox{sign}(\Delta \beta_d)  \, \Delta t \, |\dot{\beta}|_{max}$, the
actual change in pitch is

\begin{equation}
	\Delta \beta =  \left\{ \begin{array}{ll}
		\Delta \beta_d & \mbox{if} \: |\Delta \beta_d| \leq  |\Delta \beta_{max}| \\
		\Delta \beta_{max} & \mbox{if} \: |\Delta \beta_d| >  |\Delta \beta_{max}|
	\end{array} \right.
\end{equation}

Therefore the change in blade pitch from timestep $n$ to $n+1$ will be

\begin{equation}
	\beta_{n+1} = \beta_n + \Delta \beta
\end{equation}

\subsubsection{Blade-generated turbulence}\label{S:Blade-turbulence}

\begin{figure}
\begin{centering}
\includegraphics[width = 0.5\columnwidth]{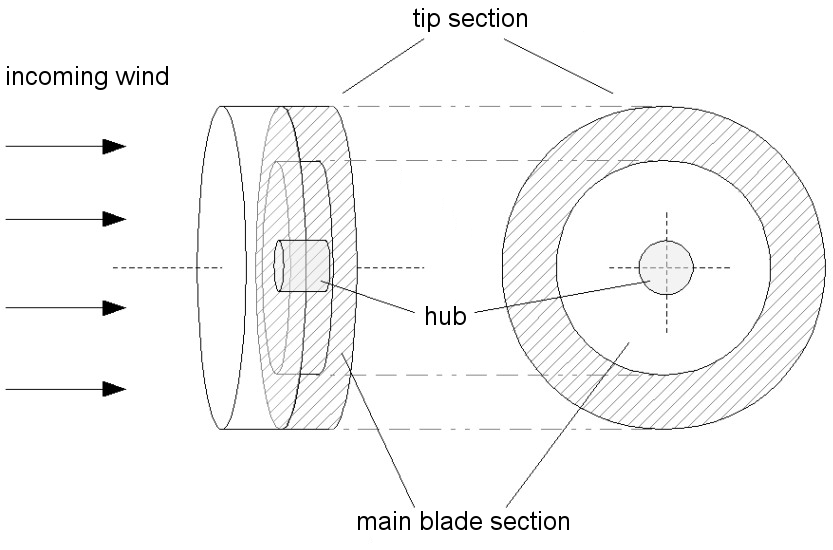} 
\caption{The turbulence-generating regions of the turbine model volume, based on \citet{CFC11}: i) the tip, where the turbulent intensity is highest, ii) the main blade section, which creates turbulence approximately half that of the tip-section, and iii) the hub section.}
\label{fig:actuator-turbulence-regions}
\end{centering}
\end{figure}

Blades in real turbines generate turbulence, particularly at the tips. However, as blades are not explicitly represented in the model, blade-induced turbulence must be described parametrically. In an approach used in previous work \citep{Creech2009,CFC11}, random fluctuations in the flow passing through the turbine volume are created by body forces, which match turbulent intensity measurements in experiment \citep{hossain2007}.  Turbulence generation in the model is divided into three sections - the tip ($r > 0.9 R$), the main blade section ($0.1 R < r \leq 0.9 R$), and the hub at $r \leq 0.1 R$, as shown in Figure~\ref{fig:actuator-turbulence-regions}.

The approach used will be briefly detailed here. A turbulence intensity function is defined

\protect\begin{equation}
\mathbf{Ti}(r, \omega, Ti_{max}) = \begin{bmatrix}
	Ti_x\\
	Ti_y \\
	Ti_z
\end{bmatrix}
\end{equation}

Which varies with $r$, $\omega$ and $Ti_{max}$ the predetermined maximum turbulence intensity, such that $\mathbf{Ti}=0$ at $\omega=0$ and at its maximum values reach at $\omega=\omega_{max}$. This is then used to calculate the random variations in velocity which statistically match the specified blade-generated turbulence intensity. In the case of the axial velocity component, this gives

\protect \begin{equation}
\Delta u_{turb} = G_x (Ti_x) u
\end{equation}

Where $G_x$ is a coherent Gaussian-noise algorithm taken from \protect \citet{fox2007}. $\Delta v_{turb}$ and $\Delta w_{turb}$ are similarly defined. These fluctuations are then translated into body force terms which are then added to the body forces defined in section \ref{sub: liftdrag}. Further details on this approach and its validation with wake data can be found in \citet{CFC11}.

It should be noted that experimental analysis has shown that the hub/root region of the turbine generates vortices, and thus significant levels of turbulence \citep{zhang2012, iungo2013}. This is to be expected due to the interaction of the flow with the blade root and the hub, a bluff body. We do not actively generate turbulence within the hub volume here, but nonetheless increased levels of vorticity near the blade root are present in simulations. Including the solid structure of the hub is at present prohibitively expensive, as it requires a very fine mesh resolution to resolve the hub geometry and the flow within the hub's boundary layer. However, we hope to include it in future work to assess its contribution to wake recovery.
\section{Turbine parameterisation} 
\label{S:turbine-parameterisations}

In this section, we detail the techniques used to create the model parameters for the turbines at Lillgrund wind farm. As complete specifications for the Siemens SWT-2.3-93 turbines used in Lillgrund \citep{assesslillgrund2009} are not publicly available, model parameters were validated by testing candidate turbines in a virtual wind tunnel, then comparing their performance with measured power and thrust data. The final parameters with which the turbine model was configured are shown in table~\ref{tab:siemens-params}. The rationale for the choice of aerofoil section and details of the blade geometry are explained in sections~\ref{S:Aerofoil}  and~\ref{S:Blade} respectively

\begin{table}
\centering
\caption{General model turbine specifications for Siemens SWT-2.3-93.}
\begin{tabular}{ll}
\hline\noalign{\smallskip}
Property & Value \\
\hline\noalign{\smallskip}
Rotor radius & 46.5 m \\
Hub height & 65 m \\
Rotor tilt & 6\degree \\
Aerofoil type & NACA 63-415 \\
Hub fraction ($r_H / R$) & 0.1 \\
Blade material density & 100 kg/m$^3$ \\
Cut-in wind speed & 4 m/s \\
Cut-out wind speed & 25 m/s \\
Design tip-speed ratio & 6.2329 \\
Maximum power & 2.3 MW \\
Wind speed at which max. power occurs & 14 m/s\\
\hline\noalign{\smallskip}
\end{tabular}
\label{tab:siemens-params}
\end{table}

\subsection{Aerofoil}\label{S:Aerofoil}

The aerofoil types used by the SWT-2.3-93 turbine are specified in \citep{planning-app} as 
`NACA 63.xxx, FFAxxx'. The FFAxxx series are thick aerofoils designed to bear high loads in the inboard part of the turbine blade \citep{risotech1998}. No information was available as to which FFA blade was used in the Siemens turbine, nor the extent of the blade that used it. Because of this, and because the inboard section generates a small portion of the total thrust, the same aerofoil type - NACA 63 series - was used across the whole blade. 

There were many candidate NACA 63 aerofoils, but eventually NACA 63-415 was chosen, as shown in figure \ref{fig:naca-63415}. This was based upon several factors: indication in literature that this is a common aerofoil used in modern turbines \citep{risotech2001}, desirable lift characteristics, and visual comparison of the NACA 63-415 profile with photographs of B45 blades.

The modelled lift and drag characteristics are a compound of various sources.  Initially, XFOIL \citep{xfoilpaper,drela1987,xfoilwebsite} plots of $\alpha$ versus $C_L$ and $C_D$ were used over the range $-10\degree < \alpha < 20\degree$.   When these were compared to the Ellipsys2D and measured results in the Airfoil Catalogue \citep{risotech2001}, major discrepancies were found even at low angles of attack, and in particular at and above $\alpha_{opt}$.

\begin{figure}
\begin{centering}
\includegraphics[width = 0.5\columnwidth]{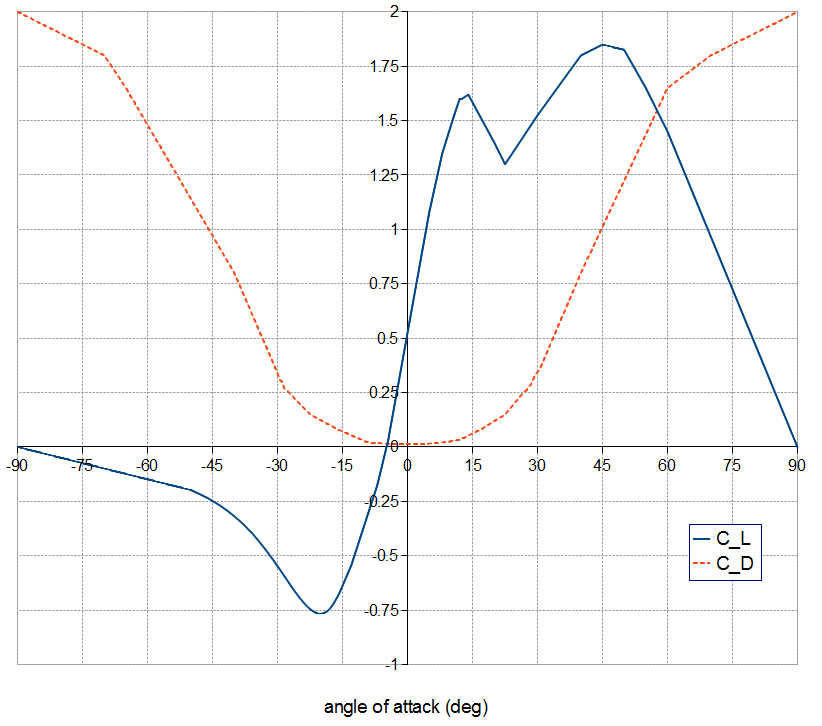} 
\caption{The lift (blue) and drag (red) coefficients as function of angle of attack for the NACA 63-415 aerofoil at $\operatorname{Re}=3 \times 10^6$.}
\label{fig:aerofoil-lift-drag}
\end{centering}
\end{figure}

It was theoretically possible that the model may experiences angles of attack outwith this range, and so the modelled aerofoil could not be based solely upon the data in the Airfoil Catalogue, nor indeed the original NACA sources. Extreme values of attack are not experienced during normal operation, but lift and drag coefficients are nonetheless required for all possible values of $\alpha$ to prevent unpredictable behaviour in the model. Firstly JavaFoil \citep{javafoilwebsite} was used to plot both lift and drag for extreme angles of attack within $-90\degree<\alpha< 90\degree$ for $Re = 3 \times 10^6$. A secondary source was a report into aerofoil characteristics at extreme angles of attack \citep{sandia2001} providing data for $180\degree$ for NACA symmetrical blades; whilst these have rather different aerodynamic properties, the same report concludes that at high angles of attack ($\alpha \gg 30 \degree $) aerofoils effectively behave as flat plates. This means they can be modelled similarly. After several iterations of aerofoil parametrisation and verification of modelled turbine performance, the lift/drag graphs in figure \ref{fig:aerofoil-lift-drag} were found to be the most accurate.

\subsection{Blade geometry}
\label{S:Blade}

The SWT-2.3-93 uses Siemens' own B45 blade with active pitch correction.  From Siemens' brochure \citep{siemens-brochure} and a technical specification published in a planning application \citep{planning-app}, rotor diameter,  maximum RPM and rated wind speed were noted in order to calculate the optimum tip-speed ratio, as shown in table~\ref{tab:siemens-params}. These gave a plausible value of 6.2329.

\subsubsection{Twist angle}

To calculate the blade twist angle, we start with the predicted flow angle $\phi$ as defined in \citet[\S~3.7.2]{windenergyhandbook}:

\begin{equation}
	\tan \phi(r) = 
	\frac{1-\frac{1}{3}}{\lambda \mu \left( 1+ \frac{2}{3\lambda^2\mu^2}\right)}
\end{equation}

where $\lambda$ is the design tip speed ratio, and $\mu = \frac{r}{R}$. Using this with the 
optimum angle of attack $\alpha_{opt}$, gives the ideal blade twist $\beta_{ideal}$:

\begin{equation}
	\beta_{ideal} = \tan^{-1} \left[ 
		\frac{1 - \frac{1}{3}}
			{\lambda \mu \left( 1+ \frac{2}{3\lambda^2\mu^2}\right)}
	 \right]
	 - \alpha_{opt}
\end{equation}

In practice, this equation gives $\beta(r) \approx \beta_{ideal}(r)$ for $r > 0.2\,R$, twice the hub fraction (table \ref{tab:siemens-params}). Below this value of $r$ however,  $\beta$ was iteratively increased in test simulation, until the model maintained optimum angles of attack for $r_H < r < R$ in test simulations, giving the final twist angles shown in figure \ref{fig:b45-chord-twist}.

\subsubsection{Chord length}

\begin{figure*}
\centering
\begin{minipage}{.56\textwidth}
  \centering
  \includegraphics[width=.98\linewidth]{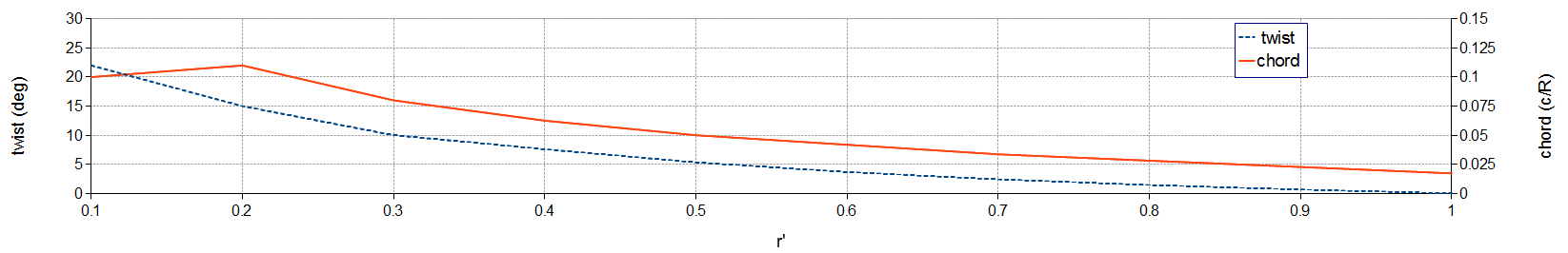} 
  \caption{Chord length and twist angle of the B45 blade as a function of r'=r/R.}
	\label{fig:b45-chord-twist}
\end{minipage}
\begin{minipage}{.01\textwidth}
\hfill
\end{minipage}
\begin{minipage}{.41\textwidth}
  \centering
  \includegraphics[width=.98\linewidth]{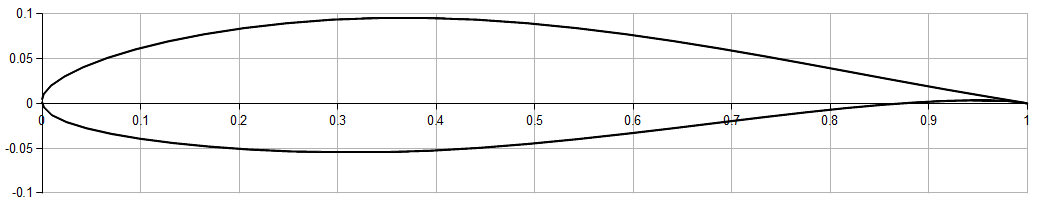} 
  \caption{The NACA 63-415 profile.}
	\label{fig:naca-63415}
\end{minipage}
\end{figure*}

An exact specification for the chord length as it varies from hub-to-tip was not available; however the chord lengths at the hub and tip were given in \citep{planning-app}. Further information on chord length was taken from \citet{laursen2007}, and a near-linear tapering blade was assumed, shown in figure \ref{fig:b45-chord-twist}.

\subsection{Turbine validation}
\label{S:Turbine-validation}
A strong indication that the turbine model is working effectively is that it will generate thrust and power values for different wind speeds that match measured data. Being entirely reactive, the model has an algorithm that continually changes blade pitch in response to wind conditions, so that at lower speeds it will aim to maximise lift. In turn, this will affect the dynamically changing values for rotor RPM, power output, and other turbine diagnostics. In theory, this means by altering the inflow wind speeds only, the model should produce equivalent performance to that of the real turbine in similar conditions.

By taking the manufacturer's CT and CP curves for the SWT-2.3-93 and extrapolating thrust and power as functions of the upstream hub-height wind speed $u_0$, we can directly compare the time-averaged values for power and thrust from the model, when both the wake and turbine itself are dynamically stable.

\begin{figure}
\begin{centering}
\includegraphics[width=0.6\columnwidth]{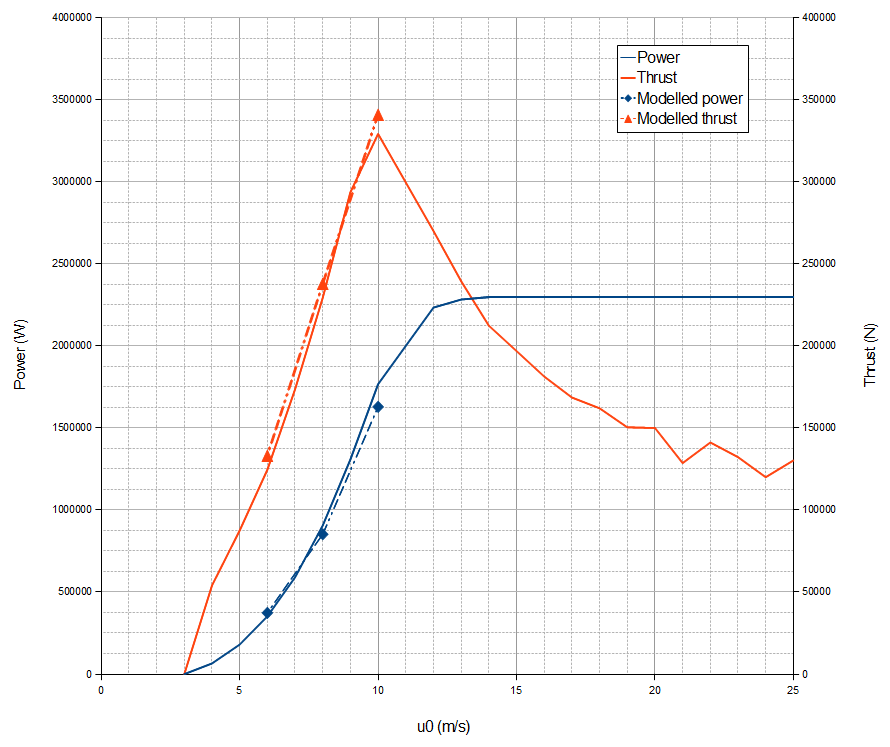} 
\caption{The power (red) and thrust (blue) of the modelled and actual turbines, as a function of wind speed. The solid lines represent published turbine performance data, the dotted lines the time-averaged model diagnostics.}
\label{fig:power-thrust-comparison}
\end{centering}
\end{figure}

The model was run in a simulated wind tunnel 1 km long, with a cross-section of 250 m x 250 m.  It had with a logarithmic inlet velocity profile, which was specified as a function of hub-height wind speed $u_0$: simulations were run at $u_0= 6, 8 {\rm~and~} 10\rm~m/s$, to cover typical wind speeds experienced at Lillgrund. The turbine was set to an initial RPM of 0, and to a blade pitch of 90\degree. The simulations were run until at least 300s of simulation time had passed with relatively stable power and thrust values. The average of the power and thrust over the final 300s are plotted against calculated averages from \citet{assesslillgrund2009} in Figure~\ref{fig:power-thrust-comparison}.  It is clear that both, the modelled power and thrust, closely follow the given specifications. The relative errors between the model and given values are shown in table \ref{tab:power-thrust-errors}. Especially considering that the precision in the reference values provided is relatively low and that the wind turbine response is very sensitive to the wind speed, the agreement between the turbine model and the manufacturer's specification are well within the uncertainty expected from the specifications. Therefore the agreement between the modelled SWT-2.3-93 turbine and the observations is acceptable for our purpose.

\begin{table}
\caption{Comparison of relative errors between actual and modelled turbine power (P) and thrust (T).}
\centering
\begin{tabular}{lllllll}
\hline\noalign{\smallskip}
$u_0$ (m/s) & $P_{actual}$ (kW) & $P_{model}$ (kW) & Relative error & $T_{actual}$ (kN) & $T_{model}$ (kN) & Relative error \\
\hline\noalign{\smallskip}
6 & 352 & 373 & 5.6 \% & 125 & 133 & 6.0 \% \\
8 & 906 & 852 & 6.0 \% & 229 & 234 & 3.8 \% \\
10 & 1767 & 1629 & 7.8 \% & 329 & 341 & 3.6 \% \\
\hline\noalign{\smallskip}
\end{tabular}
\label{tab:power-thrust-errors}
\end{table}

\section{Empty domain} 
\label{S:Empty}

Before modelling the wind farm, an empty domain without wind turbines was run for two hours of simulation time. This allowed fully turbulent flow to evolve across the entire volume, which would then be checked for correctness. At the end of the run a checkpoint was created, acting as a starting point for the full wind farm simulations; here, the problem was remeshed to accommodate finer resolution near the modelled wind turbines. This was a relatively straightforward process due to Fluidity's hr-adaptive meshing techniques and check-pointing capability.  As the present simulations concerned a neutrally stable atmosphere, buoyancy effects do not need to be included (e.g., \protect\citet{Wu:2013gb}) and a standard logarithmic velocity profile can be used for the inlet conditions with matching lower boundary conditions.

\subsection{Simulation volume}

\begin{figure}
\begin{centering}
\includegraphics[width=0.75\columnwidth]{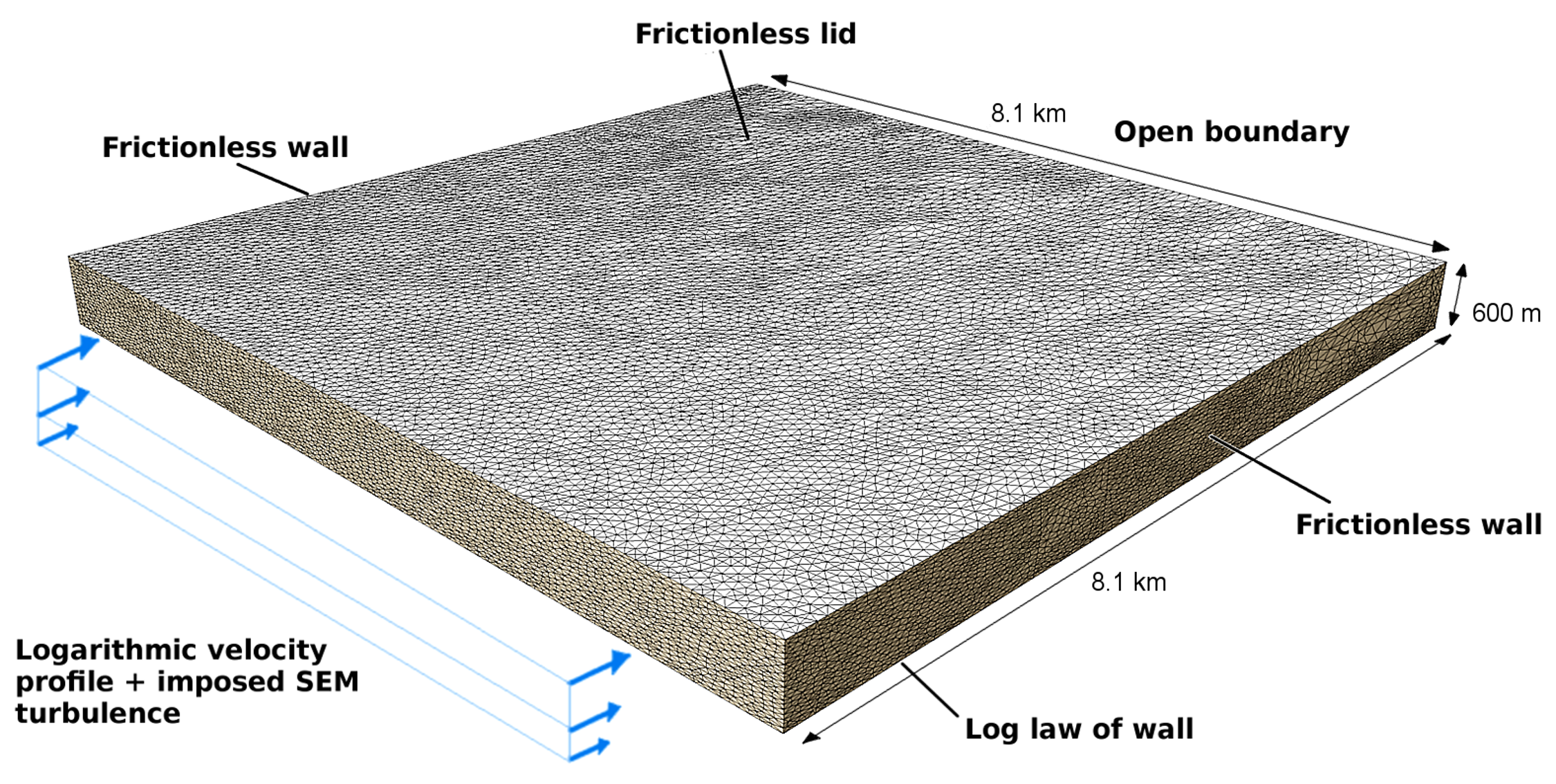} 
\caption{Empty simulation domain showing boundary conditions, measuring 8.1 km x 8.1 km x 600 m. Mesh cell dimensions were approximately 75 m x 75 m x 25 m.}
\label{fig:lillgrund-empty-domain}
\end{centering}
\end{figure}

The maximum extent of Lillgrund windfarm is 2.7 km from east to west. To ensure that no blockage effects would occur, the horizontal dimensions of the simulation domain were chosen to be 8.1 km in both horizontal directions.  This would ensure a large extent of open sea on each side of the wind farm, as well as sufficient space downwind for wake effects to be modelled. For the domain height, \protect\citet{fitch2013} presented depths of the atmospheric boundary layer ranging from around 100~m for stable conditions up to over 1000~m for unstable conditions.  To ensure a sufficient domain height, while working within the constraints of the available computing resource, wind engineering reference guidelines \protect\citep{cabezon2011} which would be appropriate for neutral conditions were used. \protect\citet{cabezon2011} suggested $5H$, where $H$ is the height of any obstacle obstructing flow.  In the Lillgrund simulations, the obstacle height would be the height of the wind turbine hubs plus the radius, so that $H=111.5\rm~m$.  To leave an acceptable margin for error, a height of 600~m was chosen, which meant the simulation domain was 8.1 km x 8.1 km x 600 m, as shown in Figure \protect\ref{fig:lillgrund-empty-domain}. While \protect\citet{calaf2010}, \protect\citet{Churchfield2012} and \protect\citet{Archer2013} adopted the compromise to resolve more of the unstable atmospheric boundary layer with domain heights of 1000~m at the expense of a much more constricted horizontal extent, one of our goals was to include more of the wind farm wake which required a larger horizontal extent. Observations reported by \protect\citet{iungo2012} as well as experiments by \protect\citet{chamorro2011}, simulated by \protect\citet{Wu:2013gb}, suggested that this compromise would be acceptable.

\subsection{Boundary and initial conditions}

\subsubsection{Sea surface}
\label{S:lillgrund-sea-surface}

The sea surface was specified as a rough wall boundary condition, with a roughness height $z_0$, which represented the drag induced by the surface's roughness. In reality this surface has waves, whose composition and frequency is affected by parameters such as mean wind speed, gusting, and wave age. This, in turn, has a reciprocative effect on air flow over the waves. However, for the sake of simplicity a single time-independent value of $z_0$ was chosen, which was cross-checked against published data for similar wind speed regimes \citep{makin1995, mahrt1996}, as shown later in this section.

\begin{figure}
\centering
  \includegraphics[width=0.6\textwidth]{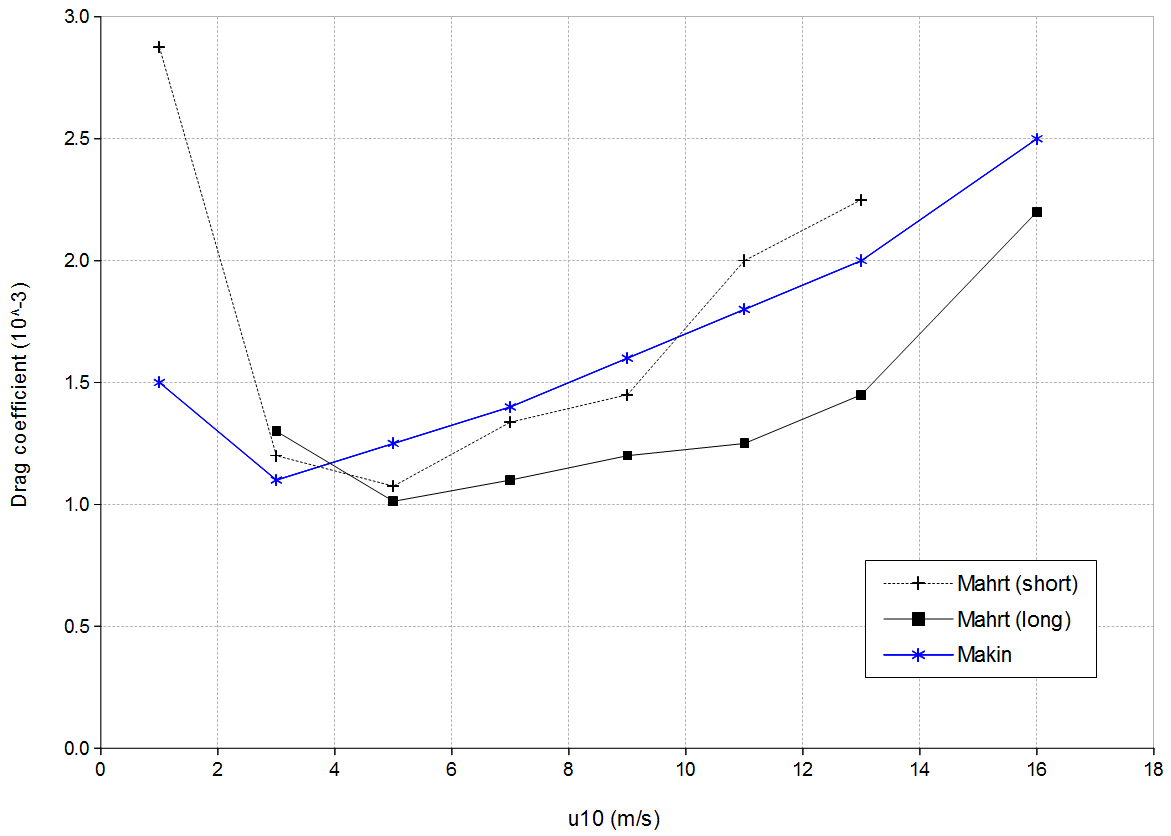} 
  \caption{Drag coefficients of the sea surface as a function of $u_{10}$, the wind speed at 10~m above sea level. The solid black line and the dotted line show the drag coefficients across long and short fetches, using data derived from \citet{mahrt1996}. The blue line represents the drag coefficients for a fully-developed sea, i.e. with an extremely long fetch, from \citet{makin1995}.}
	\label{fig:drag_coeff}
\end{figure}

The waves were considered to be in relatively open sea, which given the long fetch (approx. 10~km or greater) towards coastlines shown in figure \ref{fig:lillgrund-location} is a reasonable assumption. This is an important choice as fetch, along with wind speed, has been shown to affect the surface drag \citep{mahrt1996} and, with it, $z_0$. From \citet{makin1995}, the surface drag coefficient $C_{D, sea}$ can be related to the roughness height by
\begin{equation}
C_{D, sea} = \left[ \frac{K}{\ln(z_{R}/z_0)} \right]^2
\label{eqn:cd-sea}
\end{equation}
using the standard reference height of $z_{10} = 10 \operatorname{m}$, where $K\approx 0.41$ is the von Karman constant. The information from \citet{mahrt1996} and \citet{makin1995} is collated in Figure~\ref{fig:drag_coeff}.  

To determine the correct equivalent 10 m reference wind speed, $u_{10}$, the log law for turbulent flow was used as a starting point, ie.
\begin{equation}
	u=\left(\frac{u_{\tau}}{K} \right) \, \operatorname{ln} \left( {\frac{z}{z_0}} \right)
	\label{eqn:u10}
\end{equation}
The frictional velocity, $u_{\tau}$, can be calculated by substituting in $\bar{u}_H$ and $z_H$:
\begin{equation}
u_{\tau} = \frac{\overline{u}_H K}{ln \left(\frac{z_H}{z_0}\right)}
\end{equation}
where $z_H = 65\, \textrm{m}$ is the hub height, and $\overline{u}_H = 10\, \mathrm{m/s}$ was specified as the mean freestream wind-speed at hub height; $\overline{u}_H$ is discussed further in the next section.  If a roughness height of $z_0=2 \times 10^{-4} \, \mathrm{m}$ is chosen, $u_{\tau}$ is defined and can substituted into (\ref{eqn:u10}) to give the mean speed at 10 m as $u_{10}=8.524 \, \mathrm{m/s}$. 

Using equation (\protect\ref{eqn:cd-sea}) this leads to a surface drag coefficient of $C_{D,sea}=1.436 \times 10^{-3}$ which is in good agreement with \citet{mahrt1996} and \citet{makin1995}, as can be seen from Figure \protect\ref{fig:drag_coeff}.

\subsubsection{Inflow wind conditions}
\label{S:Inflow}

\begin{figure}
\begin{centering}
\includegraphics[width=0.55\columnwidth]{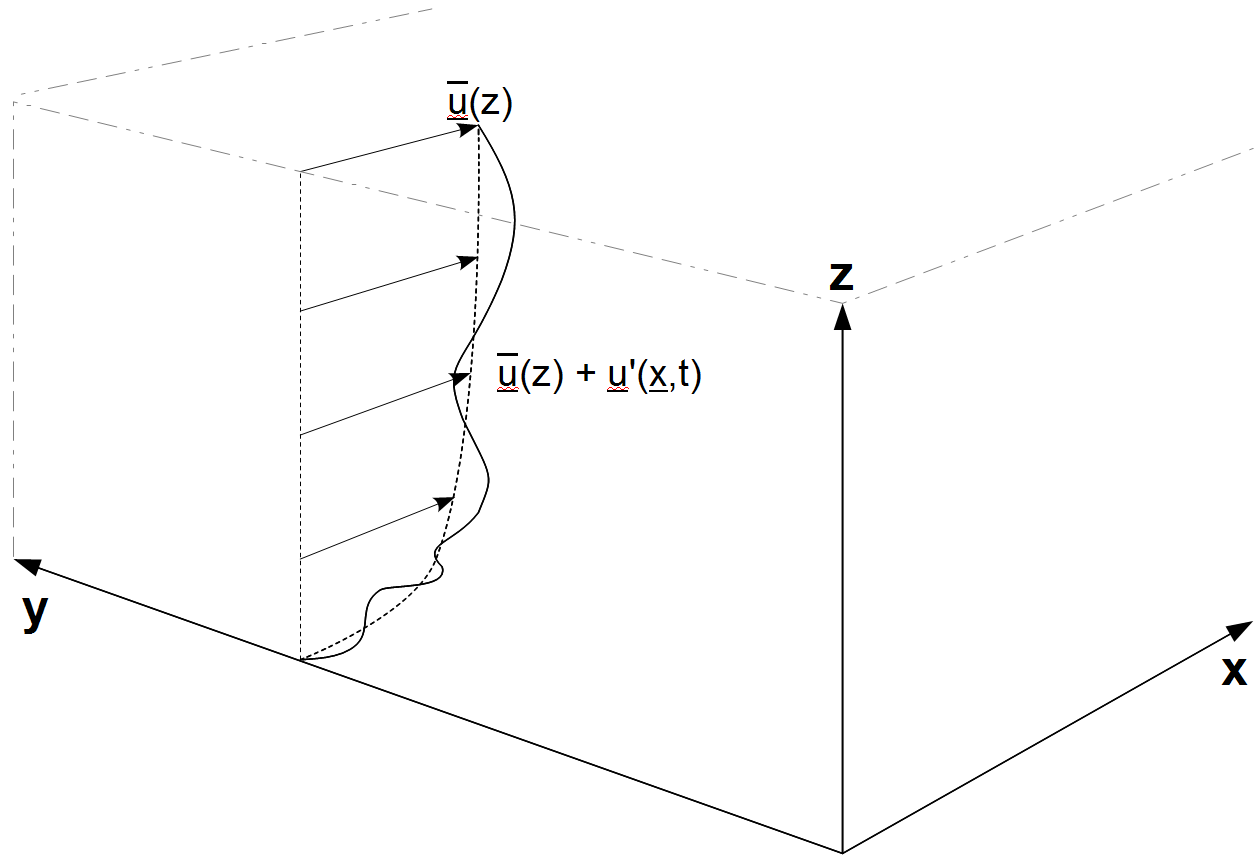} 
\caption{A synthetic-eddy method (SEM) generated velocity profile at the inflow boundary. The dotted curved line represents the mean logarithmic profile; the irregular solid line represents the velocity profile with SEM fluctuations superimposed.}
\label{fig:lillgrund-sem-profile}
\end{centering}
\end{figure}

At the start of each simulation, the wind velocity is set to $0~\mathrm{m/s}$ across the domain. The inflow conditions were specified as a mean velocity profile with a fluctuating component applied to it, as shown in Figure \ref{fig:lillgrund-sem-profile}. The mean velocity profile was specified as

\begin{equation}
\label{eqn:mean-profile}
\overline{\mathbf{u}}(z) = \left[ \begin{array}{c}
\left(\frac{u_{\tau}}{K} \right) \, 
		\operatorname{ln} \left( {\frac{z}{z_0}} \right) \\
		0 \\
		0
\end{array}
\right]
\end{equation}

To calculate the profile, the mean wind speed at hub height $u_H$ was taken as fixed at $\overline{u}_H=10\rm~ m/s$ for each simulation.  The key choice for this was to operate the turbines at a substantial power output but below the power curve knee at 12 m/s (cf.~Figures~\ref{fig:PerformanceCurve} and~\ref{fig:RelPerformance_U}).  With $u_{\tau}$, $K$ and $z_0$ already known from \S~\ref{S:lillgrund-sea-surface}, the profile for $\overline{\mathbf{u}}(z)$ is now completely specified.

For the fluctuating component, as the model used wall-adapted local eddy (WALE) LES \citep{nicoud1999} to model turbulence, the turbulence at the inlet had to be explicitly generated through the synthetic eddy method \citep{jarrin2006} at the inflow boundary, shown in figure \ref{fig:lillgrund-sem-profile}. There were two main sets of parameters  which controlled this turbulence generation, namely the turbulence length scales and the Reynolds stress profiles. 

\begin{figure}
\begin{centering}
\includegraphics[width=0.5\columnwidth]{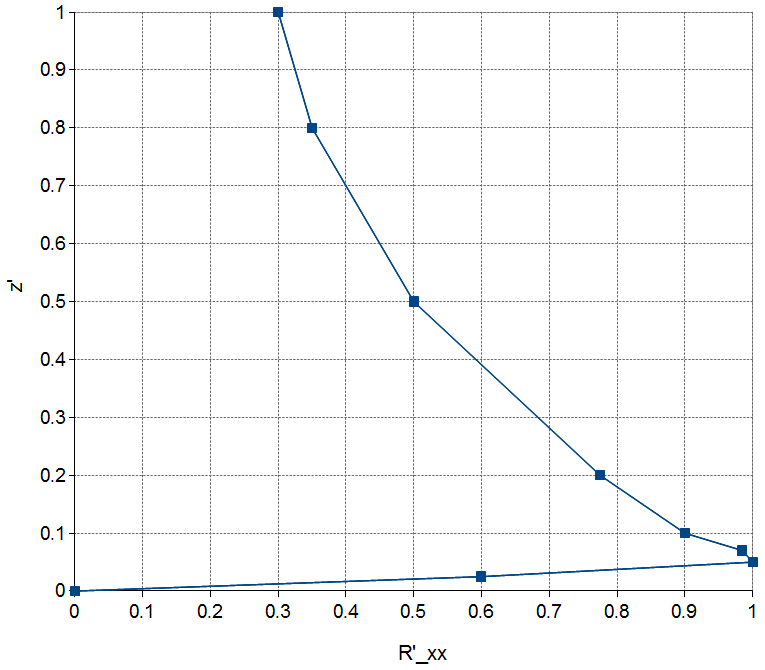} 
\caption{The normalised Reynolds stress profile $R'_{xx}$, as a function of normalised height $z'$, derived from \citet{pavlidis2010}. The squares represent the specified data points, and linear interpolation was used to reconstruct a continuous profile.}
\label{fig:reynolds-stress-profile}
\end{centering}
\end{figure}

The Reynolds stress tensor profiles were based on \citet{pavlidis2010}, with the diagonal components $R_{xx}$, $R_{yy}$ and $R_{zz}$ components specified; the normalised profile for $R_{xx}$ is shown in Figure \protect\ref{fig:reynolds-stress-profile}. According to the same paper, the non-diagonal components of the stress tensor are impractical to specify accurately, but only have a minor influence on flow far downstream and can be omitted.

The mean lengthscale components, $L_{1u}$, $L_{1v}$ and $L_{1w}$ were taken from the Danish standard DS 472, as specified in \protect\citep[p.24]{windenergyhandbook}, which gave these as:

\begin{equation}
\begin{array}{lcl}
L_{1u} & = & \left\{ \begin{array}{ll}
5 z & \mbox{for } z < 30\rm\,m\\
150{\rm\, m} & \mbox{for } z \geq 30\rm\,m\\
\end{array}\right.\\
L_{1v} & = & 0.3\,L_{1u} \\
L_{1w} & = & 0.1\,L_{1u}
\end{array}
\label{eqn:turblengths}
\end{equation}

This gave the mean length scales as a function of height from above the sea surface.

\subsection{Domain validation}
\label{s:domain-validation}

Validating the empty domain represented a challenge, its main purpose to provide realistic wind conditions at the site of the wind farm. Those conditions would be sensitive to sea surface boundary conditions, inflow conditions, mesh resolution and turbulence parameters, and arriving at the appropriate combination of these was a process of successive testing and refinement. Several criteria were formulated in order to demonstrate whether the empty domain simulation was working correctly, and that it had been run for long enough. These were limited by constraints on both time and computing resource, due to the volume of data involved.

\begin{figure}
\centering
  \includegraphics[width=0.6\textwidth]{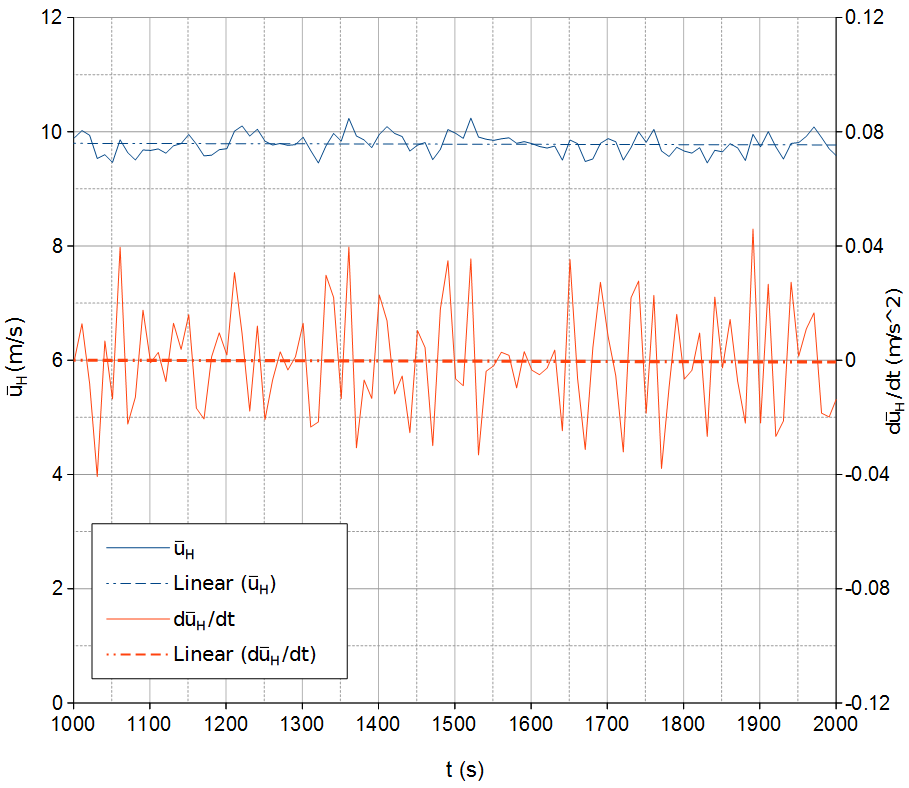} 
  \caption{Plots of spatial average $\bar{u}_H$ (blue line) and $d\bar{u}_H / dt$ (red line) From t=1000s to 2000s. The dashed lines are linear regression fits. Over this time period, $\bar{u}_H$ can be seen to fluctuate around $9.825 \mathrm{ms^{-1}}$, and that the trend for $d\bar{u}_H / dt$ is close to $0 \, \mathrm{ms^{-2}} $.}
	\label{fig:uH_convergence}
\end{figure}

Firstly, to show that the flow was fully-developed, the mean flow speed was calculated as an instantaneous spatial average at hub-height by slicing through the velocity field every 10 time-steps, and this was plotted as a function of simulated time, along with its derivative and linear regression fits, in Figure \ref{fig:uH_convergence}. This graph is plotted from t=1000~s to t=2000~s of simulation time, and it can be seen that  $\bar{u}_H$ has converged towards a constant value, since the linear regression of its temporal derivative $d\bar{u}_H / dt$ over this period is effectively 0. Moreover, the linear regression of $\bar{u}_H$ gives a value of $\bar{u}_H=9.825 \, \mathrm{ms}^{-1}$, which is within 2\% of the intended value of $10\, \mathrm{ms}^{-1}$.  Further to this, calculations of the turbulent intensity near where the wind farm would be showed a turbulence intensity at hub-height of 8\%, which is close to that measured upwind of comparable offshore windfarms \citep{Hansen2012}. Lastly, there was a degree of overlap between the empty domain and full wind farm simulations. As a final test of the empty domain conditions, a preliminary full farm simulation at a wind direction of 223\degree was run, where the rows are aligned with the wind and wake effects would be dominant. The turbulence lengthscale and Reynolds stress profiles were tuned in the precursory empty domain simulations, and the full-farm re-run until there was good agreement with SCADA data in Row D. This transpired to be an important test, as too little upstream turbulence resulted in overly pronounced wake deficits and reduced wake recovery.

\section{Full farm model}
\label{s:full-farm}

Once the turbulent air flow across the empty domain had fully developed and was statistically stable, the 48 modelled Siemens wind turbines were placed within the simulation domain, with their RPM set to 0. For practical considerations, only one hub-height wind speed was considered for the eight different wind speed directions, 198\degree, 202\degree, 207\degree, 212\degree, 217\degree, 223\degree, 229\degree, and $236\degree$, as specified in Table~\ref{tab:Cases}. 

Each modelled wind farm was run for 20 minutes of simulation time beyond the empty-domain spin-up, with the first 10 minutes considered as a secondary spin-up period with the turbines in place. For the last 10 minutes the air flow across the domain had fully evolved, and the modelled turbines' diagnostics were statistically stationary, although their instantaneous values were continually fluctuating.

The actual process of putting in the turbines involved remeshing the domain, then changing some of the parameters of the simulation to accommodate the change in flow conditions due to the turbines' presence. These were, specifically: i) anisotropic mesh ranges set as function of distance from turbines, and ii) velocity interpolation errors changed to vary with distance from the  turbines, so that the hr-adaptive meshing algorithm within the CFD code was more sensitive to steep velocity gradients closer to the turbines, and would resolve spatial velocity fluctuations in more detail.

\subsection{Turbine positioning}

Rather than rotate the domain to match the prevailing wind direction, it was decided that it was simpler to rotate the wind farm to effect the same change in oncoming flow relative to the turbines. The process was as follows.

Before rotating the wind farm, its centre, $\mathbf{p}_c$,  had to be determined from the spatial coordinates of the Lillgrund wind turbines, which were given in geographic Cartesian coordinates (Easting and Northing). This was calculated as

\begin{equation}
\mathbf{p}_c= \frac{1}{N} \sum_{j=1}^N \mathbf{p}_j
\end{equation}

where $\mathbf{p}_j$ is the position of turbine $j$, and $N$ is the number of turbines, in this case $N=48$.  The coordinates of turbine $i$ relative to this centre are then 

\begin{equation}
\mathbf{p}'_i = \mathbf{p}_i - \mathbf{p}_c
\end{equation}

\begin{figure}
\begin{centering}
\includegraphics[width=0.55\columnwidth]{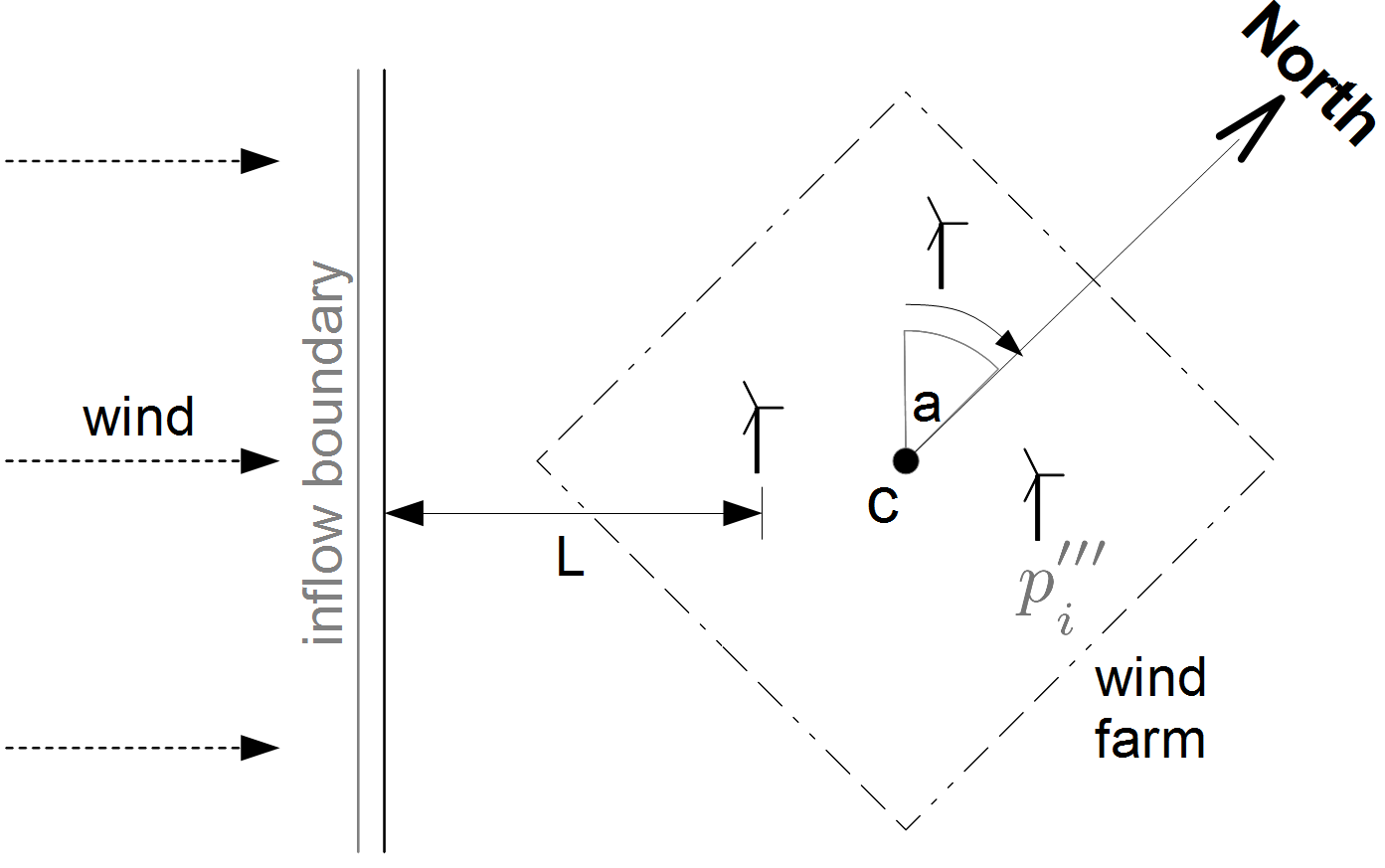} 
\caption{The translated and rotated turbine coordinates, with respect to domain origin and wind direction. $c$ represents the centroid of the wind turbines' coordinates; $L$ the minimum distance between the edge of the inflow boundary; $a$ the rotation necessary to account for wind direction.}
\label{fig:rotated-wind-farm}
\end{centering}
\end{figure}

Taking the inlet wind in the $x$-direction as specified in section~\ref{S:Inflow}, a westerly wind ($270\degree = 3\pi/2\rm\,rad$) requires no rotation and a south westerly wind (225\degree) would require a clockwise rotation of wind farm about their centroid of $45\degree$, and so on. The rotation can be written as
\begin{equation}
	\mathbf{p}''_i = \mathbf{R}_w \, \mathbf{p}'_i \mbox{\hspace{3em}with\hspace{1em}} \mathbf{R}_w =\begin{bmatrix}
	\cos a & \sin a & 0 \\
	-\sin a & \cos a & 0 \\
	0 & 0 & 1
\end{bmatrix} 
\end{equation}
where $\mathbf{R}_w$ is the rotation matrix for wind direction $w$ (in radians) and $a=\frac{3 \pi}{2} - w$ as illustrated in Figure~\ref{fig:rotated-wind-farm}.

Lastly, the turbines' coordinates are translated so that there is at least $L=2\rm~km$ from the furthest upwind turbine to the leftmost boundary, and put them in the centre of our domain laterally, which is $W=8.1\rm~km$ across. Therefore, by finding $x_{min} = \min(p''_{x,i})$, we do one final translation to get the three-dimensional coordinates of the turbines' rotors as

\begin{equation}
	{\mathbf p}'''_i = {\mathbf p}''_i +
			\begin{bmatrix} x_{min} + L \\
				W/2 \\
				z_H
			\end{bmatrix}
\end{equation}
where $z_H$ is the hub height of the turbine.

By positioning and rotating them thus (see figure \ref{fig:rotated-wind-farm}), the same empty domain could be used, while at the same time ensuring that enough space was left between the farm and the edges of the domain such that no unrealistic accelerative effects would occur on the other side of the domain, and that the wakes behind the farm would be given sufficient space to develop. This process would have to be undertaken for each different wind direction.

\subsection{Remeshing}

\begin{figure}
\begin{centering}
\includegraphics[width=0.96\columnwidth]{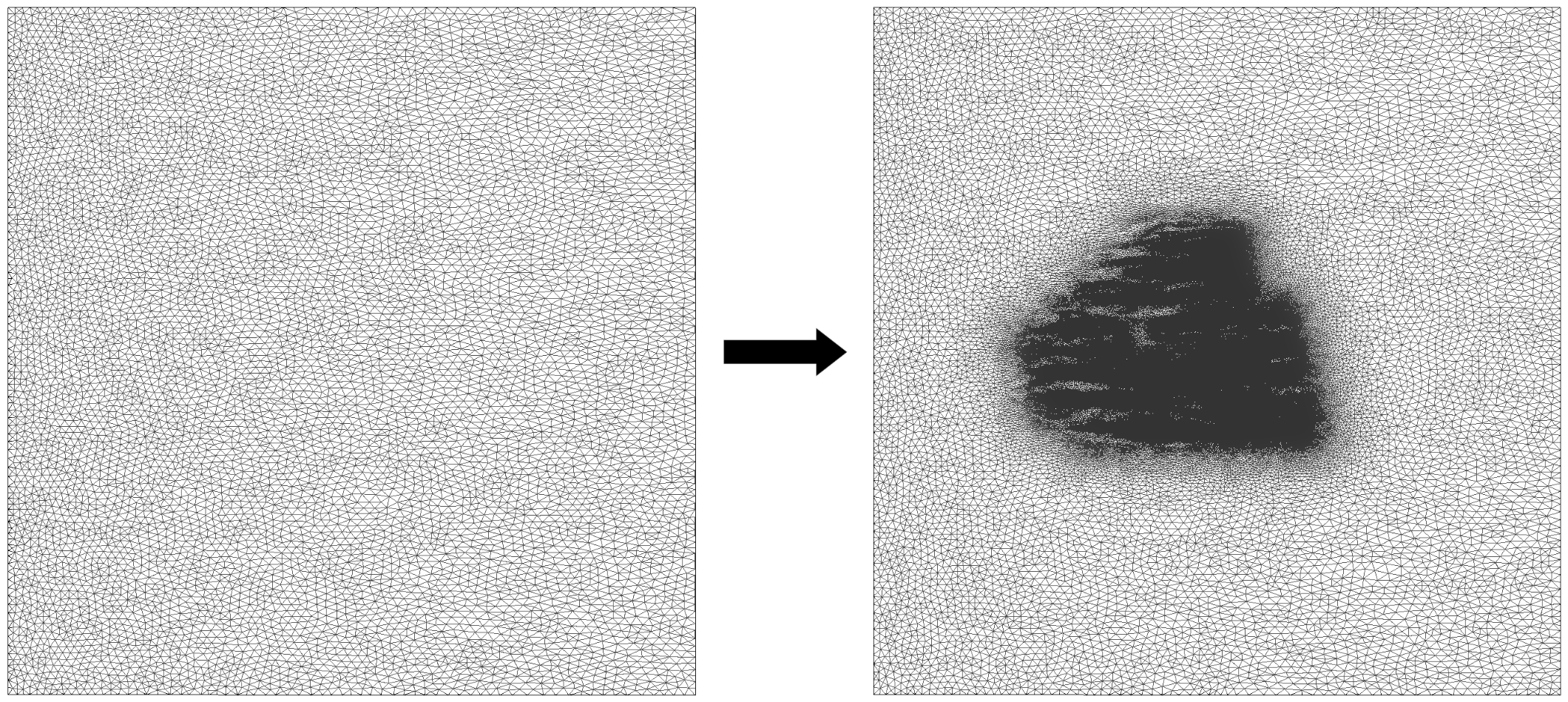} 
\caption{Two horizontal slices through the Lillgrund model mesh, showing how adding the turbines to the model increases the mesh resolution near the wind farm. The elements far away from the turbines are highly anisotropic and measure approximately 75~m horizontally; closer to the turbines, the mesh resolution becomes isotropic, with elements measuring 5~m across vertically and horizontally.}
\label{fig:mesh-adding-turbines}
\end{centering}
\end{figure}

With the turbine rotor positions within the simulation calculated, the finite element domain mesh was adapted (or remeshed), so that the mesh resolution was sufficient to resolve the flow through the rotors. Typically, this meant that resolution would have to increase from 75~m horizontally and 10~m vertically, to nearer 5~m isotropically in the vicinity of a turbine rotor and within the turbine volumes. This was done by creating a non-advective, non-diffusive field within Fluidity, to which Fluidity's hr-adaptive algorithms were sensitive; this field was a cubic function of distance extending for a distance of $2D$ from the nearest turbine. The hr-adaptivity would detect the gradient in this field, and increase the mesh resolution to resolve the solution, as Figure \ref{fig:mesh-adding-turbines} shows.

\section{Results} \label{S:Results}

\subsection{Computational model}\label{S:Results:Model}

This section gives an overview of the results from the computational model of Lillgrund. Instantaneous slices through the velocity field are used, together with the power outputs of selected turbines, to illustrate features of the wind farm flow dynamics and performance. To this end three wind directions are examined, namely 198\degree and 236\degree, which as table \ref{tab:Cases} shows, present a staggered arrangement to the oncoming wind, so that downwind turbines are relatively exposed, and 223\degree where the rows of turbines are aligned with the mean wind direction. Turbines in row D are studied in more detail; this row crosses the gap in the array at positions D05 and E05, shown in Figure \ref{fig:Map2}.

\begin{figure}
a) 198\degree at t=15~min\hspace{0.355\columnwidth} b) 198\degree at t=20~min\\
\includegraphics[width = 0.49\textwidth]{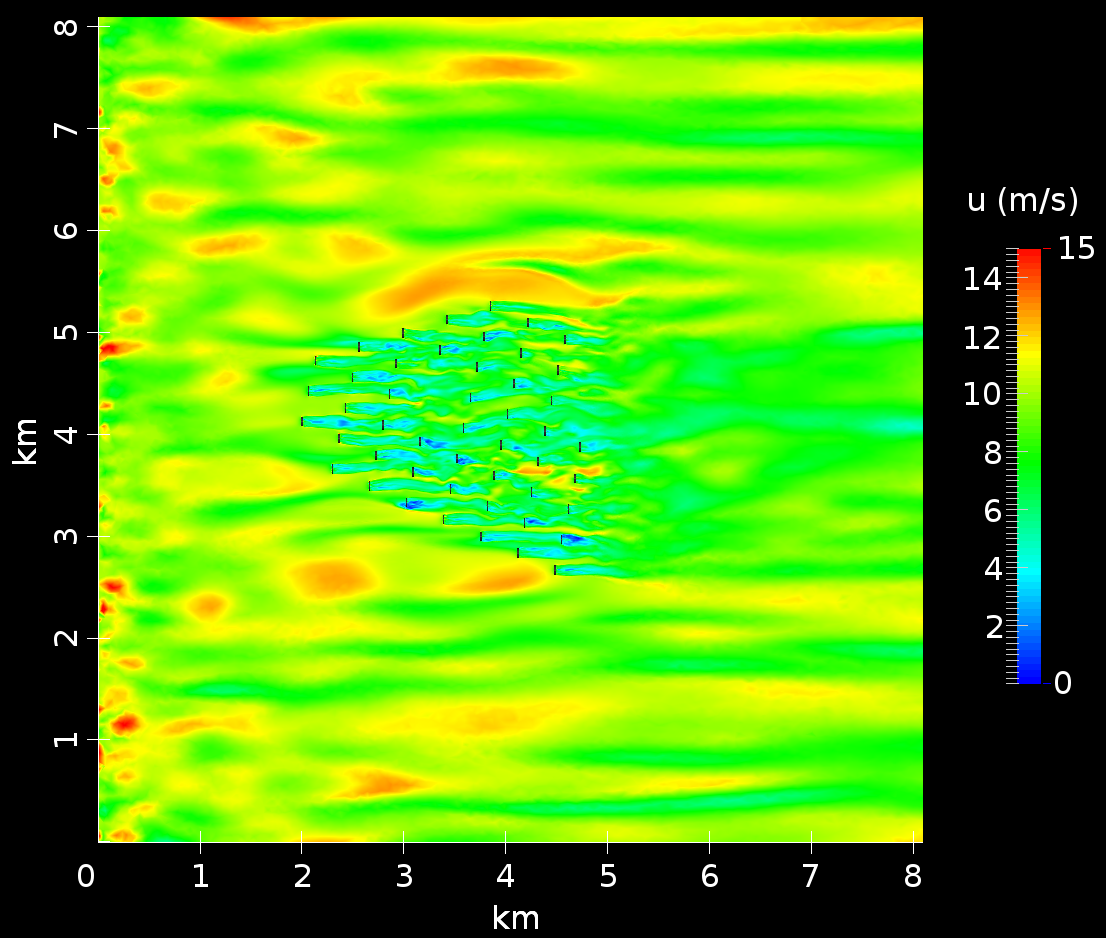} \hfill
\includegraphics[width = 0.49\textwidth]{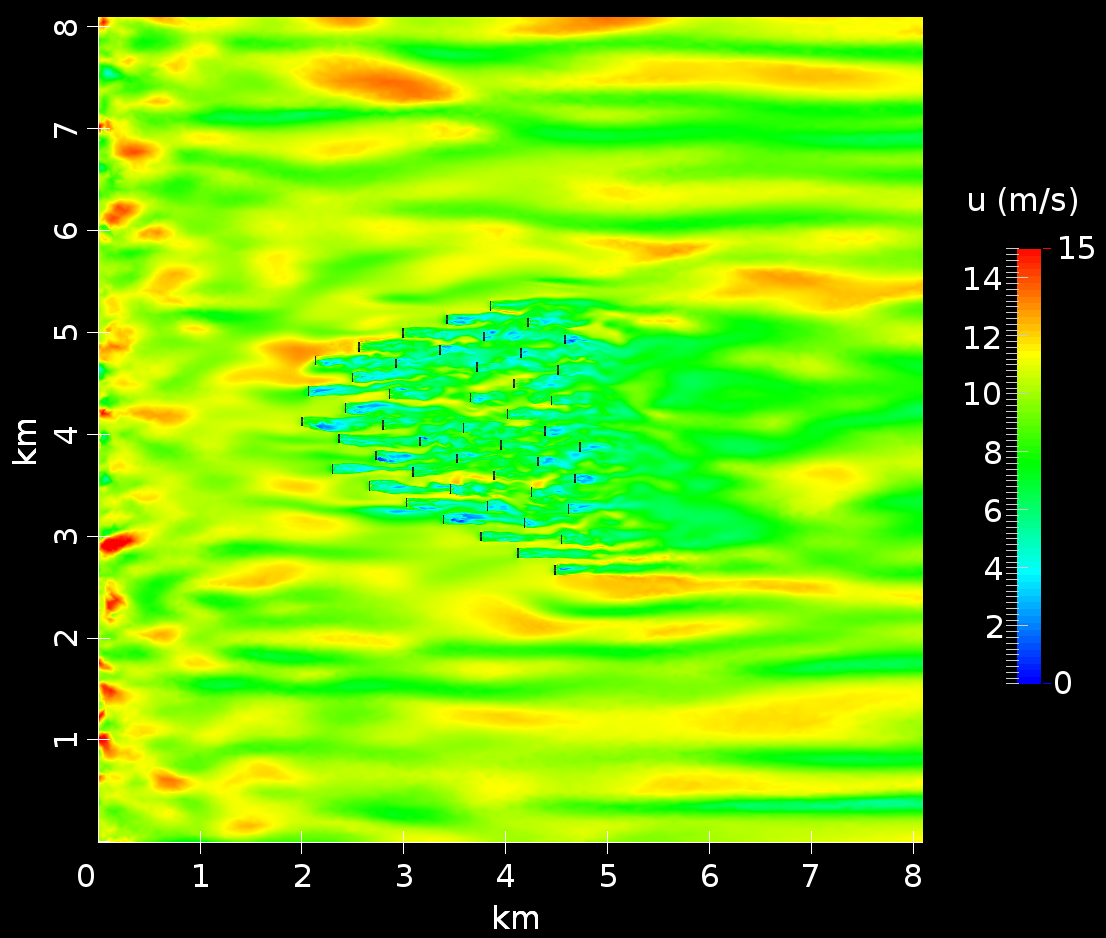} \\
c) 223\degree at t=15~min\hspace{0.355\columnwidth} d) 223\degree at t=20~min \\
\includegraphics[width = 0.49\textwidth]{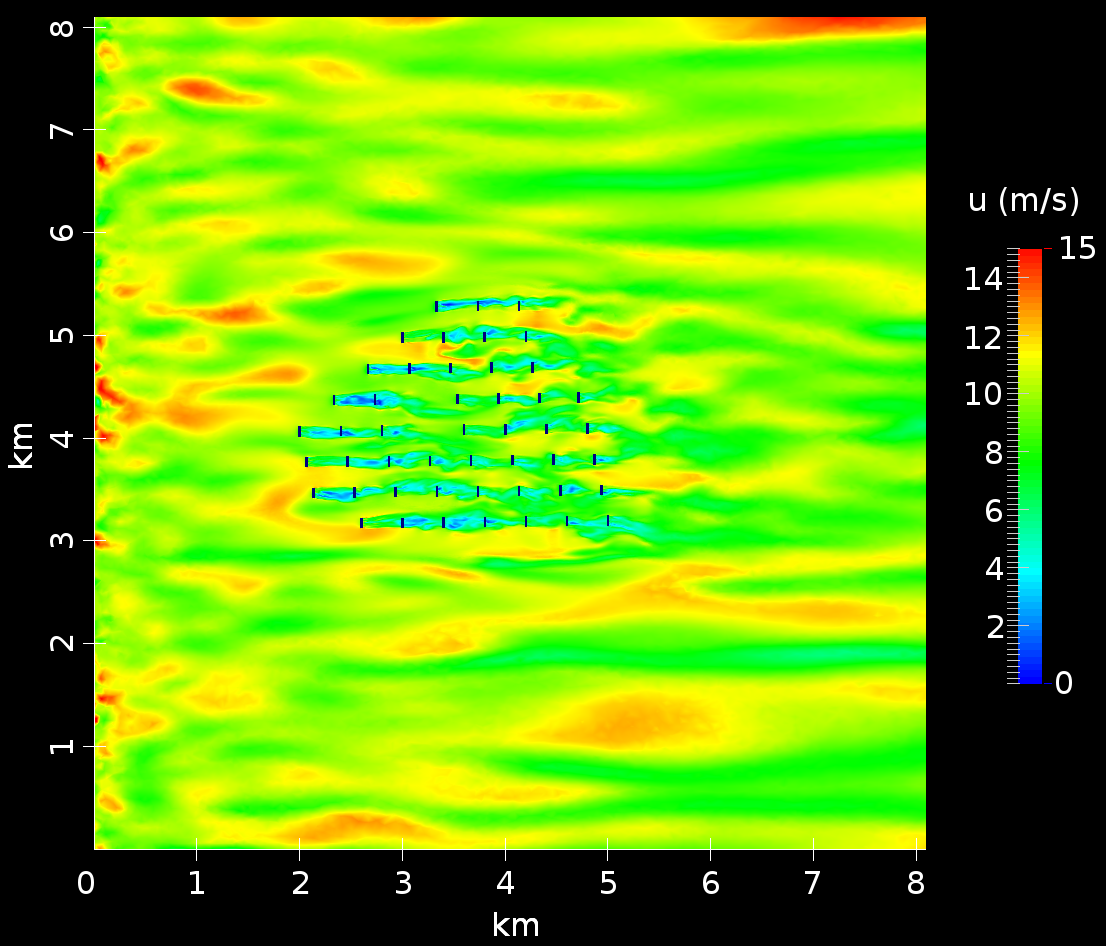} \hfill
\includegraphics[width = 0.49\textwidth]{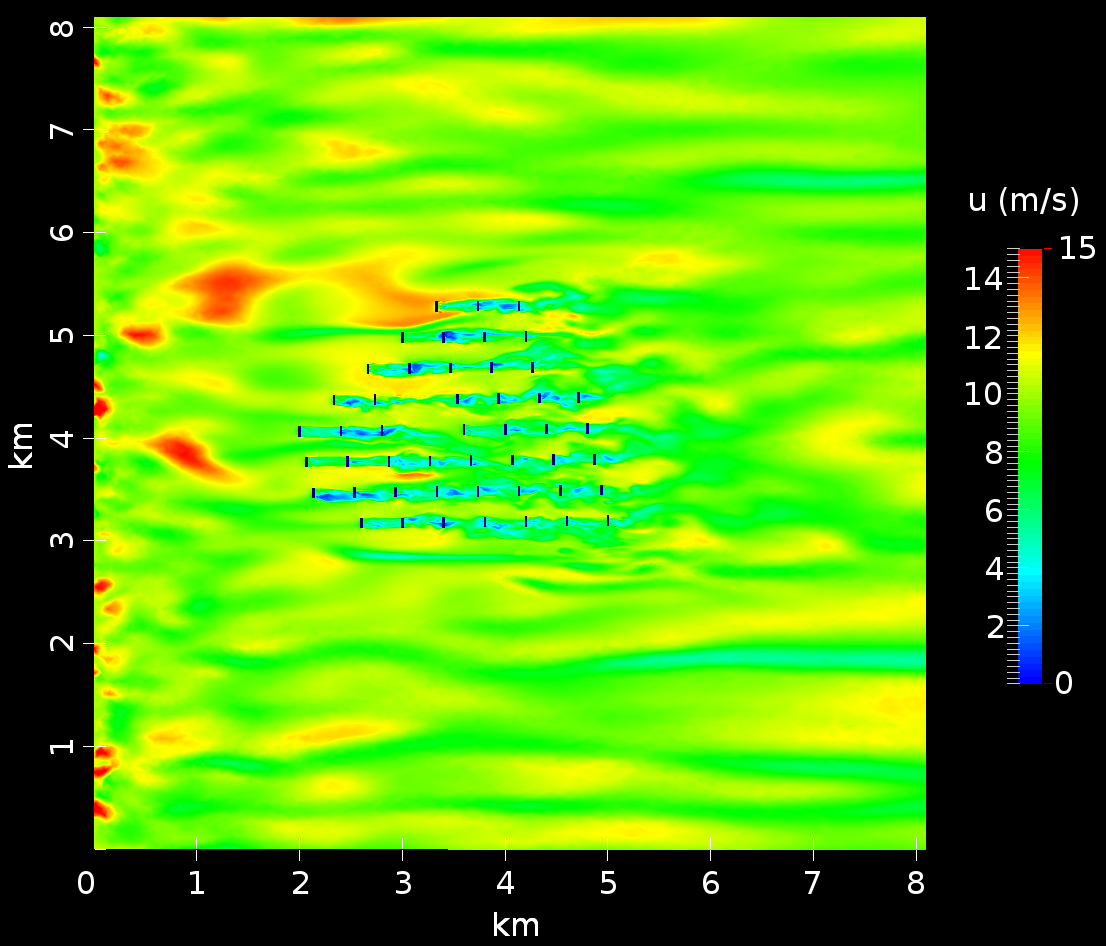} \\
e) 236\degree at t=15~min\hspace{0.355\columnwidth} f) 236\degree at t=20~min\\
\includegraphics[width= 0.49\textwidth]{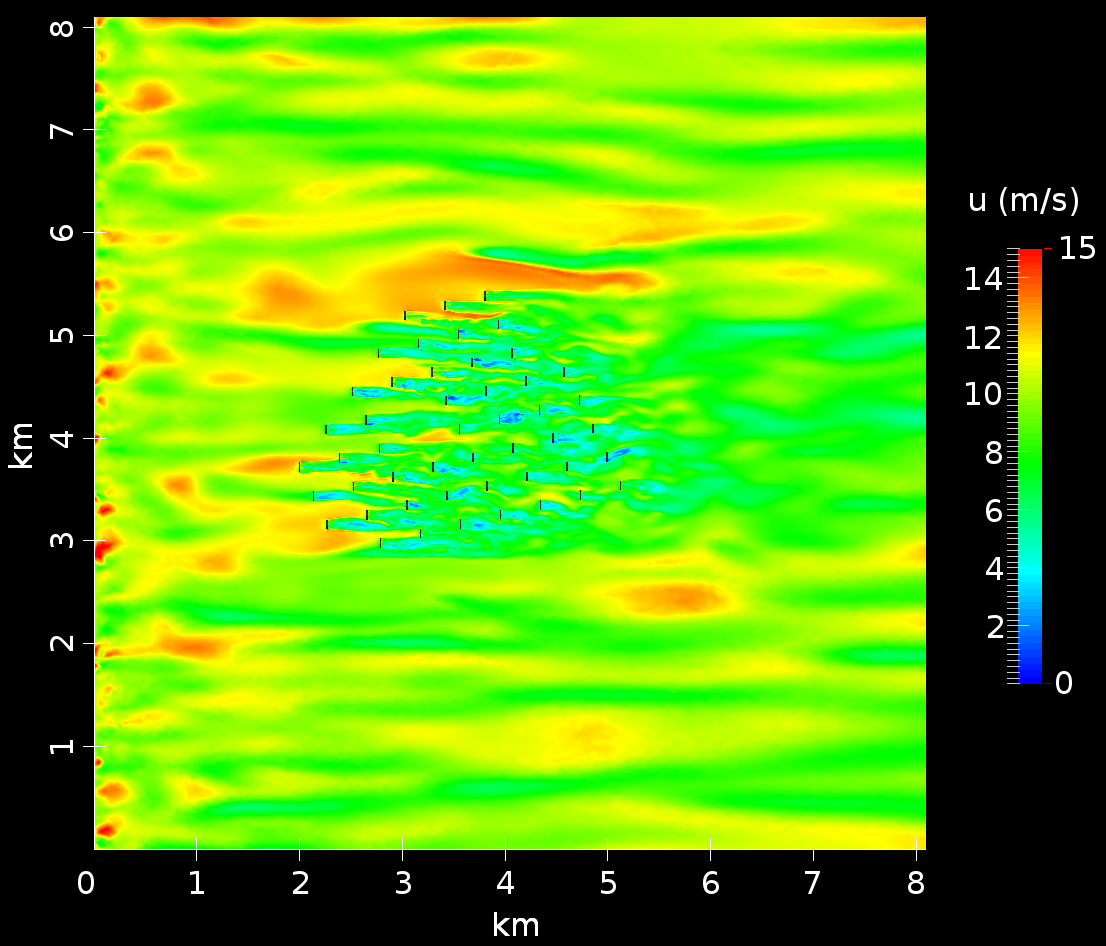} \hfill
\includegraphics[width= 0.49\textwidth]{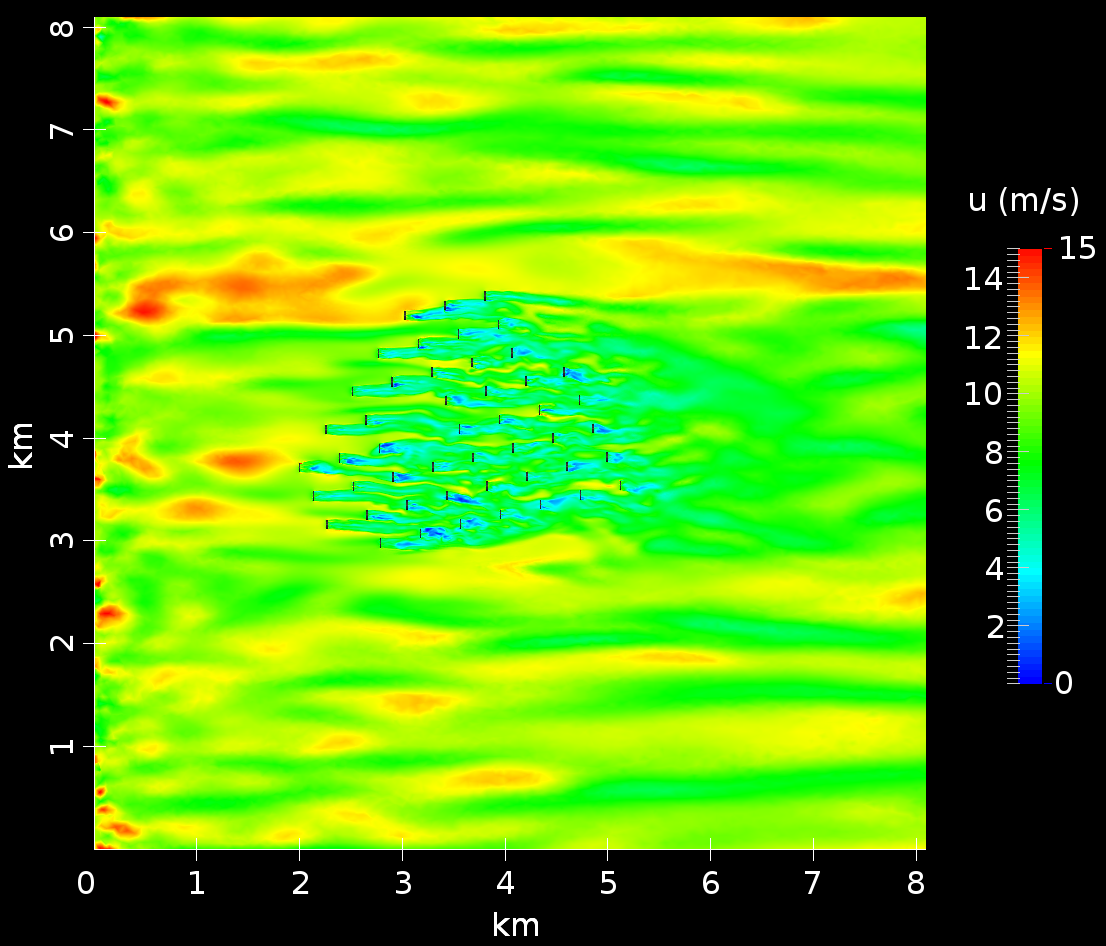} 
\caption{Horizontal slices through the instantaneous velocity field at hub height for wind directions of 198\degree, 223\degree and 236\degree.}
\label{fig:horiz-vel-slices-198-223-236}
\end{figure}

In Figure~\ref{fig:horiz-vel-slices-198-223-236} we can see horizontal slices through the instantaneous velocity field, two for each wind direction spaced 5 minutes apart. The flow is perpetually unsteady in all cases, as expected from Large Eddy Simulation CFD simulations with the SEM inlet boundary conditions described in ~\S~\ref{S:lillgrund-sea-surface}. The eddies through the domain range widely in size, from 100~m to over 1~km, and the turbulence is highly anisotropic, with those eddies typically 5-10 times longer (streamwise) than they are across (laterally). This results in varying flow speeds, ranging from approximately 6-15~m/s outwith the farm, and gusts can be seen passing through the wind farm, leading to higher wind speeds within. Turbine wakes are evident, with dark blue patches behind the turbines, indicating the regions of highest wake deficit; these wakes meander considerably. Wind farm wakes are also visible in Figure~\ref{fig:horiz-vel-slices-198-223-236}(a)-(d), extending downwind of the farm by approximately 3 km.

\begin{figure}
a) \\
\includegraphics[width = 0.99\columnwidth]{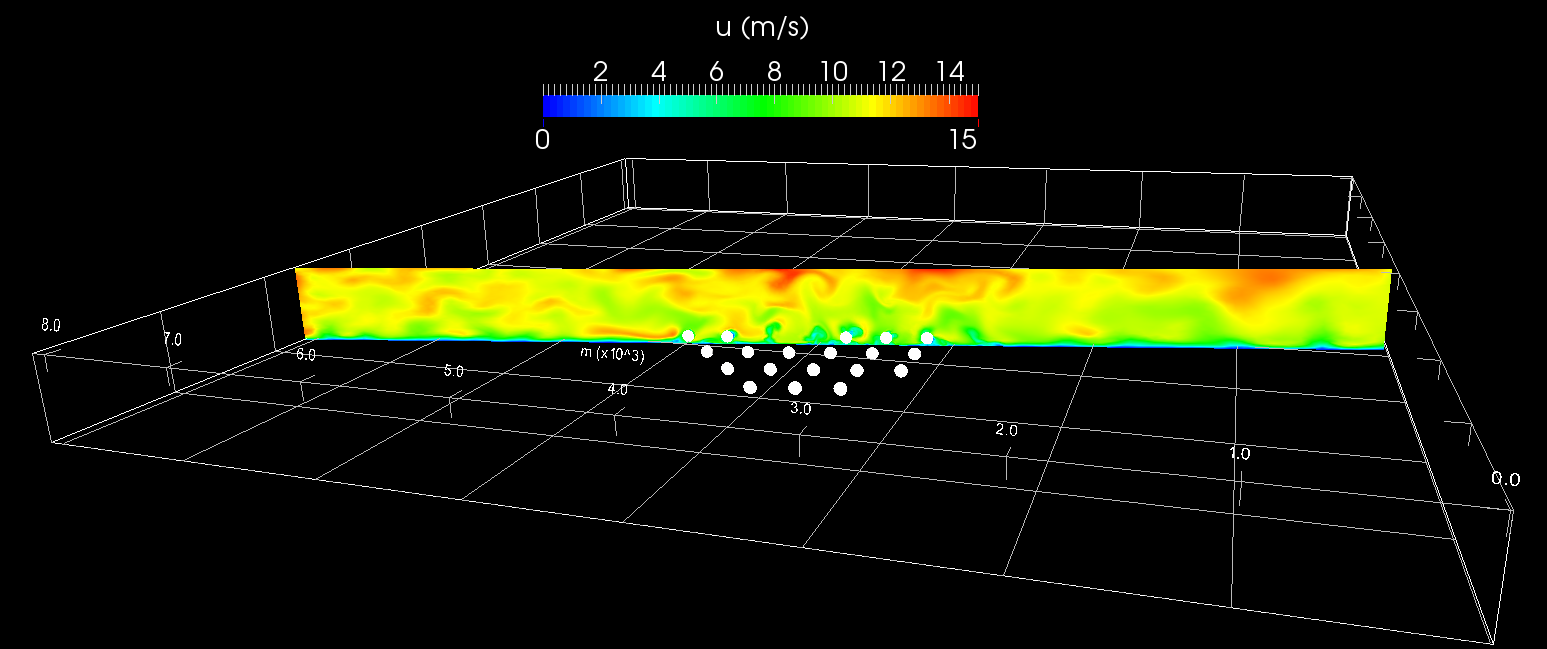} \\\\
b) \\
\includegraphics[width = 0.99\columnwidth]{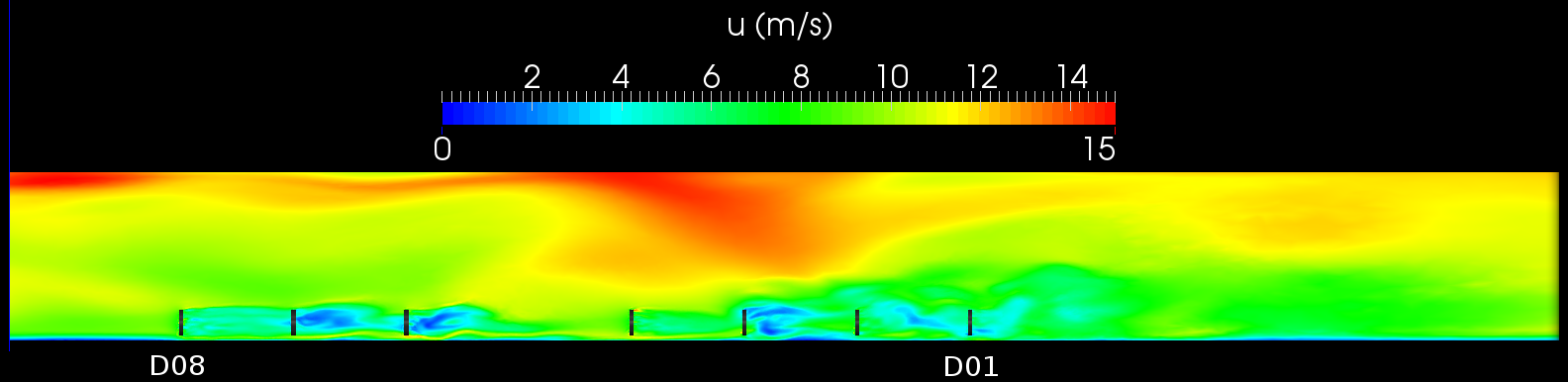} 
\caption{Vertical slices through the instantaneous velocity field at t=20 minutes, for a wind direction of 223\degree: a) cross-stream slice through the fifth column of turbines, and b) zoomed-in streamwise slice of instantaneous velocity field through row D.}
\label{fig:vert-slices-223}
\end{figure}

Large scale turbulence structures particularly above and upwind of the turbine array can be seen in Figure \ref{fig:vert-slices-223}. However, a qualitative comparison between Figure \ref{fig:vert-slices-223}(b) and similar figures from other LES simulations in \citet{Churchfield2012} shows that the latter has higher frequency turbulent features especially near the turbine blades. This is not surprising, given that their simulations use a minimum cell dimension of 1~m near the turbines, whereas here the minimum is 5~m, therefore smaller eddies are resolved in the former. On the other hand, the large-scale turbulence structure seen in our results are not present in \citet{Churchfield2012}, who relied upon a log-law velocity profile passing over an empty domain to create turbulent inlet conditions. A better comparison can be made with \citet[Figure 1]{calaf2010} where periodic boundary conditions were used to create sufficient upstream turbulence; the work presented here shows similar turbulent flow features. This suggests that the SEM boundary conditions strongly influence the aerodynamics around the wind farm, and the turbine wakes within it.


The acceleration of flow between turbines due to the blockage effect, known as jetting, is noticeable in the results. Figure \ref{fig:horiz-vel-slices-198-223-236}(b) shows a gust of wind hitting the foremost turbines, B08, C08, D08 and E07, and a jet appears to pass around D08 and E07, before encountering turbines E06, D07 and D06. Figure \ref{fig:horiz-vel-slices-198-223-236}(e) also shows this, with a jet passing between B08 and A07 towards turbine A06; between B08 and C08 towards B07; and where a 3km-long gust encounters turbines H04, H03 and H02 at the north end, the jet is turned inward of the farm towards G02. The jetting has a more consistent pattern in the aligned case of 223\degree, as the gaps between rows A to H in Figures \ref{fig:horiz-vel-slices-198-223-236}(c) and (d) all show evidence of accelerated flow. Moreover, both figures also indicate that air in these regions can exceed the average upstream hub-height wind speed, implying that jetting is an important method for injecting kinetic energy into the internal farm flow, affecting wind farm performance, and is highly dependent upon the alignment of the prevailing wind to the rows.


\begin{figure}
\begin{centering}
\includegraphics[width = 0.75\columnwidth]{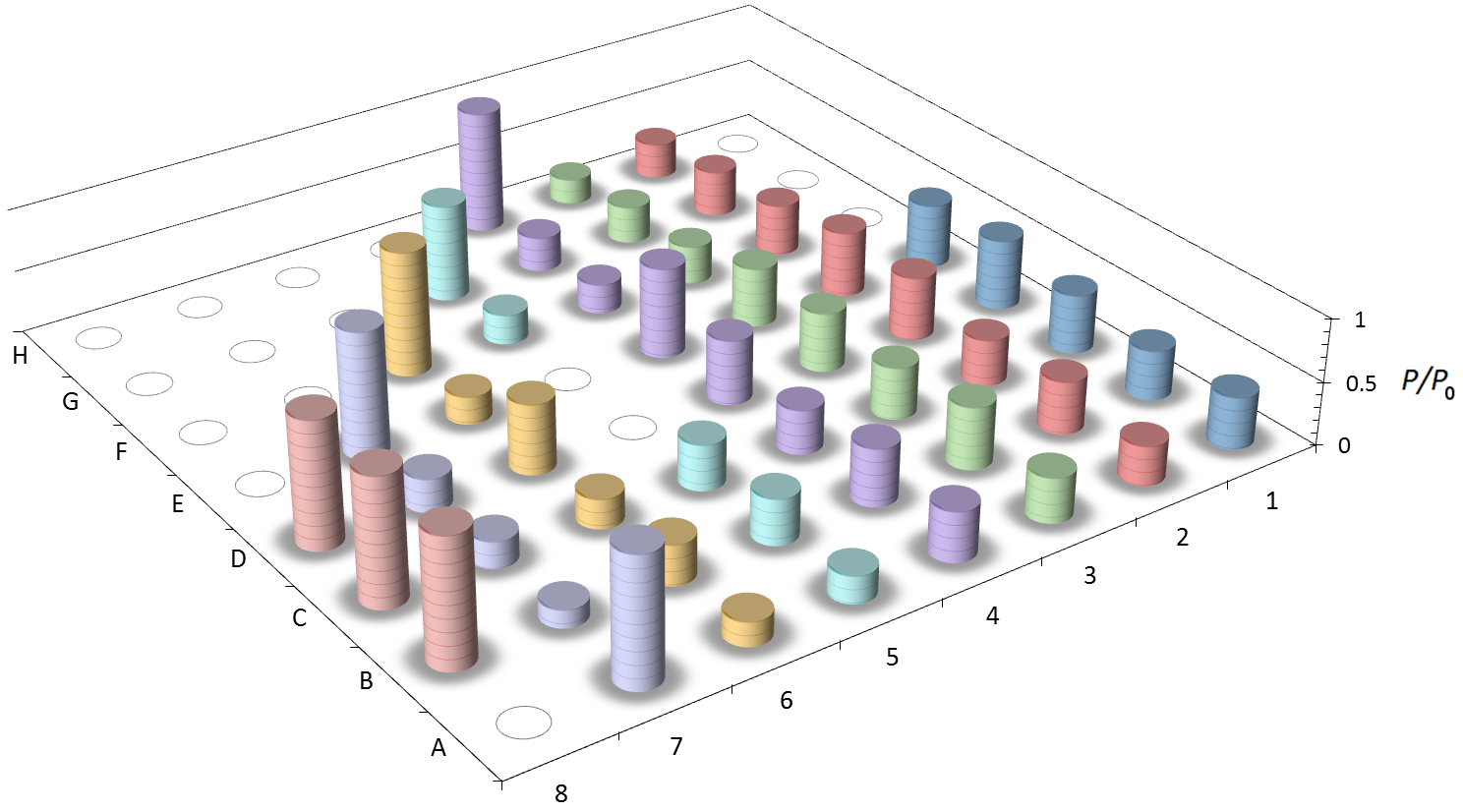} 
\caption{Time-averaged normalised power plot for wind direction of 223\degree.}
\label{fig:power_averages}
\end{centering}
\end{figure}

The wind farm is visualised as an array of time-averaged power plots for the wind direction of 223\degree in Figure \ref{fig:power_averages}. These averages were computed over the last 10 minutes of simulation, by which point the flow had fully developed. The leading turbines all have an average power close to $P_0$, the median calculated from B08, C08 and D08. Immediately downwind, the performances of turbines B07, C07 and D07 drop to 20-30\% of this value. Surprisingly the turbines in column 6 with two turbines upwind show a mild increase in power, on average 37\%. After the empty space in column 4, D04 is over 50\% of $P_0$ while E04 rises to 72\%. This increase can be explained by looking at Figure \ref{fig:vert-slices-223}(b), where the wind speed increases in the gap behind D06 and E06, as faster air flowing over the wind farm is entrained downwards and mixed with the wake of upwind turbine. Beyond this, the turbines' performance remains at around 30\%, before decreasing slightly below this in column 1\@. It should be noted there is a large difference in the mean power between D06 and E06; this is also seen between D04 and E04. There is no obvious reason for this unusual behaviour. It may be due to particular eddies passing those turbines and, were additional computing time available, longer simulations with greater averaging periods could be reduce these disparities in mean power output.


\begin{figure}
a) \\
 \includegraphics[width = \columnwidth]{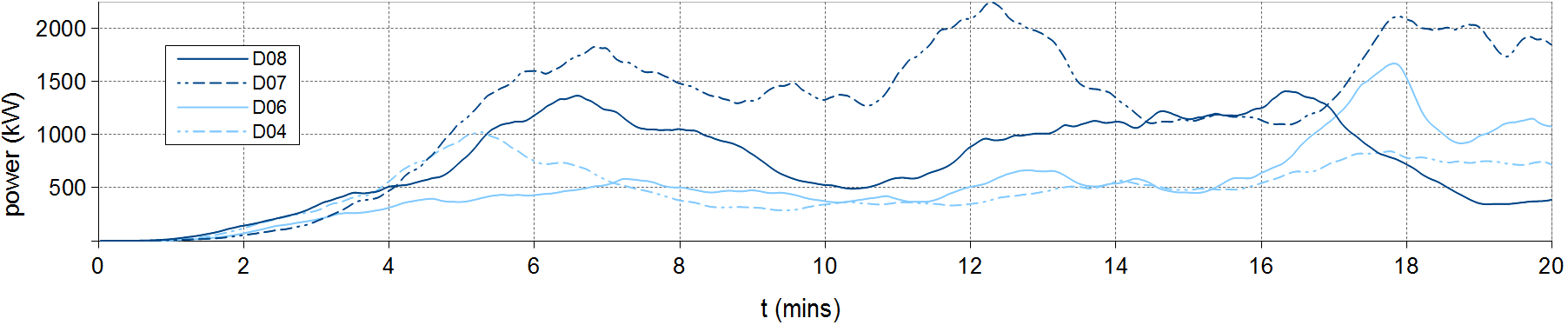}\\
b) \\
\includegraphics[width = \columnwidth]{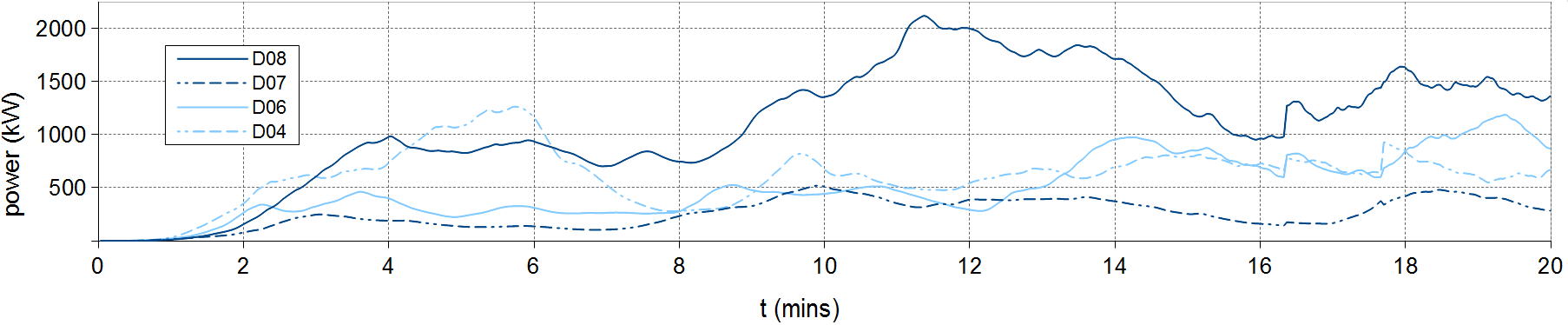}\\
c) \\
\includegraphics[width = \columnwidth]{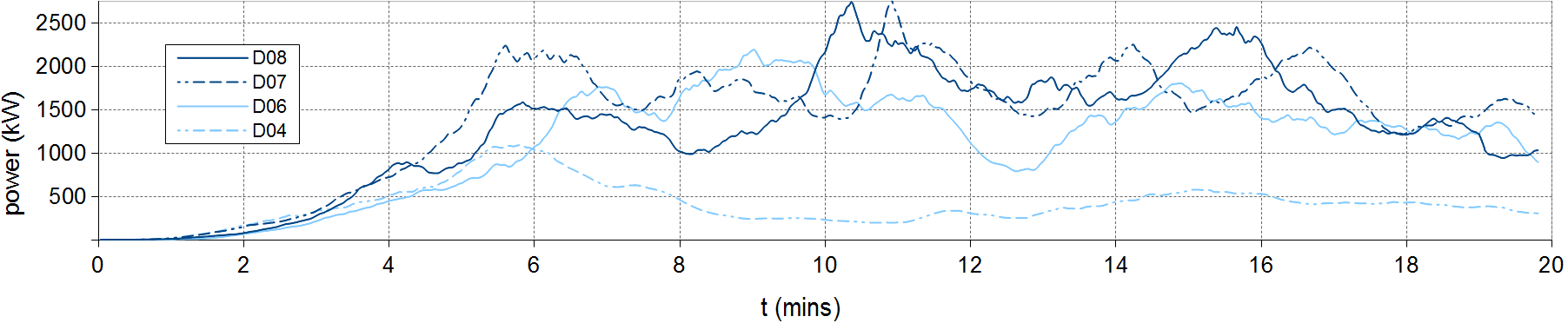}
\caption{Power time-series for selected modelled turbines in row D, at wind directions of  a) 198\degree, b) 223\degree and c) 236\degree. The mean power is averaged over the final 10 minutes of simulation.}
\label{fig:instant-power-row-D}
\end{figure}

For wind directions 198\degree, 223\degree and 236\degree, the time series of power output from selected turbines in Row D are shown in Figures \ref{fig:instant-power-row-D}(a), (b) and (c) respectively. As the rotors start from a stationary position, the power increases predictably for the first 4-6 minutes, before achieving statistically stable values after 10 minutes. The variability of power output is clear: while D08 for 223\degree and 236\degree appear to fluctuate about a value close to that shown in Table \ref{tab:power-thrust-errors}, the power can peak at 2250 kW or higher in both cases, as well as drop down to almost 1000 kW. This can be attributed to the passage of gusts of wind (and associated lulls) through the wind farm, causing the turbine rotor to speed up and slow down accordingly, and indeed these long, slow variations have a period of 3-4 minutes, which equates approximately to a distance of 2-2.5 km for a hub-height wind speed of 10 m/s. This observation agrees well with the size of the flow features shown in Figures \ref{fig:horiz-vel-slices-198-223-236}~(c)-(f). 

Comparing the aligned case in Figure \ref{fig:instant-power-row-D} (b) with the non-aligned cases in (a) and (c), it is clear that the second and third turbines, D07 and D06, experience higher performance when non-aligned due to increased exposure to the wind. This effect is enhanced by jetting particularly at a prevailing wind direction of 236\degree, where their power outputs are comparable to the leading turbine. Indeed for 198\degree in Figure \ref{fig:instant-power-row-D} (c), D07 spends the majority of its time outperforming the leading turbine. For this particular case, D08 is mostly underperforming, possibly due to insufficient hub-height wind speed; with jetting as a mechanism for accelerating the flow it would be possible for D07 to experience a wind speed greater than that upwind of the wind farm.

\subsection{SCADA data}\label{S:Results:SCADA}

In this section, the relevant results from the SCADA data are extracted to find episodes of at least 10 minutes' duration in which the wind speed was within the specified range, and the reference wind direction from the leading turbines was within a 3\degree-sector of the wind direction corresponding to Table~\ref{tab:Cases}.  Considering the consistent bias in wind direction recorded by the met mast and the nacelle, the relative performance of a few turbines against wind direction is analysed before focussing on the response at the selected key wind directions.

\begin{figure}
a) \hspace{0.46\columnwidth} b)\\
\includegraphics[width = 0.45\columnwidth]{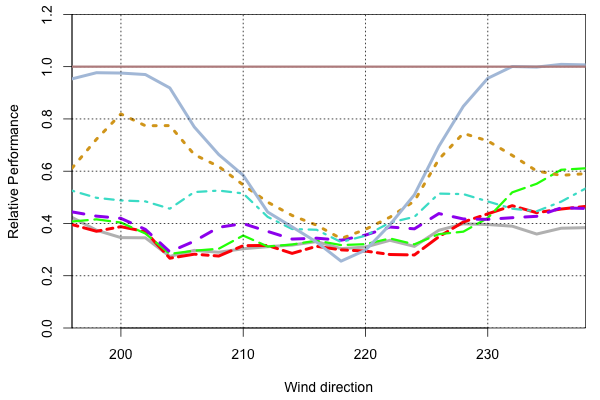} \hfill
\includegraphics[width = 0.45\columnwidth]{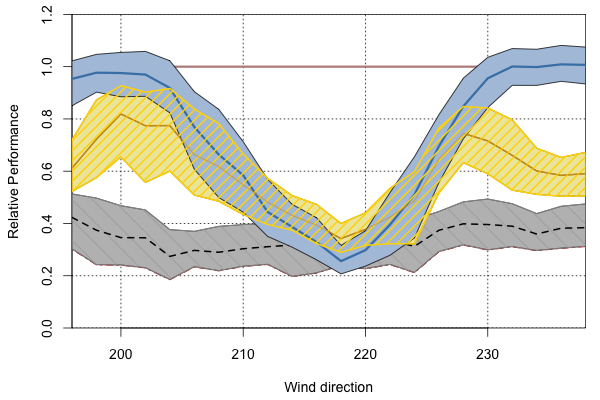}
\caption{Relative turbine performance of turbines in row C against wind direction; a) The median performance of all seven turbines C07 to C01 (pale blue solid line, gold narrow-dashed line, turquoise dash-dotted line; purple medium-dashed line, green wide-dashed line, red double-dot-dashed line, respectively);  b) the median and interquartile ranges for turbines C07 (pale blue), C06 (gold with narrow hatching), C01 (grey with wider hatching). }
\label{fig:Turb_C3_8}
\end{figure}

The median of the relative performance over the front turbines (B08, C08, and D08) against the wind direction in Figure~\ref{fig:Turb_C3_8}~(a) for row C shows clearly  that the performance of each turbine is affected as one would expect from the geometric shading of one turbine by another.  The relative performance of C07 in second row shows a clear minimum at around 30\% when the wind is aligned with the turbines, and clear maxima around 100\% when C07 is between two front-row turbines, namely B08 and C08 for around 198\degree, and C08 and D08 for around 236\degree.   On the other hand, the turbines in fifth row and beyond never show more than 30\% to 40\% of the front turbine's output;  these turbines are in the `deep array' wake.  The somewhat increased performance of C02, C03, and C04 above 230\degree~can be explained by the fact that the wind is coming from the gap in the array, which allows for some wake recovery.  Those in the third and fourth rows still perform better than the deep array with geometrically favourable wind directions, but they do not rise above 80\% and 50\%, respectively.   

The observation that the turbine in the second row produces less power than those further into the array was also noted by \citet{BarthSmith12}, but they could not reproduce it in any of their computational models.  This strong power deficit is only apparent when the data are taken over 3\degree~bins or narrower.   To our knowledge, a deficit in the second row stronger than  in the third row is seen in wind farms where the turbine spacing in the streamwise direction is less than about five rotor diameters ($5D$).  To put the turbine-by-turbine observations into the context of the overall variability of the power output, the range around the median is shown as the extent of the interquartile range for three selected turbines, namely the second, third and last row in Figure~\ref{fig:Turb_C3_8}(b).   This not only shows that the reduction in the second row turbine is significantly lower than that of the third turbine but it also shows that the variability across all wind directions is higher in the third row compared to that of the second row, which can be interpreted as resulting from a higher turbulence level created by the interaction of the turbulence generated by the first and second turbines.  

While the geometry of the wind farm suggests the strongest power deficit at 223\degree, the observations plotted against the front turbine's yaw direction puts that minimum at 218\degree.  Plotting the same results against the wind direction measured at the met mast upstream of the wind farm would put the minimum at 229\degree (cf.~\S~\ref{S:Lillgrund:FarmPerf}).  Considering the presence of this systematic error in the directional data, a yaw direction of 218\degree~is fully consistent with a true wind direction of 223\degree.  In the following, the yaw direction is adjusted by that possible bias of 5\degree~and the results are presented according to their nominal true wind direction.

\begin{figure}[p]
\begin{centering}
$198^\circ$ \hfill ~\\
\includegraphics[width = 0.32\columnwidth]{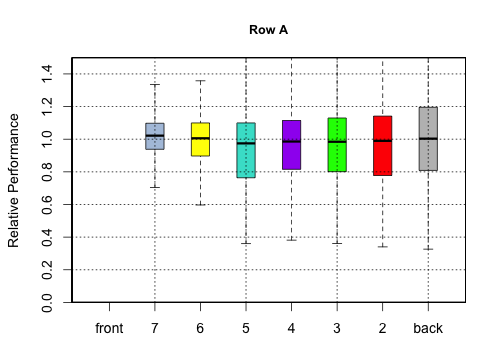} 
\includegraphics[width = 0.32\columnwidth]{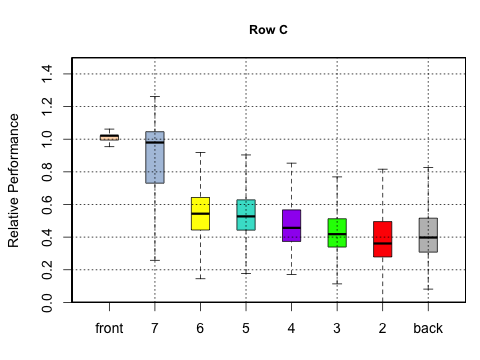} 
\includegraphics[width = 0.32\columnwidth]{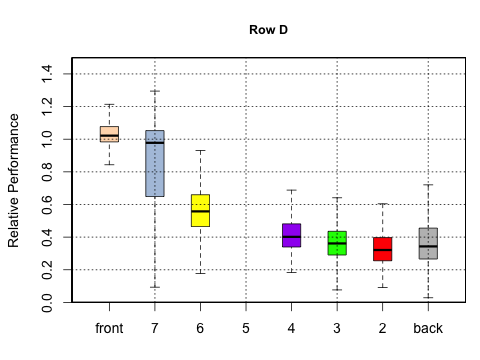} \\
$212^\circ$ \hfill ~\\
\includegraphics[width = 0.32\columnwidth]{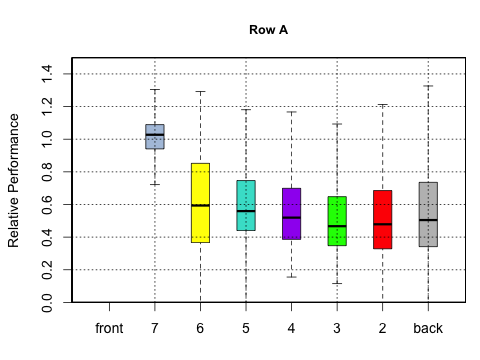} 
\includegraphics[width = 0.32\columnwidth]{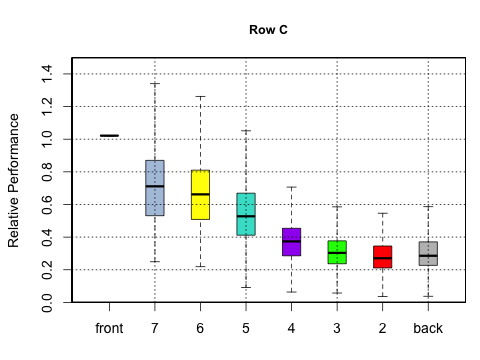} 
\includegraphics[width = 0.32\columnwidth]{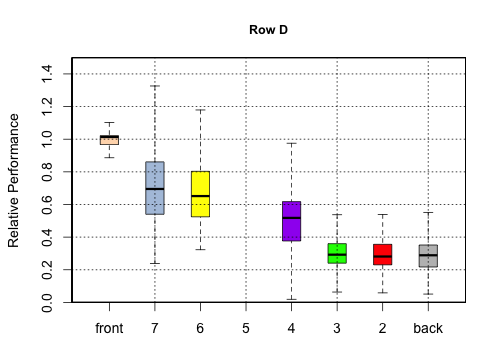} \\
$223^\circ$ \hfill ~\\
\includegraphics[width = 0.32\columnwidth]{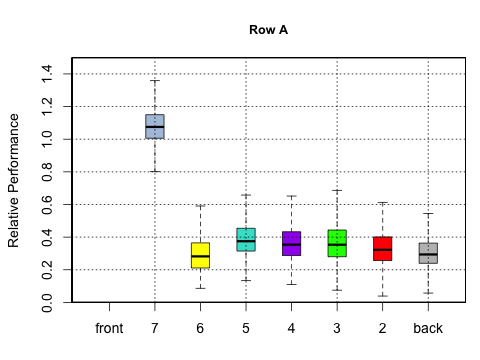} 
\includegraphics[width = 0.32\columnwidth]{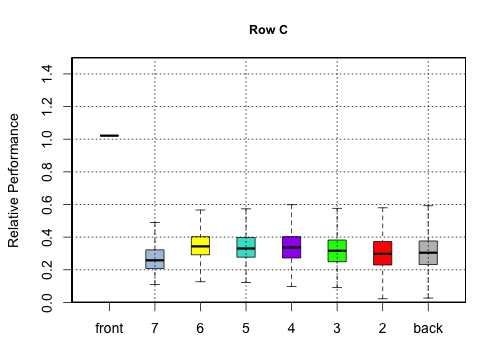} 
\includegraphics[width = 0.32\columnwidth]{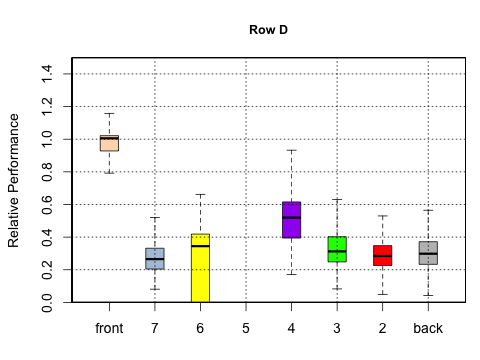} \\
$236^\circ$ \hfill ~\\
\includegraphics[width = 0.32\columnwidth]{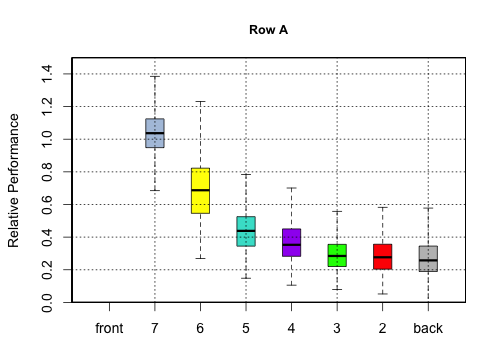} 
\includegraphics[width = 0.32\columnwidth]{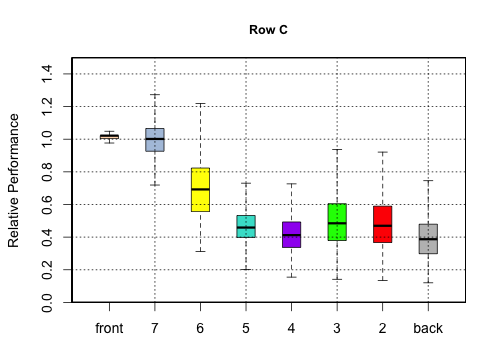} 
\includegraphics[width = 0.32\columnwidth]{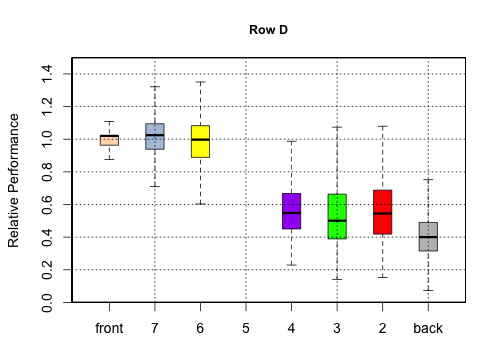} \\
\caption{Boxplots for rows of turbines at different wind directions. }
\label{fig:RowBoxplots}
\end{centering}
\end{figure}

To illustrate the variability for each case, we make use of the standard plot of power deficit of turbines within a row, as used by others \citep{Churchfield2012,Lillgrund15,Hansen2012}, but  add the information about the variation around the mean or median value through the use of box-and-whisker plots \citep{RFAQ2011} instead of the more common single-valued charts.   The cases shown in Figure~\ref{fig:RowBoxplots} show the three rows A (at the edge of the array), C (a full set of turbines through the centre) and D (a set with a gap) for four selected wind directions of 198\degree, 212\degree, 223\degree, and 236\degree~(cf. Table~\ref{tab:Cases}).  At a wind direction of $198^\circ$, each turbine in row A is nominally fully exposed to the wind which is reflected in a uniform median relative power output around 100\%.  However, the variation around the median increases progressively towards the back of the row, from around $\pm 20\%$ at the front (turbine A07) to around $\pm 80\%$ at turbine A01 at the back.  This suggests that each turbine adds variability or turbulence to the wind even outside the typical wake direction, possibly due to wake meandering.  A similar observation is made for the second turbine in row C, turbine C07, which according to Table~\ref{tab:Cases} is expected to be exposed to the wind and situated between the wakes of B08 and C08.   As expected, the average relative power output of C07 is around 100\% but with a substantial variability.  Deeper into the wind farm, C06 would be partially in the wake of C08 and, as expected from this, the performance of C06 is reduced to around 60\% which deepens further towards the back of the array to around 40\%.  A similar behaviour is seen in the adjacent row D\@.  The second row in Figure~\ref{fig:RowBoxplots} shows the case of $212^\circ$ where turbines in the second and third row are not directly shielded but expected to be affected by wake expansion.

At 223\degree~the full shading of all turbines is evident, including the very strong deficit in second row followed by a slight recovery in third row.  The clearly enhanced performance of turbine D04 can be explained by the gap in the row leading to an effective turbine spacing between D06 and D04 of $8.6D$.  At 236\degree, finally the behaviour for columns A and C is qualitatively similar to that at 212\degree, while column D benefits from the shape of the wind farm where column E terminates at turbine E07 and column F and F06.

This section has presented the result from the computer simulations and the observations in turn. Section \ref{S:Validation} combines these two sets of results for a qualitative and quantitative validation of the model.

\section{Model Validation}\label{S:Validation}
\subsection{Validation Methodology}
In this section, the computational model results are directly compared to those from the actual wind farm SCADA data where the SCADA yaw direction was adjusted as in section~\ref{S:Results:SCADA}.

\begin{figure*}[t]
a) \hspace{0.45\columnwidth} b) \\
\includegraphics[width = 0.45\columnwidth]{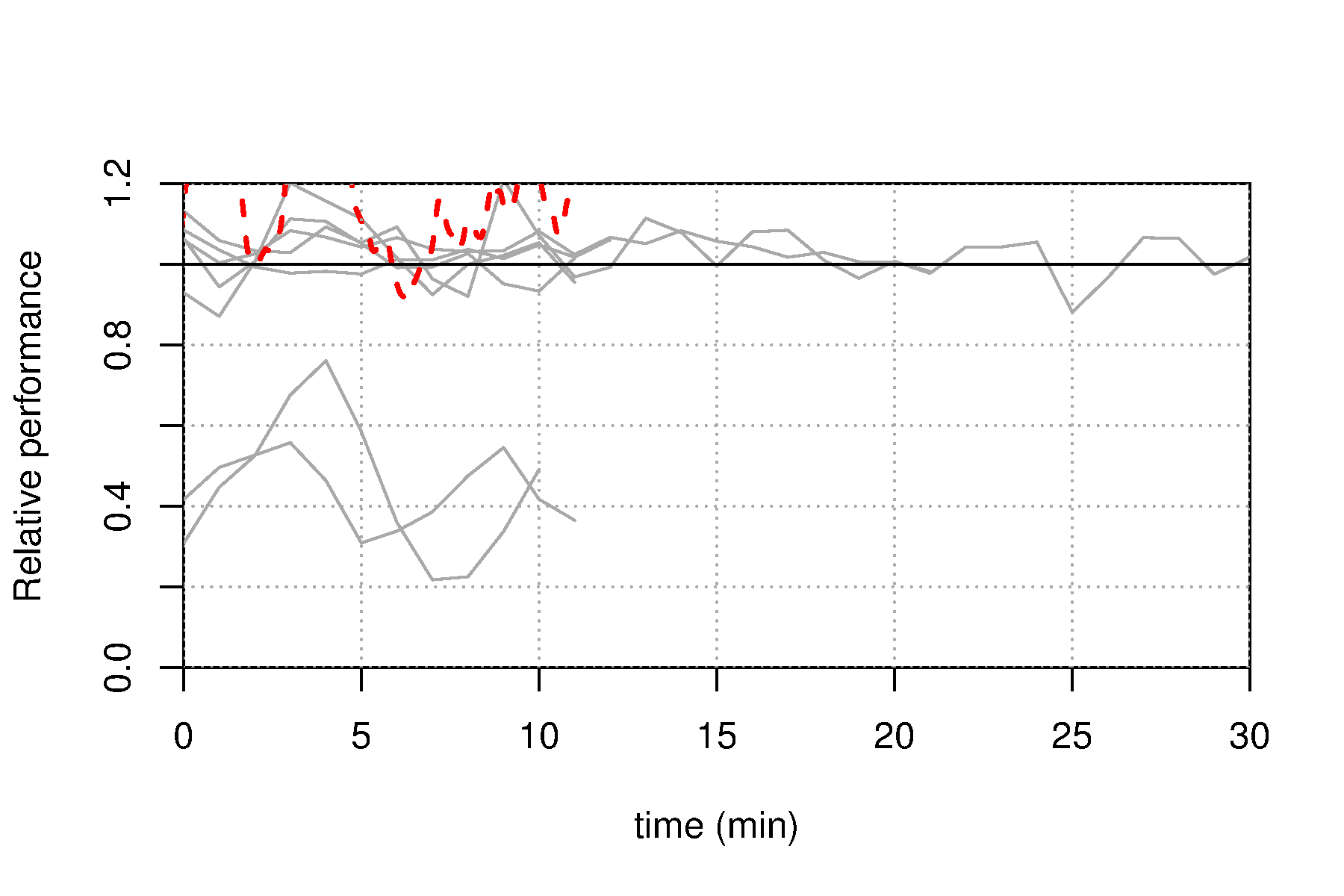} \hfill
\includegraphics[width = 0.45\columnwidth]{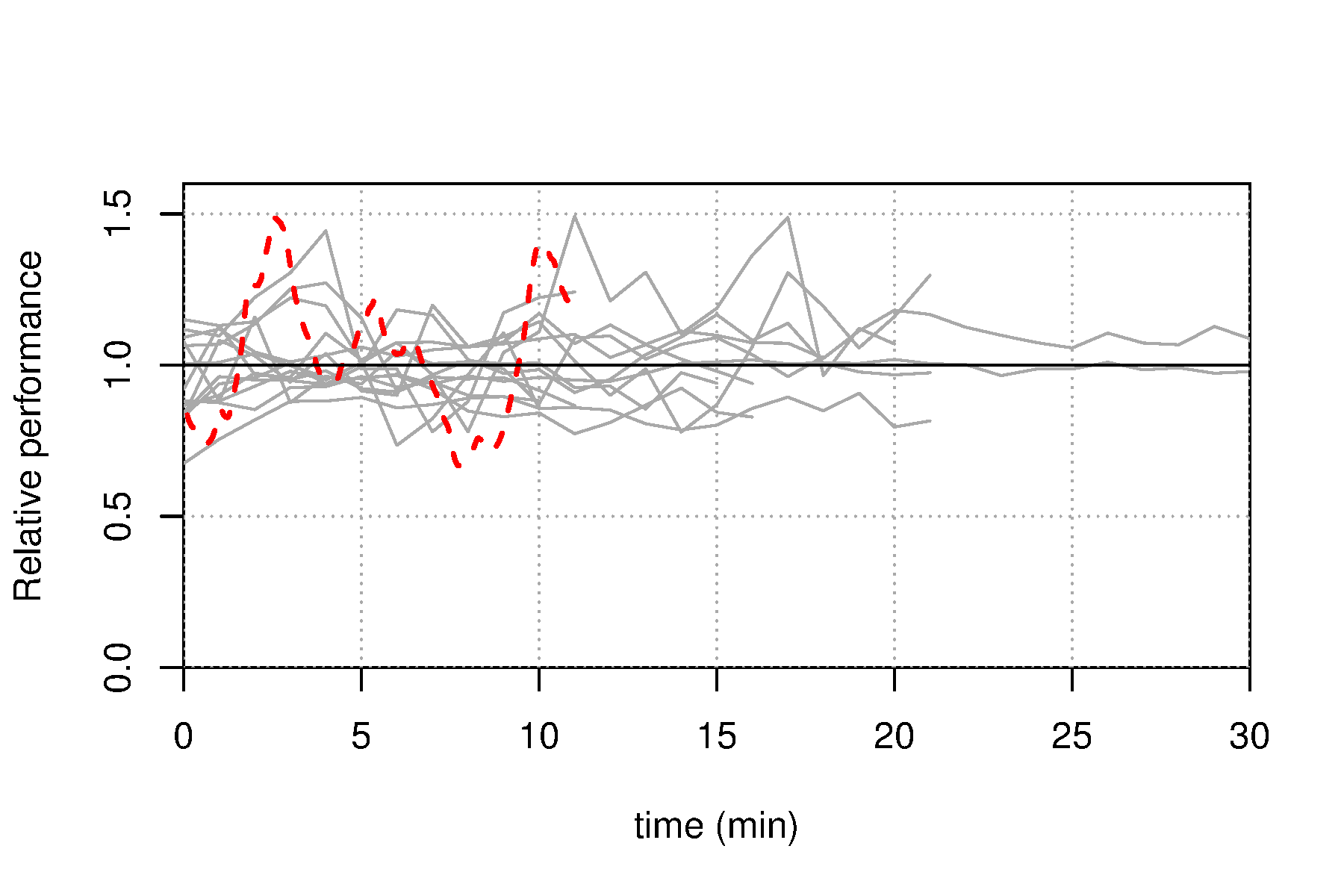} \\
c) \hspace{0.45\columnwidth} d) \\
\includegraphics[width = 0.45\columnwidth]{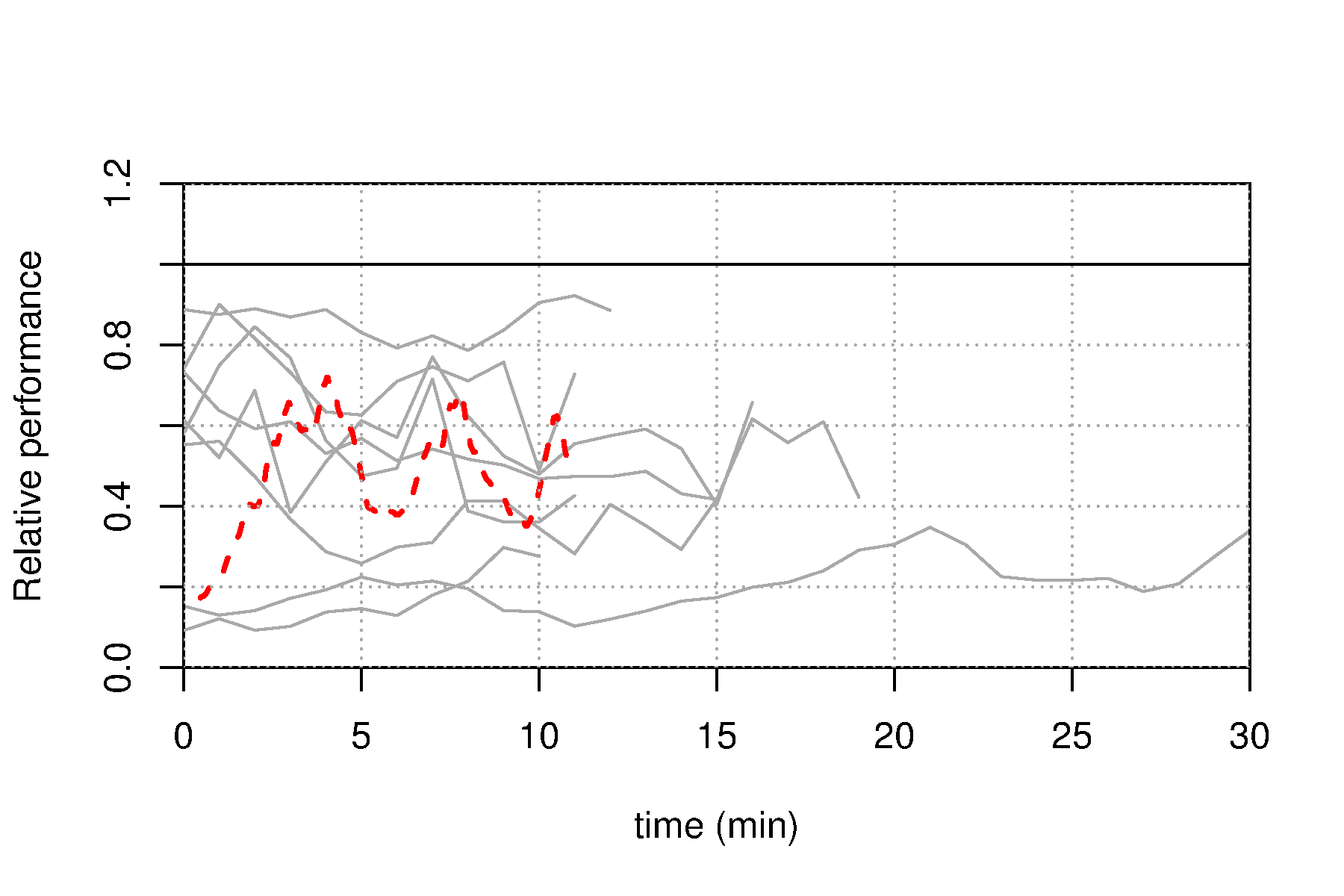} \hfill 
\includegraphics[width = 0.45\columnwidth]{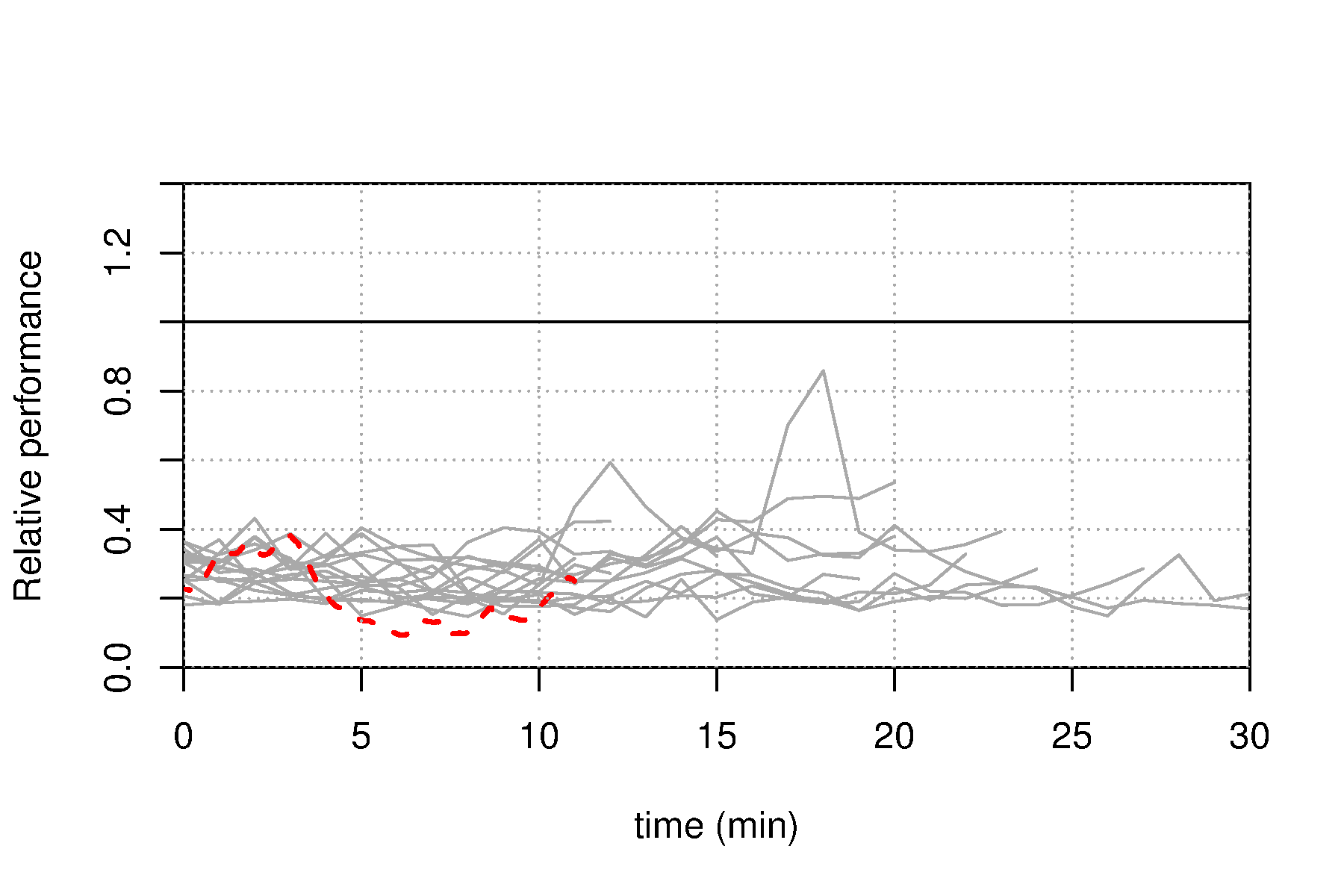} \\
\caption{Time series of the relative power output from turbine C07 from the SCADA data, where each grey line represents a valid observation period, and the computational model (red dashed lines) for the following wind directions: a) 198\degree, b) 207\degree, c) 217\degree, and d) 223\degree. }
\label{fig:Validation_T1}
\end{figure*}

The simulations with turbines covered a period of 20 minutes. The initial 10 minutes were considered a `spin-up' period, with fully-developed flow and stable turbine performance in the final 10 minutes.  This resulted in a section of equilibrated flow of 10 minutes which was used for the comparison with the SCADA data.  To ensure that the validation was based on truly comparable observed conditions, the comparison was made only with those sections from the SCADA data for which both, the wind speed was within the range of 5.5 to 10.5 m/s and the wind direction within a range of $\pm 1\degree$ either side of the wind direction (corrected for the 5.5\degree~bias) for a duration of at least 10 minutes.  The computational results and all corresponding valid observed periods are shown together for four representative cases of the eight wind directions in Figure~\protect\ref{fig:Validation_T1}.  Shown are time series of the relative power output from the numerical simulation as the dashed red line and each matching SCADA observation as a grey line.  While there were very few periods of the wind speed and direction remaining within the specific range for over 20 minutes, there are ten or more instances where the conditions were met for at least 10 minutes. 

In most cases, the computational results are well aligned with the ensemble of observations both, in terms of their time-averaged power output and the magnitude of their fluctuation around that mean.  In most cases, such as Figure~\ref{fig:Validation_T1}~(b) for 207\degree (and similarly for 202\degree~and 236\degree - not shown here) as well as for 223\degree , all valid SCADA episodes are very similar across each other as well as similar to the numerical simulation.

\begin{figure}
a) \hspace{0.45\columnwidth} b) \\
\includegraphics[width = 0.47\columnwidth]{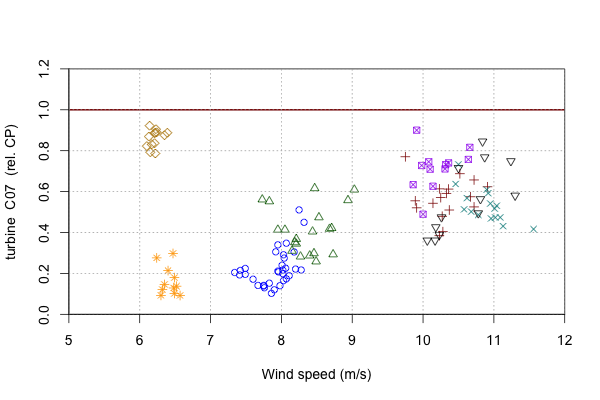} \hfill
\includegraphics[width = 0.47\columnwidth]{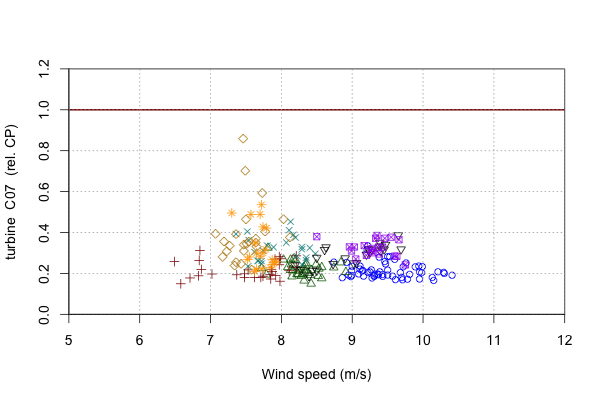}
\caption{Relative performance of turbine C07 against the wind speed measured by the front turbine, (a) wind direction 217\degree~and (b) 223\degree.  Each symbol/colour represents one of the episodes covering that wind direction. }
\label{fig:relP_check}
\end{figure}

In some cases, such as turbine C07 for 198\degree~in Figure~\ref{fig:Validation_T1}~(a), a few episodes were very different from the behaviour of all others or, as the same turbine for 217\degree~in Figure~\ref{fig:Validation_T1}~(c) (as well as 212\degree~and 229\degree), the observed episodes cover an extensive range without a clear representative behaviour.  One concern was that the spread of the mean relative performance was an artefact, caused by wrongly assuming that the plateau in the relative performance shown in Figure~\ref{fig:RelPerformance_U} did not hold for all wind directions. Although this assumption is clearly supported by Figures~\ref{fig:Validation_T1}~(b) and (d), we checked this assumption for all eight wind directions discussed here.  Figure~\ref{fig:relP_check} shows two examples of this check, in both cases for turbine C07, (a) for the most variable wind direction sector around 217\degree, corresponding to Figures~\ref{fig:Validation_T1}~(c), and (b) for the wind direction of 223\degree~when C07 is directly downstream of C08, corresponding to Figures~\ref{fig:Validation_T1}~(d).  The individual symbols show the relative performance of C07 against the wind speed. In the case of 217\degree, there might be at first sight a systematic change between wind speeds of 7~m/s and 11~m/s, but the two episodes with the lowest wind speeds cover the entire range observed in the power deficit.  Inspecting corresponding plots for other turbines and other wind directions did not support the evidence of any persistent systematic bias in the relative performance with wind speed, as illustrated by Figure~\ref{fig:relP_check}~(b).  As a result of this, we are confident that the relative performance within the investigated wind speed range is a robust performance indicator for wind farms, and that the variability observed in Figure~\ref{fig:Validation_T1}~(c) is caused by the actual variability of the flow induced by both, the inlet conditions and the upstream turbines.

An unusual case is Figure~\ref{fig:Validation_T1}~(a) for 198\degree, where most episodes are very close to each other at the 100\% mark except for two episodes which fluctuate around a level of 40\%.  Those two exceptional episodes are also those with the largest within-episode fluctuations.  For the computational results, we do not have any measure of variation between different realisations (initial conditions) which would be equivalent to the different episodes, but the mean power level is consistent with the actual observations.  Furthermore, the magnitude and time scales of the fluctuations found in the simulations results are consistent with those observed within the observed episodes.  The agreement in the mean performance confirms the initial model validation in Figure~\ref{fig:power-thrust-comparison} that the power extraction is modelled correctly in the turbine model.  The agreement in the time scale suggests that the relaxation time scales used in the control of the turbine parameters (mainly rotation rate and blade pitch) was appropriately set.

To demonstrate the correspondence between the computer simulations and the SCADA data, we will combine the box plots for the SCADA of relative performance (e.g.,~Figure~\ref{fig:RowBoxplots}) with corresponding plots from the model results.   In the comparison figures, the coxes and whiskers from the combined SCADA episodes are replaced by shaded regions indicating the interquartile range with dotted lines showing the median, while the results from the computer simulations are superimposed as standard box-and-whiskers plots.

\subsection{Validation Results}\label{S:Validation:Results}
\begin{figure}
\begin{centering}
$198^\circ$ \hfill ~\\
\includegraphics[width = 0.32\columnwidth]{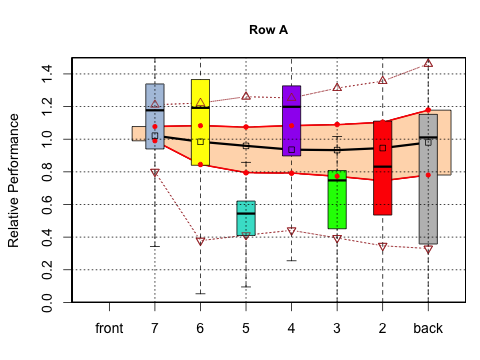} 
\includegraphics[width = 0.32\columnwidth]{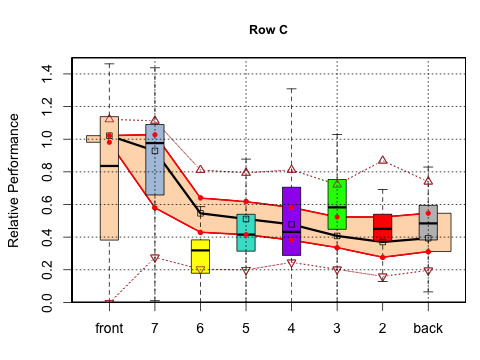} 
\includegraphics[width = 0.32\columnwidth]{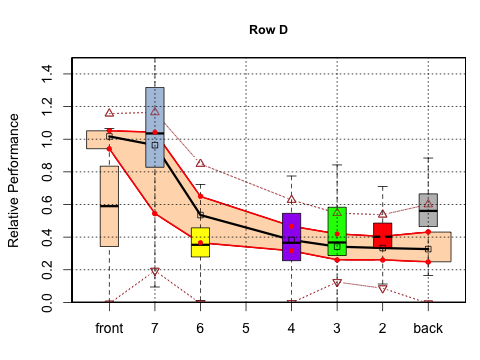} \\
$212^\circ$ \hfill ~\\
\includegraphics[width = 0.32\columnwidth]{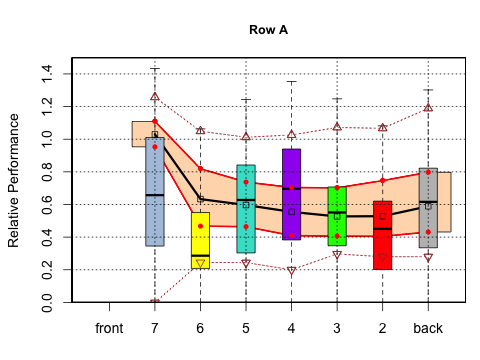} 
\includegraphics[width = 0.32\columnwidth]{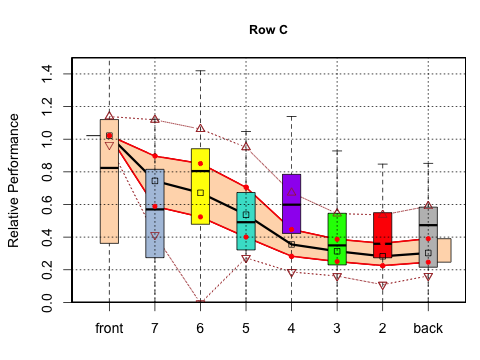} 
\includegraphics[width = 0.32\columnwidth]{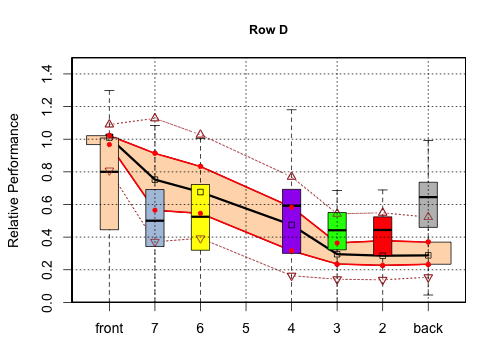} \\
$223^\circ$ \hfill ~\\
\includegraphics[width = 0.32\columnwidth]{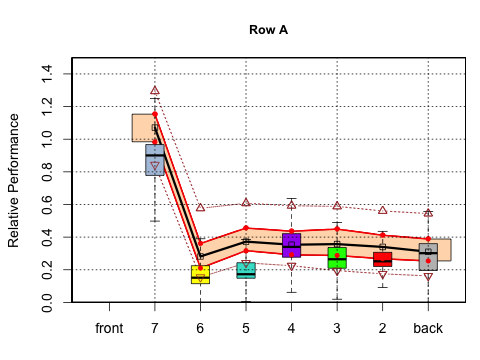} 
\includegraphics[width = 0.32\columnwidth]{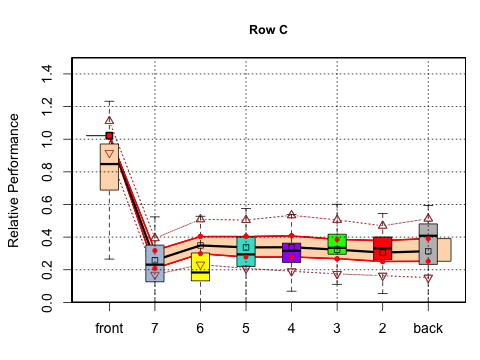} 
\includegraphics[width = 0.32\columnwidth]{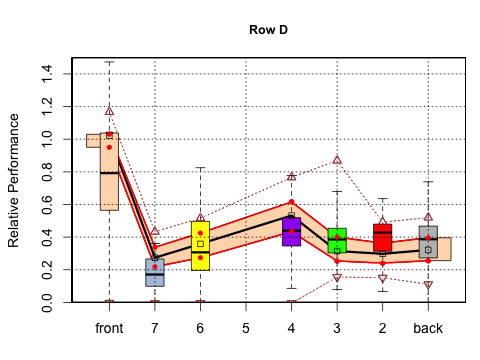} \\
$236^\circ$ \hfill ~\\
\includegraphics[width = 0.32\columnwidth]{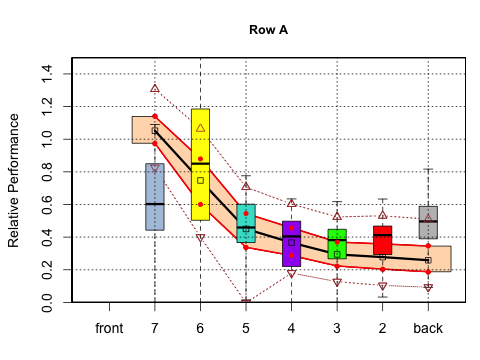} 
\includegraphics[width = 0.32\columnwidth]{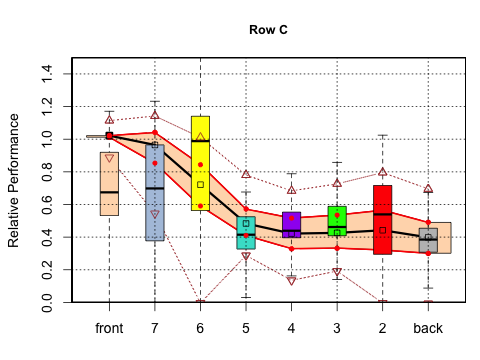} 
\includegraphics[width = 0.32\columnwidth]{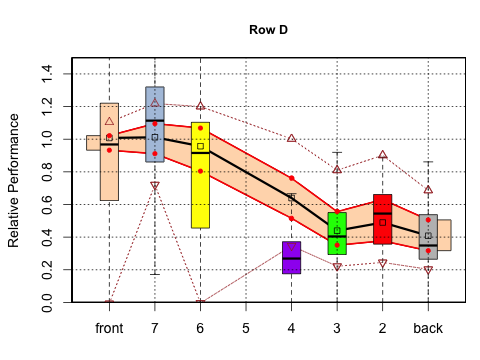} \\
\caption{Comparison of the observed relative wind turbine performance with that from the CFD simulations for the turbines in three selected rows (A, C, and D) at different wind directions (198\degree, 212\degree, 223\degree, 236\degree).  The shaded area indicates the two centre quartiles and the dotted lines the 5\% and 95\% quantiles.  The box plots show the quartiles from the equilibrated part of the CFD simulations. }
\label{fig:SC_RowBoxplots}
\end{centering}
\end{figure}

Figure~\ref{fig:SC_RowBoxplots} reproduces Figure~\ref{fig:RowBoxplots} of the relative performance for the three selected turbine rows A, C, and D where the SCADA data now are the lines and shaded regions in the background.  Superimposed are the CFD results as the box-and-whisker plots using the same colour convention as the original Figure~\ref{fig:RowBoxplots}.  An overview over the figure suggests that there is good agreement between observations and simulations with a few isolated discrepancies and very few systematic differences.  One consistent feature across all panels is that the front turbines, A07, C08 and D08, show a much larger range than the SCADA data suggest. The other consistent feature is that the back turbines, A01, C01 and D01, show in most cases a slightly better performance in the model than the observations.

At 198\degree,  the overall pattern of nearly 100\% performance in row A, and good performance from the first two turbines but reduced performance to around 40 - 60\% in rows C and D is well reproduced by the model, but the model shows substantial variation from the individual turbines in row A against the relative uniform observations from the SCADA data.  In rows C and D, the front turbines show differences in the mean performance although the ranges are very large, so that the CFD and SCADA are still consistent with each other. The main difference is in the substantially elevated performance of the second turbines, C06 and D06, in the model.  

At 212\degree, the model results are largely consistent with the SCADA results except for turbine D01.  Even though the ranges are very large for the simulations and the observations, the drop-off from the front turbine to the deep array appears to be faster in the model than the simulations as the median in boxes for the CFD results for all second-row turbines, A6, C7 and D7, is below the median from the observations.  The correspondence at 223\degree~is very good, but here one can also see that the second row turbines and to some degree the third appear to be reduced more strongly in the model than the observations.

\begin{figure}
\centering
\begin{minipage}{0.75\columnwidth}
a) A07 \hspace{0.4\columnwidth} b) C07 \\
\includegraphics[width = 0.48\columnwidth]{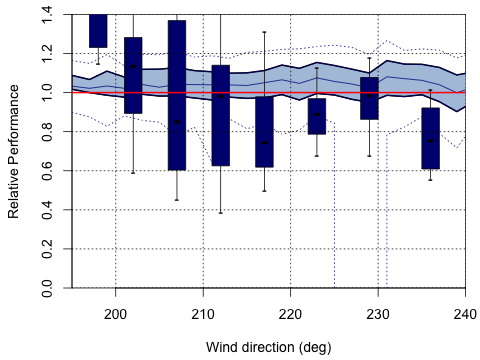} \hfill
\includegraphics[width = 0.48\columnwidth]{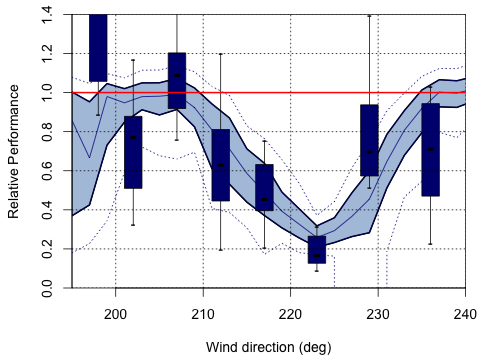} \\
c) A06 \hspace{0.4\columnwidth} d) C06 \\
\includegraphics[width = 0.48\columnwidth]{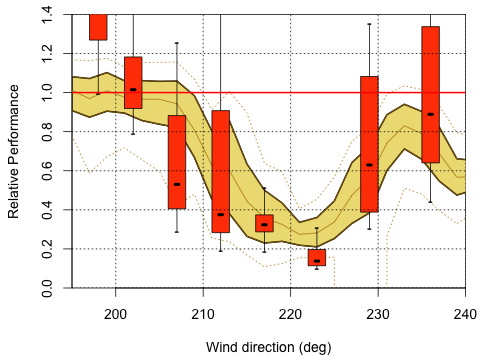} \hfill
\includegraphics[width = 0.48\columnwidth]{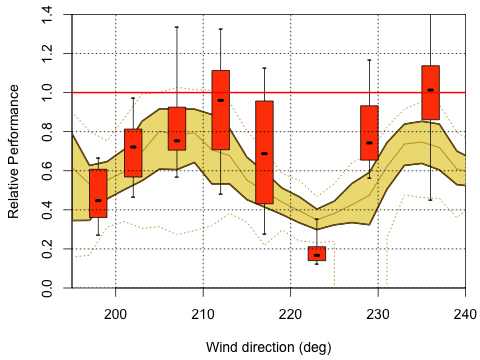} \\
e) A04 \hspace{0.4\columnwidth} f) C04 \\
\includegraphics[width = 0.48\columnwidth]{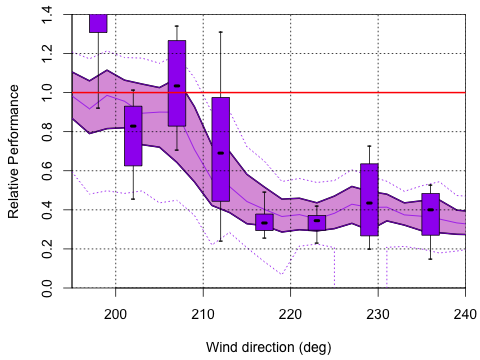} \hfill
\includegraphics[width = 0.48\columnwidth]{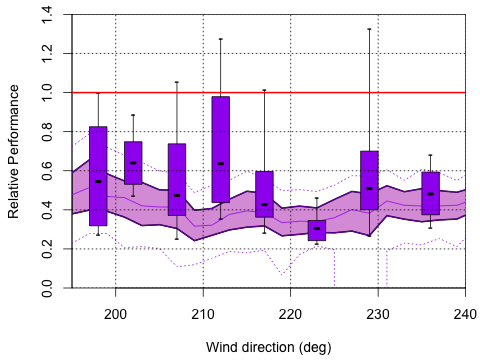}  \\
g) A01 \hspace{0.4\columnwidth} h) C01 \\
\includegraphics[width = 0.48\columnwidth]{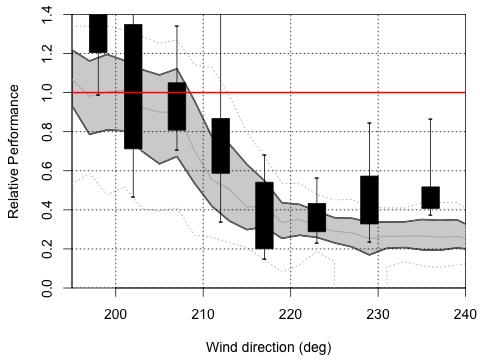} \hfill
\includegraphics[width = 0.48\columnwidth]{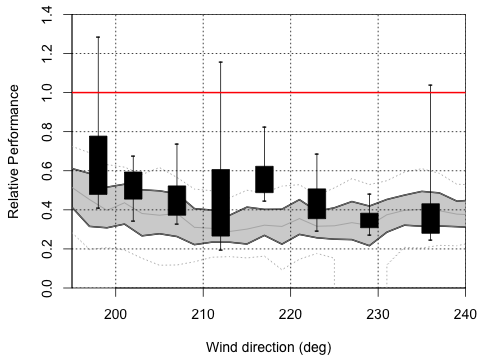}
\end{minipage}
\caption{Comparison of the observed relative wind turbine performance with that from the CFD simulations against wind direction for four selected turbines (01, 04, 06 and 07) in two selected rows (A and C).  The shaded area indicates the two centre quartiles and the dotted lines the 5\% and 95\% quantiles.  The box plots show the quartiles from the equilibrated part of the CFD simulations. }
\label{fig:Validation_T2}
\end{figure}

\begin{figure}
\begin{center}
\includegraphics[width = 0.6\columnwidth]{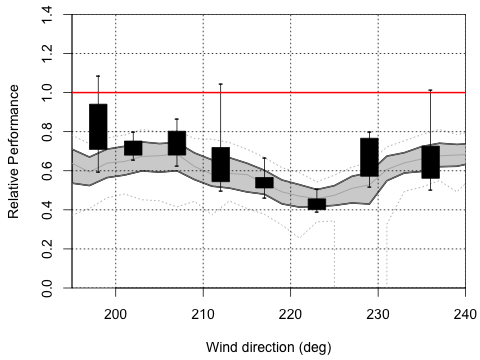}
\end{center}
\caption{Comparison of the observed relative wind farm performance with that from the CFD simulations against wind direction.  The shaded area indicates the two centre quartiles and the dotted lines the 5\% and 95\% quantiles.  The box plots show the quartiles from the equilibrated part of the CFD simulations. }
\label{fig:ValFarm}
\end{figure}

Changing perspective from the response of a row of turbines for a specific wind direction to the response of a single turbine against changing the direction, we turn to Figure~\ref{fig:Validation_T2} for which we have selected four turbines each from rows A and C\@.  The structure of the figures follows the previous convention that the shaded areas show the interquartile ranges of the SCADA data, while the box-and-whisker plots represent the quartiles from the CFD simulations.  As above, there are cases where the agreement between observations and simulations are extremely good but also some where there are substantial difference.  The first impression is that the wind direction of 198\degree~ shows substantial differences between observations and simulations in all eight panels.  Putting that aside, the overall pattern of variation appears to be well captured by the model.   In addition to the overall performance against wind direction, the model also appears to generate a larger variability (larger boxes) in places where the observed range in the SCADA data is also wider.  Aggregating all turbines into the total wind farm output is shown in Figure~\ref{fig:ValFarm} against wind direction.  Except for the unusual case of 198\degree, the agreement between model and SCADA data is very good to a degree where the boxes from the model overlap substantially with the shaded region from the SCADA data.

\begin{figure}
a)\hspace*{0.3\columnwidth} b)\hspace*{0.3\columnwidth} c) \\
\includegraphics[width = 0.3\columnwidth]{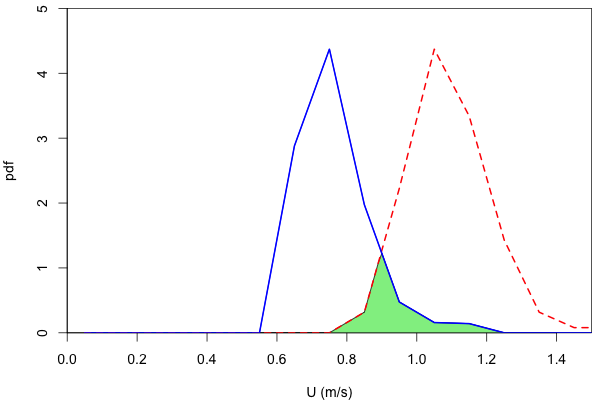}
\hfill
\includegraphics[width = 0.3\columnwidth]{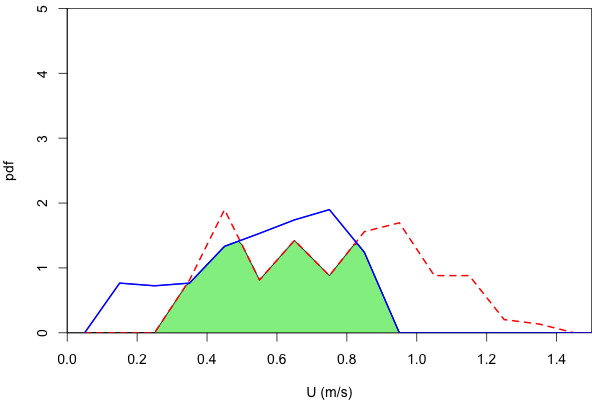}
\hfill
\includegraphics[width = 0.3\columnwidth]{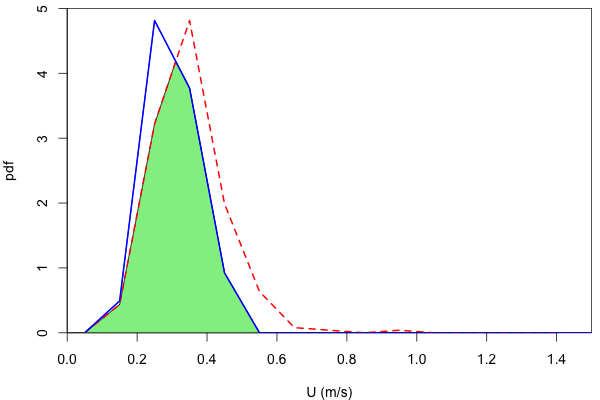}
\caption{Probability density function of relative performance from model (solid blue line) and SCADA data (dashed red line).  The common area is shaded. (a) for turbine A07 at 236\degree, b) turbine C07 at 212\degree, and c) turbine C05 at 223\degree. }
\label{fig:Validationpdf}
\end{figure}

To quantify the agreement between model and observations, we can calculate the area under the normalised distribution of the model performance of a particular turbine from a selected model integration shared with the distribution from the corresponding SCADA events.  This is illustrated in Figure~\ref{fig:Validationpdf} for three representative cases.  In Fig.~\ref{fig:Validationpdf}~(a) both show a relatively narrow range around the mean performance but at different levels. As a result, the common area is only 15\% of the area of each of the two distributions.  In Fig.~\ref{fig:Validationpdf}~(b) both show a broad distribution around somewhat different mean values, and the overlap is 68\%.  In the last example, the distributions are very close, with an overlap of 87\%.

\begin{figure}[htp]
\centering
\includegraphics[width = 0.50\columnwidth]{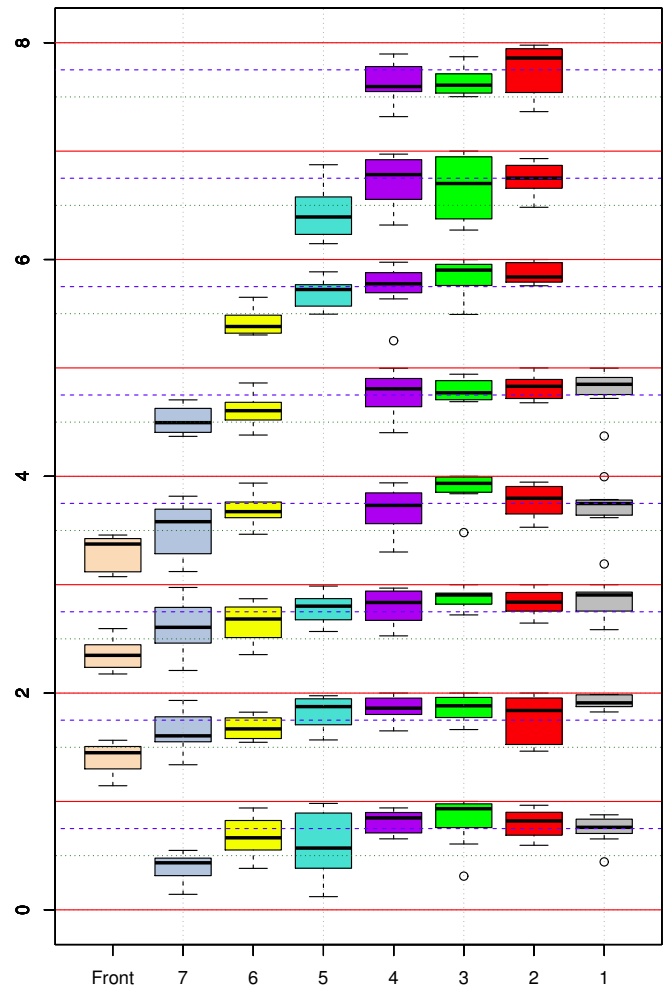}
\caption{Overview over agreement between CFD results and SCADA data for each turbine over the eight wind directions analysed.  The $y$-axis shows the percentage agreement as the remainder, e.g.~1.51 corresponds to an agreement of $51\%$ for the turbines in row B\@. A level of 50\% agreement is indicated by the green dotted line and 75\% agreement by the dashed blue line.}
\label{fig:summary_t}
\end{figure}

Aggregating the overlap for all wind directions into Figure~\ref{fig:summary_t} shows the overlap for each turbine.  Overall the agreement of the wind farm performance as calculated by the model compared to the selected SCADA data, using this method is $70\% \pm 20\%$.  From Figure~\ref{fig:summary_t} it is clear that the turbines exposed to the free stream show least agreement.  The agreement of the turbines from the third row onwards is as high as $78\% \pm 18\%$.

One possible cause could be the fact that only the wind speed and direction could be determined, but not the atmospheric stability or the freestream turbulence intensity.  Given the observed spread of wind shear exponents (cf~Fig~\protect\ref{fig:Rose}b), one would expect a larger range in the agreement score rather than a systematically reduced agreement.  Similarly, unless the turbulence intensity inlet conditions, which were chosen as typical for these latitudes, were systematically different from the actual ones, one would not expect this systematic difference.  A further possible cause for the mismatch between model and observation in the front turbines could be the fact that we compare instantaneous results with a 0.5 s sampling rate with 1-minute averages.  This is consistent with \citet{Poulsen2012}, who noted that increasing the averaging window eliminated local turbulence and wake meandering to a degree, producing results closer to those from standard engineering wake models. If this is the case, then it appears that the enhanced mixing and the establishing of a deep-array wake act to smooth out individual large features, so that the behaviour in the deep array tends toward more uniform flow, which is equally well described by a high time resolution or by time-averaged data. Considering that the inlet turbulence characteristics were chosen carefully to result in both a typical value at the wind farm location and realistic wake recovery, it is unlikely that the inlet turbulence conditions would result in this mismatch.

\section{Discussion}
\label{S:Discussion}

Many turbine models exist and have been applied to wind farms using computational fluid dynamics  \citep{Churchfield2012, calaf2010, BaPretal09, MCJGM07}. The model presented here differs from most in one significant way, in that the lift and drag generated by the turbine blades simultaneously apply torques to the generator, the blades, and the air which flows through the actuator volumes. This dynamic, reactive model of the turbine allows us to study deep-array wake effects in wind farms using a more physically accurate model than commonly-used methods, which rely upon estimations of the upwind wind speed to directly calculate the backthrust \citep{prospath2009}, or to calculate the angle of attack for rotors turning at a calibrated rate of rotation \citep{calaf2010, meyers2010}. Through use of Large Eddy Simulation (LES) it also permits the study of unsteady, turbulent flow effects within the wind farm, which we have shown to be a key driver in the performance of the Lillgrund model. 

The alternative of RANS, and especially unsteady RANS (URANS), has been utilised in wind turbine models elsewhere \citep{BaPretal09, MCJGM07, elkasmi2008}, but excessive wake diffusion is an issue \citep{sanderse2011, sumner2010}, and the applicability of techniques limiting this turbulent diffusion at the rotors \citep{elkasmi2008} for multiple turbines has been called into question by \citet{rethore2009}. This is problematic for wind farm modelling, as  blade-generated turbulence plays an important role in deep-array wakes. The model used here uses a previously validated technique \citep{Creech2009, CFC11} to effect blade-generated turbulence with LES CFD and, as can be seen in this paper, its effectiveness has been vindicated by the power recovery in downwind turbines. Particularly in the extreme case of the wind direction 223\degree, we can see that turbines in the second row produce low levels of power, but in the third and fourth rows we see power recovery, due to the increased wake mixing due to rising levels of turbulence within the wind farm.

The Synthetic Eddy Method (SEM) used here allowed the characteristics of the atmospheric turbulence being fed into the model at the inflow boundary to be finely tuned, and these were varied with height according to Danish turbulence standards \citep{windenergyhandbook}. The SEM boundary conditions turned out to be a secondary, but also important, source of turbulence for the model farm. In initial tentative simulations too little turbulence was fed into the model, which resulted in poor downwind wake recovery; only when the correct turbulence statistics were applied, did the model produce deep-wake effects on turbine performance which matched the SCADA data. This suggests that while the blade-generated turbulence is important for wake mixing and dissipation, so too does the atmospheric turbulence at longer length scales (10-150~m as opposed to $<$~5~m). Indeed, it is the combination of these two that produces the levels of mixing and recovery within the simulation.

Validation of the wind farm model against observations was challenging, as the model resolves time scales not accessible from available measurements.   While SCADA data may be available at a high sampling rate, the same will never be true for the required boundary conditions.  As a result it is not possible to truly reproduce a computer solution of an actual, observed situation, and one has to resort to modelling a set of typical cases and compare these with as many appropriate observations as possible.  Given the nature of atmospheric flows, it is possible that cases with similar wind conditions may lead to locally very different flows and turbine responses within the wind farm, as seen in Figure~\ref{fig:Validation_T1}(c).  Due to the computational expense of CFD, however, it is impractical to explore all possible solutions, and the solution obtained from computer simulations must be evaluated to how well it matches the distribution of possible solutions. To achieve this, we chose a small number of possible wind conditions, covering a set of wind directions which represent key geometric relationships between the upstream wind direction and the turbine positions relative to that direction.  To capture a sufficient number of observations corresponding to the simulation, cases were selected from the wind speed range between the cut-in and rated wind speeds, over which the normalised power output appeared to be constant.

Despite this careful selection, there was still the challenge to compare an ensemble of time-average observations with a single realisation of a flow sampled at a high temporal resolution.  For example, 10-minute averages presented for Horns Rev~\citep{Gaumond13} showed that the wake effect was apparently much less pronounced than predicted by standard engineering wake models when analysed over a narrow 5\degree~wind direction.  In contrast, the wake models gave extremely good results to the observations when the results were averaged over a 30\degree~($\pm 12\degree$) wind direction sector.  Their effect is also apparently much less pronounced than our observations presented here, with a power deficit for the second turbine at around 65\%, which gradually but continually decreased to around 55\% for the last turbine in the row \citep[Fig.4]{Gaumond13}.  The mismatch between their observations was attributed to uncertainties in the wind direction due to bias in the sensors as well as spatio-temporal variation.   Considering that the 10-minute averaging of the data is equal to the residence time $U/L$ ($U$ the wind speed and $L$ the length of the wind farm), this averaging will smooth out any local features within the farm, and the results would indeed be expected to conform to a broader selection of inflow situations as represented by the wider 30\degree-sector.  In our case, the residence time is 4 to 5 minutes while the SCADA data have a 1-minute SCADA resolution.  With that, there will be some spatio-temporal averaging of individual flow features noticeable but there should also be evidence of the larger of the features be visible in the data.

The substantial fluctuations demonstrated by the simulations have been observed around wind turbines by~\citet{HirthSchroeder13}.  These features include strong wake meandering, breaking-up of atmospheric eddies moving into the array, and jetting between turbines. As these features are resolved within the model, results at any time may vary considerably from the more uniform flows found in RANS CFD simulations or time-averaged observations.  A comparison between the simulation results and the corresponding SCADA data confirms this especially for the first three rows of turbines. In the deep array, the agreement between LES CFD and time-averaged observations is much better, which suggests that the turbulent mixing provided by the turbines is very effective in destroying larger coherent flow structures, while enhancing the more isotropic smaller scale turbulence.

It is also possible that the appropriate modelling of the three apparent sources of turbulence for such CFD modelling was an essential component of capturing the very strong power deficit of the turbine fully in the wake of the front turbine.  In the model, the sources of turbulence are the drag from the water surface, the turbulence created at the turbine rotor, and the free-stream turbulence advected into the domain by the SEM inflow conditions.  This means that the air flowing into the front turbine is relatively clean, only with the upstream turbulence consisting more of larger eddies and with relatively weak turbulence generated at the surface and transported upwards.  The flow structure behind that turbine is then a turbulent wake expanding in fairly quiescent air, apart from the surviving large atmospheric eddies. Therefore the wake recovery is relatively low, given the low drag coefficient of the sea surface~\citep{CFC11}, and the wake deficit is still substantial at the point of the second turbine. This second turbine generates another wake which is now located within the decaying wake of the upstream turbulence, and that latter turbulence helps to mix the wake and tip vortices more rapidly, which then leads to a wake less deep at the location of the third turbine.  Beyond those turbines, the flow becomes more and more uniformly mixed as the newly formed wakes mix with the existing turbulent wakes.

\section{Conclusions}\label{S:Conclusion}

In this paper, we have presented the use of dynamically active models of wind turbines embedded in a high-resolution Large-Eddy Simulation CFD model of the environment, with appropriate extent and resolution to represent both the response of wind turbines to the atmospheric flow, and the effect of a large array of wind turbines on the flow. The main aim of this work was to demonstrate a validation of this modelling approach as a valid tool to investigate the interaction between wind farms and the environment.  Key requirements for this were: a) to describe a turbine based on its rotor design and key operating controls of blade pitch and rotational speed, b) to simulate the response of these turbines, in a way which follows the control mechanism of actual turbines when they are placed in a naturally fluctuating wind, and c) to describe the resulting wake and its recovery within and around the wind farm. The turbine validation in section~\ref{S:Turbine-validation} has demonstrated that a model of the turbine based on best estimates of the rotor blades and the design rotor rotational speed generates a power output and thrust coefficient for such a model in 'clean' reference wind conditions produce a very good agreement with manufacturer's data and observations.  

The Lillgrund model has shown that it is possible to simultaneously resolve flow features in three dimensions over a wide range of scales, from 5~m at the rotor, to large-scale atmospheric eddies and wind farm wakes several km in length. Through its coupling of LES to the dynamic turbine model, the performance of turbines has been shown to fluctuate in response to local flow conditions, and that the turbulent flow generated by the turbines reflects that found in real turbines~\citep{HirthSchroeder13}. Our model has compared well to the SCADA data from the real wind farm; where it has not can perhaps be attributed to four reasons: i) insufficient information regarding meteorological conditions at Lillgrund, such as temperature and humidity, ii) the assumption of a neutrally stable atmosphere, iii) too short a simulation period for accurate performance statistics, and iv) the limitations of the actuator disc in the near-wake. While this could be addressed in principle by acquiring more data from the wind farm to ensure that observations and model represent the same flow conditions, the latter three are model constraints. In particular, the last reason  demands considerably more computational resources. In their simulation of Lillgrund, \citet{Churchfield2012} employed actuator lines \citep{sorensen2002} to represent the turbine blades, which require much higher resolution meshes near the rotors. In contrast to the $40\,\textrm{km}^3$ simulation domain here, which contained 30 million elements running on 256 computer cores, Churchfield used 315 million cells running across 4096 cores, for a domain less than half the size at $16\,\textrm{km}^3$. 
Finer meshes and the actuator line approach have the undoubted ability to better resolve near-wake features than we do here, but to decide whether the increased fidelity in the near-wake is significant enough to merit the trade-off in computational effort, a detailed study of both actuator disc and an actuator line wind farm models is required. The key next step, however, is to address the second point about the atmospheric condition.  At the time this work was carried out, computing power allowed either a substantial horizontal extent, covering a significant part of the farm wake as chosen here, or a substantial vertical extent, covering a significant part of the atmospheric boundary layer as chosen by \citet{Archer2013}.  Computing power has progressed to a degree where our methodology can be applied to a larger domain covering both the horizontal extent to resolve the farm wake, and the vertical extent to cover the unstable atmospheric boundary layer.  While buoyancy effects were not considered here, following the choice to start with a neutrally stable atmosphere, the LES CFD software used here has already been used for convective flows \citep{mactavish2013}, and allows to incorporate convectively unstable conditions as a refinement to the methodology rather than a step change.

Now that the methodologies for wind farm characterisation are validated, with fully resolved atmospheric boundary layer and convective processes included, it is possible to apply this modelling methodology to both engineering and atmospheric sciences applications.

In the engineering context, large scale simulations of wind farms are now practical using state-of-the-art computational fluid dynamics, and can be used to inform wind farm design. Once a turbine has been parameterised, it can be placed in a variety of real or imagined scenarios. The modelled turbines can be turned off to simulate failure, and the impact on surrounding turbines can be examined. Will the turbines downwind experience greater power output? Do they experience higher levels of turbulence? Such questions could be applied to control strategies, optimising the balance between turbine loading and maintenance costs on one side, and overall energy yield on the other. Alternatively, by adding, removing or moving turbines, we could alter array layouts, and observe the resulting change in power output and electricity yield of the simulated farm.

In the atmospheric science context, this modelling methodology will allow for a full dynamic simulation of the interaction between a large wind farm and the atmospheric boundary layer at horizontal length scales of tens of kilometres, whilst resolving the key length scales of the fluid-rotor interactions without requiring excessive computing power. Through the multiscale wind farm modelling shown in this paper, we can investigate the transport and decay of turbulence induced by the turbines, the wind farm wake dynamics and decay, and the impacts of large wind farms on local weather as much as for parameterisation in climate models.
\section*{Acknowledgements}
We would like to thank Vattenfall AB for funding this research, along with Adrian Jackson at EPCC and Ian Chisholm at the Institute of Petroleum Engineering, Heriot Watt, for technical support that made this project possible.

\bibliography{Wind}
\bibliographystyle{spbasic}

\end{document}